%% file: main.tex
\documentclass[preprint,10pt,a4paper]{elsarticle}

\usepackage{preamble}

\newcommand{\gridrowsep}{\par\vspace{-4pt}}

\begin{document}

\begin{frontmatter}

\title{Efficient regime boundary detection in discontinuous simulation models}

\author[a]{Federico Betti\corref{cor1}}
\ead{bettif@ethz.ch}

\cortext[cor1]{Corresponding author}

\author[a]{Eleni Chatzi}
\ead{chatzie@ethz.ch}

\author[b]{Paolo Tiso}
\ead{ptiso@ethz.ch}

\address[a]{Department of Civil, Environmental and Geomatic Engineering,
ETH Zürich, Wolfgang-Pauli Strasse 5, 8049, Zürich, Switzerland}

\address[b]{Department of Mechanical and Process Engineering,
ETH Zürich, Leonhardstrasse 21, 8092, Zürich, Switzerland}

\begin{abstract}

\input{sections/abstract}

\end{abstract}

\begin{keyword}
regime boundary detection, discontinuous simulator output, adaptive sampling, support vector machines, Hamiltonian Monte Carlo
\end{keyword}

\end{frontmatter}

%%%%%%%%%%%%%%%%%%%%%%%%%%%%%%%%%%%%%%%%%%%%%%%%%%%%%%%%%%%%%%%
\section{Introduction}

\input{sections/intro}

%%%%%%%%%%%%%%%%%%%%%%%%%%%%%%%%%%%%%%%%%%%%%%%%%%%%%%%%%%%%%%%%%%%%%
\section{Background} \label{sec:background}

\input{sections/background}

%%%%%%%%%%%%%%%%%%%%%%%%%%%%%%%%%%%%%%%%%%%%%%%%%%%%%%%%%%%%%%%%%%
\section{Active learning for excursion set estimation} \label{sec:al-lse}

\input{sections/al_lse}

%%%%%%%%%%%%%%%%%%%%%%%%%%%%%%%%%%%%%%%%%%%%%%%%%%%
\section{Our proposed approach} \label{sec:ingredients-proposed}

\input{sections/proposed}

%%%%%%%%%%%%%%%%%%%%%%%%%%%%%%%%%%
\section{Numerical examples} \label{sec:examples}

\input{sections/examples}

\section{Conclusions} \label{sec:conclusions}

\input{sections/conclusions}

\section*{Acknowledgments}

The authors acknowledge the funding of the ETH Center for Sustainable Future Mobility project “ESTRA -- Efficient and Safe Train Dynamics”,
within the context of ETH Mobility grants. Federico Betti acknowledges Dr. Nora Lüthen
for fruitful discussions at the conceptualization stage of this work.

\bibliographystyle{unsrt}
\small
\bibliography{references}

\end{document}

%% file: sections/abstract.tex
\noindent Many computational models exhibit multiple operating regimes separated by abrupt or non-smooth transitions as the input parameters vary. Such regime transitions complicate the construction of global parametric reduced-order models, often necessitating large training datasets to capture the distinct system behaviors. Accurate regime boundary detection instead enables localized modeling strategies by partitioning the parameter space into regions of qualitatively distinct behavior. In many cases, regime switching corresponds to the excursion set of a simulator output. When such a quantity is discontinuous or non-smooth, regression-based algorithms for level set estimation can struggle to correctly approximate the boundary. In this work, we show that solving a resulting binary classification problem can improve sample efficiency and we further propose a novel adaptive sampling strategy for support vector machines. Unlike uncertainty-based sampling schemes, the proposed approach uses a continuous search space and efficiently explores the currently estimated boundary in its tangential directions by gradient-based dynamic sampling. Because the decision function admits analytical gradients, this exploration incurs negligible computational overhead relative to the underlying model evaluations. Then, a space-filling subset of the generated samples is selected as the next batch to be queried. We demonstrate improved sample efficiency compared to existing active learning algorithms on analytical and numerical examples. The proposed method provides a sample-efficient framework for regime boundary detection in a broad class of discontinuous simulation-driven problems.

%% file: sections/intro.tex
Abrupt changes in the dominant system dynamics arise in a wide range of parameter-dependent computational models, ranging from fluid dynamics \cite{crawford1991symmetry, dijkstra2014numerical} and fluid-structure interaction \cite{khamlich2022model}  to neurosciences \cite{milton1989complex, izhikevich2007dynamical}, as well as systems with friction \cite{li2004bifurcation, ivanov2009bifurcations}. A sudden change in the dominant features of the dynamics poses challenges to the efficient construction of both intrusive and non-intrusive reduced-order models for downstream tasks such as optimization, control, and uncertainty quantification. A large class of projection-based and non-intrusive reduced-order models relies on low-dimensional representations extracted from high-fidelity solution data, with Proper Orthogonal Decomposition (POD) being one of the most widely used approaches. In such a case, the reduced basis may be constructed for each training sample in the parameter space or globally over the entire parameter domain. In the former case, standard manifold-based interpolation schemes \cite{amsallem2008interpolation} become ill-conditioned when large principal angles arise due to abrupt changes in the dynamics; in the latter case, the number of POD modes required to capture dominant dynamics across regimes may be very large, making the application of hyper-reduction techniques particularly challenging \cite{grimberg2021mesh, hernandez2024cecm} and more generally hindering the efficiency of the reduced-order model.

To address these limitations, when multiple regimes coexist within a parametric space, treating each regime separately enables the construction of accurate and efficient localized surrogate or reduced-order models. Beyond enabling regime-specific models, inferring the regime boundary is of standalone interest, as it can provide valuable insight into the underlying system behavior for safety assessment or reliability analysis. Central to this divide-and-conquer type of approach is a \,\emph{discriminant} that predicts, a priori, the regime associated with a given set of input parameters, thereby partitioning the parameter space into regime-specific regions. In many real-world engineering systems, the regime transition occurs across a codimension-one surface in parameter space, corresponding to an observable or function thereof exceeding a critical threshold value. In correspondence of such a critical value, the system response switches between qualitatively distinct behaviors. For example, in aeroelasticity, combinations of flow and structural parameters for which the aerodynamic damping crosses zero define a boundary separating stable from unstable dynamics.

In practice, this separating surface is rarely known in closed form and must be inferred from experimental observations or high-fidelity simulations. When individual model evaluations are computationally expensive, exhaustive sampling of the parameter space is infeasible, making the accurate reconstruction of the regime boundary from a limited number of queries a central challenge. This motivates adaptive sampling strategies that concentrate new evaluations in regions that are most informative for identifying and refining the separating surface, rather than relying on static space-filling designs.

In this work, we consider a computational model that depends on input parameters $\mu \in \mathcal{D} \subset \mathbb{R}^n$, where $\mathcal{D}$ is a bounded domain, and a scalar simulator output $g: \mathcal{D} \mapsto \mathbb{R}$ that exhibits discontinuities or non-smoothness with respect to the parameters. We are interested in efficiently approximating the boundary $\partial \mathcal{D}_\theta^+$ of the excursion set
\begin{equation}
\label{eq:boundary-definition}
    \mathcal{D}_\theta^+ = \{ \mu \in \mathcal{D}: g(\mu) \geq \theta \},    
\end{equation}
for a prescribed threshold $\theta \in \mathbb{R}$ that determines the regime boundary. When $g$ is sufficiently regular, the boundary of the $\theta$-excursion set corresponds to the $\theta$-level set of $g$. A common approach to level set estimation is then to construct a regression surrogate for $g$, often using Gaussian processes (GPs), and exploit the associated uncertainty estimates to guide adaptive sampling. When $g$ is continuous but non-smooth (for example, when it features a kink), accurately reconstructing the response and its level set may require a large number of model evaluations. In the presence of discontinuities, the difficulty is more fundamental, since the boundary of an excursion set need not coincide with the corresponding level set of $g$; indeed, the latter may be empty, despite a well-defined transition between regimes. Moreover, GPs with smooth kernels tend to regularize discontinuities and sharp variations, potentially leading to over-smoothed predictions and inaccurate localization of the regime boundary.

These considerations motivate addressing regime boundary detection directly as a classification problem rather than through regression of the quantity of interest of the underlying computational model. Accordingly, in this work we formulate the identification of $\partial \mathcal{D}_\theta^+$ in \eqref{eq:boundary-definition} as a binary classification problem and approximate the separating surface using the decision function $f(\mu)$ of a support vector machine (SVM).

Building on this classification formulation, we propose a novel strategy for active learning of $f(\mu)$: unlike standard margin-based sampling schemes that select informative points from a discrete candidate pool based on their proximity to the current decision boundary, we consider a continuous search space for the next points to be queried. This is achieved by using $f(\mu)$ to naturally induce a potential energy proportional to $f^2(\mu)$, that is minimized on the $(n-1)$-dimensional decision surface $\{ \mu \in \mathcal{D}: f(\mu) = 0 \}$; the currently estimated boundary can then be efficiently explored in its tangential directions using Hamiltonian Monte Carlo. While this requires gradient evaluations of the potential, it remains computationally cheap owing to the sparse representation of support vector machines and the availability of analytical gradients. Subsequently, we select a space-filling subset of the generated samples, naturally promoting underrepresented parts of the estimated boundary and improving the global approximation of the separating surface. This mitigates the sample redundancy that is typical of margin-based schemes and improves sample efficiency; we show in particular that the gains are increasingly pronounced for boundaries with higher curvature and in higher-dimensional parameter spaces.

The remainder of this manuscript is organized as follows. In Section \ref{sec:background} we clarify the notation, give the problem statement and review existing adaptive sampling and localized approaches for surrogate and reduced-order models. In Section \ref{sec:al-lse} we first introduce active learning strategies for level set estimation using Gaussian processes; we then turn to support vector machines by motivating their use in the discontinuous model response setting and reviewing the corresponding active learning techniques. Section \ref{sec:ingredients-proposed} describes the algorithmic building blocks of our proposed approach; in this section, we detail the boundary exploration mechanism and compare it with other popular choices, and we show on an analytical example that the additional computational overhead for sample generation is negligible for real-world scenarios where labeling cost dominates. In Section \ref{sec:examples} we assess the performance of the proposed algorithm. First, we show that it often provides improved sample efficiency, compared to other active classification algorithms, on a variety of analytical examples; second, we consider increasingly complex simulation models with discontinuous response, where we show the failure mode of regression-based algorithms and that our approach generally yields superior results. Finally, in Section \ref{sec:conclusions} we draw conclusions, discuss strengths and limitations, and highlight future directions.

%% file: sections/background.tex
\subsection{Problem statement}
We consider a bounded parametric domain $\mathcal{D} \subset \mathbb{R}^n$ and assume a scalar function $g(\mu): \mathcal{D} \mapsto \mathbb{R}$ that can be evaluated anywhere in $\mathcal{D}$ and returns a simulator output of an underlying computational model. We treat $g$ as a deterministic black-box quantity of interest; extensions to stochastic or noisy model evaluations and multiple model fidelities are left for future work. The quantity of interest $g$ depends non-smoothly on $\mu$ and may exhibit a finite number of discontinuities within the parametric domain $\mathcal{D}$. We further assume that there exists a critical value $\theta \in \mathbb{R}$ for which the behavior of the system changes fundamentally. That is, let us define
\begin{equation}
\label{eq:regime-volume-definition}
    \mathcal{D}_\theta^+ = \{ \mu \in \mathcal{D} : g(\mu) \geq \theta \},
\end{equation}
with $|\mathcal{D}_{\theta}^+| > 0$; then, the regime boundary is defined by the interface
\begin{equation}
\label{eq:regime-interface-definition}
    \Gamma = \partial \mathcal{D}_\theta^+.
\end{equation}
Our goal is to construct an accurate approximation $\widetilde{\Gamma}$ of $\Gamma$ while minimizing the number of queries to the underlying computational model represented through the simulator output $g(\mu)$.

\subsection{Adaptive sampling and localized modeling}
To construct an accurate approximation of the regime boundary $\Gamma$, adaptive sampling strategies, often referred to as active learning schemes, are often preferred to static learning due to their better sample efficiency. In its broader scope, which is not limited to the regime boundary detection setting we consider in this work, active learning works by repeated construction of an approximate solution to the original problem, coupled with an oracle that, given such a solution, provides valuable information on where to query next to expand the training dataset \cite{di2024active, settles2009active}. This procedure can be repeated iteratively until the computational budget is exhausted or the solutions computed in two successive active learning steps do not differ considerably. For the construction of surrogate models, \cite{marelli2018active} proposed using bootstrap-based local error estimates of a PCE surrogate to decide where to enrich the experimental design. Recent work has proposed to balance exploration and exploitation to improve active learning algorithms in the context of reliability analysis \cite{hong2022novel, guo2024active}. In the context of discontinuous function approximation, treed Gaussian processes \cite{bilionis2012multi, gramacy2008bayesian} or multi-element generalized polynomial chaos (gPC) \cite{wan2006multi} perform recursive partitioning of the input space or element-wise domain decompositions, respectively. Finally, \cite{sargsyan2012uncertainty} maps the obtained subdomains to distinct hypercubes where polynomial chaos expansions are used for approximation. These approaches can be combined with adaptive sampling strategies exploiting for example the GP posterior uncertainty, however they inherit the poor scaling of tensor-product PCE and Gaussian processes in higher dimensions. Moreover, their computational cost increases with the number and complexity of the discontinuities, eventually rendering them computationally intractable. For reduced-order models, \cite{paul2015adaptive} introduced a greedy sampling strategy coupled with a cheap error estimator to query high-fidelity model evaluations where the error of a global reduced-order model is likely to be high. Similar approaches have been proposed for contact mechanics \cite{balajewicz2016projection}. However, due to interpolation and because the error indicator is residual-based, it is not obvious why such an error indicator would query points close to the regime boundary. Recently, there have been a number of attempts to investigate subspace distances between parameter-dependent POD subspaces to capture sensitive areas of the parameter space in an active learning fashion \cite{varona2017automatic, kapadia2025subspace}. While such schemes were successfully applied to structural and CFD problems, the identification was only applied in low-dimensional parameter spaces, and the sensitivity measure was not used to guide a parameter space partitioning prior to the construction of localized reduced-order models. Attempts in this direction exist \cite{vlachas2021coupling, vlachas2024parametric}, but the regime-awareness of the clustering is not discussed. More generally, the majority of the work on localized models relies on static one-shot clustering \cite{amsallem2012nonlinear, geelen2022localized, chasapi2026parameterized}, either in parameter space or state space, via $k$-means \cite{ahmed2020k} or similar variance-based clustering algorithms. As highlighted and discussed in \cite{amsallem2016pebl, washabaugh2012nonlinear}, the computed cluster need not reflect the actual dynamics regimes. Finally, for surrogate models, \cite{moustapha2022learning} uses clustering and classification to detect discontinuities of a simulation output; localized Gaussian processes are then used for regression on each side of the discontinuity with a mixture of experts approach; however, this is done in a static fashion without adaptive sampling for discontinuity detection.

Therefore, this work aims to develop an adaptive sampling framework that identifies distinct regimes of the underlying computational model, thereby partitioning the parameter space into physically meaningful clusters. Within each cluster, tailored local surrogate models or reduced-order models can then be constructed. This further motivates the need to provide an accurate approximation from a limited number of model evaluations. In particular, because the quantity of interest of our computational model is discontinuous, rather than constructing a regression-based surrogate for $g$ and recovering the boundary from its $\theta$-level set, we formulate the problem directly as binary classification. This avoids the need to accurately reconstruct a potentially non-smooth model response and, in the presence of discontinuities, remains applicable even when the excursion-set boundary does not coincide with the $\theta$-level set of $g$. Specifically, we construct a classifier that approximates the labeling function
\begin{equation}
\label{eq:regime-label-definition}
    y(\mu) = \begin{cases}
        1 & \text{if $g(\mu) \geq \theta$}, \\
        -1 & \text{otherwise},
    \end{cases}
\end{equation}
The motivation for this algorithmic choice can be further supported by three considerations. First, in regime boundary identification, one is typically interested in a single threshold $\theta$, or at most in a small number of nearby threshold values. In such cases, retaining an accurate global approximation of $g(\mu)$ may be unnecessary, particularly when reconstructing the model response over the entire parameter domain is difficult because of discontinuities or sharp variations. Second, if several nearby thresholds are of interest, for example to assess the sensitivity of the inferred boundary to $\theta$, an approximation to $\partial \mathcal{D}_{\theta_1}^+$, can provide an effective initialization for the estimation of $\partial \mathcal{D}_{\theta_2}^+$ when $\theta_2 \approx \theta_1$.
Third, discontinuous function approximation remains very expensive and, unlike classification, scales badly with the dimension of the parameter space. Two-stage approaches, where regimes are firstly identified and regime-specific surrogates or reduced-order models are constructed subsequently, can therefore be more data efficient by exploiting a divide-and-conquer paradigm. This can be particularly advantageous when dealing with expensive computational models.

For the purpose of classification, in this work we employ a nonparametric approximation using \textit{support vector machines} (SVM) \cite{hearst1998support, steinwart2008support}. In Section \ref{sec:al-lse}, after a brief review on regression-based level set estimation, we detail the SVMs methodology. The adaptive sampling strategy we propose for support vector machines, which moves away from so-called \textit{uncertainty sampling} schemes and instead favors tangential boundary exploration, will be detailed in Section \ref{sec:ingredients-proposed}.

%% file: sections/al_lse.tex
In this section, we review active learning methodologies for excursion set estimation by firstly presenting the ones relying on function approximation, and then those relying on a resulting (and equivalent, in the continuous case) classification problem. For the latter case, we further describe the implicit representation of the regime boundary obtained by support vector machines, a popular supervised machine learning technique for classification. SVMs are particularly well suited to the present setting because they provide an explicit decision function defining the boundary and perform effectively in low-data regimes. We summarize the main properties and computational advantages of SVMs before reviewing established active learning strategies for adaptively refining their decision boundary.

\subsection{Gaussian process regression} \label{subsec:gpr}
Gaussian processes (GPs) have long served as a regression tool, widely known as Kriging \cite{cressie1990origins}. A GP is a collection of random variables $Y(\mu)$, indexed by the $n$-dimensional vector $\mu$, such that any subset of these variables has a multivariate Gaussian distribution. A GP is therefore a non-parametric approximation that can be fully characterized by its mean function, $m(\mu) = \mathbb{E}\left[ Y(\mu) \right]$, where $\mathbb{E}$ denotes expectation, and its covariance function
\begin{equation}
\label{eq:gpr-cov}
    K(\mu, \mu') = \mathrm{cov}\left( Y(\mu), Y(\mu') \right) = \mathbb{E} \left[  (Y(\mu) - m(\mu)) (Y(\mu') - m(\mu')) \right].
\end{equation}

In GP regression, a prior distribution is placed over the unknown response function and is subsequently conditioned on the available model evaluations. For simplicity, assuming a zero-mean GP $\xi \sim \mathcal{N}(0, K)$ with covariance kernel $K$, the unknown response function $g: \mathcal{D} \mapsto \mathbb{R}$ can be approximated from measurements $g(\mu_i) := g_i, \, i = 1, \ldots, N$, where $N$ denotes the number of training data points. This approximation is given by the regressor
\begin{equation}
\label{eq:gpr-regression}
    g^\dagger(\mu) = \mathbb{E} \left[ \xi(\mu) \left. \right\vert  \xi(\mu_i) =  g_i, \, i = 1, \ldots, N \right] = K(\mu, \mathcal{U}) K^{-1} (\mathcal{U}, \mathcal{U}) \mathbf{g}, 
\end{equation}
where $K(\mathcal{U}, \mathcal{U})$ is an $N \times N$ matrix with entries $K(\mu_i, \mu_j)$,
\begin{equation}
\label{eq:gpr-cov-def}
    K(\mu, \mathcal{U}) = \left[ \begin{array}{ccc} K(\mu, \mu_1)  & \ldots & K(\mu, \mu_N) \end{array} \right]
\end{equation}
and $\mathbf{g}$ is a vector that contains the collection of measurements $\{ g_i \}_{i=1}^{N}$. The regressor \eqref{eq:gpr-regression} represents the mean of the conditional GP, with its variance given by
\begin{equation}
\label{eq:gpr-var-def}
    \sigma^2(\mu) = \mathrm{Var} \left( \xi(\mu)  \left. \right\vert  \xi(\mu_i) =  g_i, \, i = 1, \ldots, N \right) = K(\mu, \mu) - K(\mu, \mathcal{U}) K^{-1} (\mathcal{U}, \mathcal{U}) K^T(\mu, \mathcal{U}).
\end{equation}

\subsection{Gaussian process-based active learning} \label{subsec:al-lse-gpr}
In the context of level set estimation, whenever $g$ is at least continuous and its $\theta$-level set is well defined, Gaussian processes are used as a surrogate $g^\dagger(\mu)$ for the quantity of interest $g(\mu)$. Once the regressor is trained, reconstructing $\partial \mathcal{D}_{\theta}^+$ is straightforward and provides an approximation to $\Gamma$. Due to the uncertainty estimates that come from evaluating the variance \eqref{eq:gpr-cov-def}, using Kriging as a regression tool naturally gives information about the most informative regions of the parameter space. A variety of active learning paradigms have therefore been developed in this direction. Here, we summarize representative approaches, some of which are used in Section \ref{sec:examples} as benchmarks for comparison with classification-based methods in the presence of discontinuous model response.

\begin{enumerate}
    \item \textit{Level Set Estimation}. This algorithm, proposed in \cite{gotovos2013active}, defines at every active learning iteration $t$ a \textit{confidence interval} of the form
    \begin{equation}
    \label{eq:qt-lse}
        Q_t(\mu) = g_{t-1}^\dagger(\mu) \pm \beta_t^{1/2} \sigma_{t-1}(\mu)
    \end{equation}
    for any point $\mu \in \mathcal{D}$. In practice, \eqref{eq:qt-lse} is computed on a dense grid of points. This confidence interval allows us to determine whether a point can be classified in the super-level or sub-level sets or whether such decision is still uncertain; therefore, it guides the sampling process towards points that are likely to be informative with respect to the desired level set. The next point to be evaluated is the one that maximizes the so-called \emph{ambiguity}, defined as
    \begin{equation}
        a_t(\mu) = \min \{ \max(C_t(\mu)) - \theta, \theta - \min(C_t(\mu)) \}, \label{eq:ambiguity-lse}
    \end{equation}
    where the confidence region $C_t(\mu)$ reads as
    \begin{equation}
        C_t(\mu) = \bigcap_{i=1}^{t} Q_i(\mu) = C_{t-1}(\mu) \, \cap \, Q_t(\mu). \label{eq:confidence-region-lse}
    \end{equation}
    
    \item \textit{Straddle}. This algorithm, proposed in \cite{bryan2005active}, is very similar in spirit to the aforementioned one, but it chooses $\beta_t^{1/2} = 1.96$ in \eqref{eq:qt-lse} and uses the confidence interval $Q_t(\mu)$ instead of the confidence region $C_t(\mu)$ in \eqref{eq:ambiguity-lse}. This leads to the acquisition function
    \begin{equation}
    \label{eq:ambiguity-straddle}
        a_t(\mu) = \min \{ \max(Q_t(\mu)) - \theta, \theta - \min(Q_t(\mu)) \} = \beta_t^{1/2} \sigma_{t-1}(\mu) - | g^\dagger_{t-1}(\mu) - \theta |.
    \end{equation}

    \item \textit{Maximum variance}. This approach is completely agnostic to the vicinity to the level set and simply acquires the candidate sample with the maximum posterior uncertainty, i.e.
    \begin{equation}
    \label{eq:variance-gp-acquisition-function}
        a_t(\mu) = \sigma_{t-1}(\mu).    
    \end{equation}
\end{enumerate}

The acquisition functions presented above are written for the sequential case, with $N_{\text{add}}=1$ new query per active learning iteration. In many applications, to exploit parallelization, it is better to produce a batch of size $N_{\text{add}} > 1$. To this aim, regression-based approaches use the fact that the GP posterior uncertainty in \eqref{eq:gpr-var-def} does not depend on the measurements but only on the set of points. As such, to produce a batch of size $N_{\text{add}} > 1$, the uncertainty $\sigma_{t-1}(\mu)$ is updated after every addition, while keeping the posterior mean $g^\dagger_{t-1}(\mu)$ fixed until the model evaluations for the complete batch become available. For more details, we refer again to \cite{gotovos2013active}.

\subsection{Support vector machines} \label{subsec:svm}
Some of the GP-based acquisition strategies described in the previous subsection admit theoretical convergence guarantees under appropriate assumptions on the response and the GP model. Their effectiveness, however, depends on the ability of the regression surrogate $g^\dagger(\mu)$ to adequately represent $g(\mu)$ in the vicinity of the target boundary. For continuous but non-smooth responses, accurate regression may require a large number of model evaluations, while in the presence of discontinuities an excursion-set boundary need not coincide with a level set of $g$.  As we discussed in Section \ref{sec:background}, alternative approaches for regime boundary detection when $g$ is discontinuous are generally either data demanding or sensitive to the geometry of the discontinuities, which limits their applicability when dealing with computationally expensive underlying simulation models. Furthermore, they scale suboptimally with the dimensionality of the problem and may need dimensionality reduction as a preprocessing step, e.g., by projection onto a latent space through principal component analysis \cite{abdi2010principal} or active subspaces \cite{constantine2014active, constantine2015active}. However, the amount of data required to reveal such a low-dimensional structure is a priori unclear. For such reasons, we focus on the related binary classification problem for approximation of the labeling function \eqref{eq:regime-label-definition}.

Given a dataset of measurements $\{ \mu_i, g_i \}_{i=1}^{N}$, with $\mu_i \in \mathcal{D}$, we assign to each sample a label $y_i$ according to \eqref{eq:regime-label-definition}, with $y_i \in \{-1, 1 \}$. The labeled dataset $\{ \mu_i, y_i \}$ is used to train a support vector machine classifier. This is an accurate and efficient technique used to build a boundary, between the different classes of points, that approximates the true separating surface. The idea is to obtain an explicit representation of the decision function of the form
\begin{equation}
\label{eq:svm}
    f(\mu) = \sum_{i=1}^{N} \alpha_i y_i K(\mu_i, \mu) + b.
\end{equation}
In \eqref{eq:svm}, the coefficients $\alpha_i$ are Lagrange multipliers associated with the training datapoints and are obtained from the solution to a dual problem \cite{steinwart2008support}, $K(\cdot, \cdot)$ is a positive semidefinite Mercer kernel \cite{steinwart2012mercer} and $b$ is an offset. Evaluation of the kernel yields the dot product between the inputs in a high-dimensional feature space; that is, for every valid kernel there exists a map $\Phi$ such that $K(y, z) = \Phi(y)^T \Phi(z)$. A linear classification problem is then solved in the higher-dimensional feature space. Therefore, \eqref{eq:svm} can equivalently be rewritten as
\begin{equation}
\label{eq:svm-with-w}
    f(\mu) = w^T \Phi(\mu) + b, \quad w = \sum_{i=1}^{N} \alpha_i y_i \Phi(\mu_i).
\end{equation}
For a binary classification problem, the prediction of the classifier in a new query $\widetilde{\mu}$ is determined by its sign: $y(\widetilde{\mu}) = 1$ if $f(\widetilde{\mu}) \geq 0$ and vise versa $y(\widetilde{\mu}) = -1$ if $f(\widetilde{\mu}) < 0$. The two approximate regimes are separated by the $(n-1)$-dimensional manifold
\begin{equation}
\label{eq:svm-boundary}
    \{ \mu \in \mathcal{D}: f(\mu) = 0\},
\end{equation}
which acts as an approximation $\widetilde{\Gamma}$ to the true boundary $\Gamma$ defined in \eqref{eq:regime-interface-definition}.

Importantly, for the computational cost of our proposed approach, the solution to the dual problem $\{ \alpha_i \}_{i=1}^{N}$ is generally sparse and $\alpha_i > 0$ only when the corresponding datapoint is a so-called \textit{support vector}, which means that it is close to the decision boundary between the two classes and actively determines its final representation. As a consequence, we can rewrite
\begin{equation}
\label{eq:svm-cheap}
    f(\mu) = \sum_{i \in \mathrm{SV}} \alpha_i y_i K(\mu_i, \mu) + b,
\end{equation}
where $\mathrm{SV}$ denotes the subset of support vectors, with $| SV | \ll N$ in most practical applications. Furthermore, $K(\cdot, \cdot)$ is often simple enough and amenable to parsimonious differentiation. A very standard choice when no prior information about the distribution of the two classes is available is that of a Gaussian kernel, which corresponds to a lifting map $\Phi$ to infinite dimensions. In this case $K(y, z) = \exp (-\gamma \| y - z \|^2)$ and we have
\begin{equation}
\label{eq:svm-grad-cheap}
    \nabla f(\mu) = \sum_{i \in \mathrm{SV}} \alpha_i y_i \nabla K(\mu_i, \mu) = 2\gamma \sum_{i \in \mathrm{SV}} \alpha_i y_i (\mu_i - \mu) K(\mu_i, \mu),
\end{equation}
whose computation can be efficiently vectorized. From expressions \eqref{eq:svm-cheap} and \eqref{eq:svm-grad-cheap}, we conclude that the computational cost of a function and gradient evaluation is $\mathcal{O}(|SV|) + \mathcal{O}(n | SV |)$, which is proportional to the number of support vectors rather than the total size of the training dataset. This makes repeated evaluations of the acquisition function tractable. Finally, the training cost of kernel support vector machines (that is, with a nontrivial lifting $\Phi$) ranges between $\mathcal{O}(N^2)$ and $\mathcal{O}(N^3)$. This is generally cheaper than the $\mathcal{O}(N^3)$ training cost of GPs, and is nonetheless negligible, in low-data regime, compared to the expensive model evaluations.

\subsection{Active learning for support vector machines} \label{subsec:al-svm}

Due to their well understood mathematical foundation and properties, support vector machines have been the gold standard for classification in low-data regime. Furthermore, due to the low retraining cost given new incoming data, a wide variety of active learning strategies have been proposed to refine the classification boundary. We now introduce the most popular techniques, as this will allow us to naturally set the stage for our proposed approach. To this end, we first provide some additional remarks on support vector machines.

Points for which $| f(\mu) | < \tau$ for a given $\tau > 0$ are said to be within the $\tau$-margin of the classifier, while larger magnitudes of $f$ imply an increasing distance from the boundary and an increasing confidence of the classifier in its prediction. Moreover, as we discussed previously, $K(\cdot, \cdot)$ is a positive semi-definite kernel, corresponding to a feature map $\Phi$ such that $K(y, z) = \Phi(y)^T \Phi(z)$. This allows computing distances in the feature space as
\begin{equation}
\label{eq:feature-distance}
    \| \Phi(y) - \Phi(z) \|^2 = \| \Phi(y) \|^2 - 2 \Phi(y)^T \Phi(z) + \| \Phi(z) \|^2 = K(y, y) - 2 K(y, z) + K(z, z).
\end{equation}
Equation \eqref{eq:feature-distance} shows that the kernel $K(\cdot, \cdot)$ induces a similarity distance in the input space through the feature representation. Thus, a common scale-invariant similarity measure is the cosine similarity between $\Phi(y)$ and $\Phi(z)$, corresponding to the normalized kernel
\begin{equation}
    \label{eq:kernel-normalized}
        K^*(y, z) = \frac{K(y, z)}{\sqrt{K(y, y) K(z, z)}}. 
\end{equation}
The most common active learning techniques stem from considerations about proximity to the current separating surface and appropriate variations. In this work, we consider a pool-based setting in which, at each iteration, a surrogate boundary $f(\mu)$ of the form \eqref{eq:svm} is trained on the labeled set and a batch of $N_{\text{add}}$ new points is selected from an unlabeled candidate pool $\mathcal{U}$; the pool can either be fixed or regenerated at each iteration, for example using a low-discrepancy sequence in $\mathcal{D}$. All algorithms can either select one point at a time or be executed in batch mode, by picking the $N_{\text{add}}$ points with highest score according to the acquisition function; the latter choice is usually preferred to enable parallelism and reduce wall-clock time.

\begin{enumerate}
    \item \textit{Margin/entropy sampling}. In their naive form, margin sampling \cite{schohn2000less} and entropy sampling \cite{wang2010sample} start from an unlabeled pool $\mathcal{U}$ and propose to query the points that are closest to the current separating surface. In the former case, such closeness is measured by the magnitude of $f$ as a proxy for uncertainty; that is, the acquisition rule reads
    \begin{equation}
    \label{eq:margin-sampling-update}
        \mu^* = \arg \min_{\mu \in \mathcal{U}} | f(\mu) |.
    \end{equation}
    In the latter case \cite{wang2010sample}, it uses as a proxy the entropy of the posterior probability $p_\theta(y | x)$ that can be fitted to a support vector machine boundary via Platt scaling \cite{platt1999probabilistic}. Therefore, in this case
    \begin{equation}
    \label{eq:entropy-sampling-update}
        \mu^* = \arg \max_{\mu \in \mathcal{U}} \left(  - \sum_{y} p_\theta(y | \mu) \log p_\theta(y | \mu)  \right),
    \end{equation}
    where in our case $y \in \{-1, 1\}$ but the above definition generalizes to an arbitrary number of classes. The acquisition functions \eqref{eq:margin-sampling-update} and \eqref{eq:entropy-sampling-update} rely on point-wise uncertainty estimates. A well-known issue of these approaches is their \textit{mode-collapse} behavior, meaning that they tend to produce sample redundancy by oversampling the same boundary region. They are therefore not incline to exploring unrepresented parts of the boundary, where refinement could greatly improve predictive accuracy. Besides, two other issues should be mentioned: (i) the performance can be very sensitive to the initial estimate of the boundary, and (ii) pure exploitation in normal direction to the boundary may lead to poor performance in the case of boundaries with multiple disconnected components.

    \item \textit{Local refinements}. This approach was presented in \cite{gorodetsky2014efficient} and was used for discontinuity detection as refinement strategy after the application of polynomial annihilation \cite{kayal2009complexity, surace2013use, archibald2005polynomial} to obtain a very good initial estimate of the separating surface. It proposes a new point $\mu \in \mathcal{D}$ by solving
    \begin{equation}
    \label{eq:marzouk-proposal}
        \min_\mu \left( \sum_{i=1}^{N} \alpha_i y_i K(\mu_i, \mu) \right)^2. 
    \end{equation}
     The candidate $\mu$ is accepted among the next batch to be queried if it satisfies two criteria. First, it must be sufficiently far from the current labeled set $\mathcal{S}=\{\mu_i\}_{i=1}^N$, i.e.
    \begin{equation}
    \label{eq:localref-distance}
        \min_{\mu_i \in \mathcal{S}} \|\mu - \mu_i\| \geq \varepsilon
    \end{equation}
    for some user-defined $\varepsilon > 0$. Second, it must be close to a region where class labels disagree, which is enforced by requiring that the ball $B(\mu,\delta)=\{\nu:\|\nu-\mu\|\leq \delta\}$ contains points from both classes, i.e.
    \begin{equation}
    \label{eq:localref-disagreement}
        \exists\, \mu_i,\mu_j \in \mathcal{S} \cap B(\mu,\delta) \quad \text{such that} \quad y_i \neq y_j.
    \end{equation}
    A similar conclusion to the one drawn for margin/entropy sampling emerges here: while condition \eqref{eq:localref-distance} ensures that training points are at least $\varepsilon$ away from each other, this algorithm only improves the local approximation of the boundary unless a large $\delta$ is chosen.
    
    \item \textit{Clustering approach}. The work presented in \cite{xu2003representative} promotes diversity among the generated samples by clustering the subset of $\mathcal{U}$ lying inside the $\tau$-margin of the classifier, using a k-means algorithm \cite{ahmed2020k}; the centroids of each cluster are then selected as the next batch of points to be queried. Such an algorithm has a higher computational cost compared to the previous alternatives as it requires performing clustering at every active learning iteration. Furthermore, the original work shows improvements only in the low-data regime, as the accuracy tends to plateau afterwards. As a remedy, it proposes to switch back to standard uncertainty sampling after a certain number of active learning iterations, but such a switch cannot be finetuned in advance. Finally, the effectiveness of variance-based clustering algorithms is known to deteriorate as soon as the parameter space has moderate dimension.
    
    \item \textit{Incorporation of diversity}. In a similar spirit, \cite{brinker2003incorporating} incorporates diversity in the selected samples by taking as an acquisition function a convex combination of uncertainty and diversity. More precisely, the acquisition rule in this case reads as
    \begin{equation}
    \label{eq:diversity-update}
        \mu^* = \arg \min_{\mu \in \mathcal{U}_\tau} \left( \lambda | f(\mu) | + (1 - \lambda) \max_{j \in \mathcal{S}} K^*(\mu, \mu_j) \right),
    \end{equation}
    where $\mathcal{U}_\tau = \{ \mu \in \mathcal{U}: | f(\mu) | < \tau \}$ denotes the set of unlabeled points that lie within the classifier margin, $\mathcal{S}$ is the set of labeled points, and the diversity measure $K^*(\cdot, \cdot)$ is the cosine similarity defined in \eqref{eq:kernel-normalized}. Finally, $\lambda$ is a parameter that must be tuned to balance uncertainty and diversity. Note that the case $\lambda = 1$ reduces to the margin sampling acquisition rule \eqref{eq:margin-sampling-update}. The numerical experiments in \cite{brinker2003incorporating} have shown the superior sample efficiency obtained when taking $\lambda = 0.5$. This algorithm works well and conceptually is the closest to our proposed approach; the main disadvantages are that adjusting $\lambda$ may still be cumbersome and that, because it still relies on a discrete search space, its effectiveness can degrade in higher dimensions.
\end{enumerate}

%% file: sections/proposed.tex
Our proposed approach stems from ideas that have been explored for rare-event estimation and reliability analysis \cite{chakraborty2025affine}. The fundamental paradigm is to view the current estimate of the level set as a geometric object and to prioritize its exploration with mechanisms that (i) keep samples close to the estimated boundary and (ii) allow coherent motion along it. Given a currently available training dataset at active learning iteration $t$, and a trained SVM $f_t(\mu)$, we define the estimated regime boundary
\begin{equation}
\label{eq:estimated-interface}
\widetilde{\Gamma} := \{\mu \in \mathcal{D}: f_t(\mu) = 0 \},
\end{equation}
and the potential energy
\begin{equation}
\label{eq:hmc-boundary-potential}
U_t(\mu) := \frac{\beta}{2} f_t^2(\mu), \qquad \nabla U_t(\mu) = \beta f_t(\mu) \, \nabla f_t(\mu),
\end{equation}
where the inverse-temperature parameter $\beta$ controls the growth of the potential energy \eqref{eq:hmc-boundary-potential} in the normal direction to the boundary $\widetilde{\Gamma}$. The potential energy $U_t(\mu)$ is minimized at the estimated boundary; furthermore, since we know that for support vector machines the margin between the two classes is maximized, defining such a potential to be symmetric with respect to $\widetilde{\Gamma}$ is thorough. The task of generating candidate samples that are well distributed along the tangent directions to $\widetilde{\Gamma}$ can be tackled with a variety of gradient-based methods that efficiently explore low-energy landscapes in the proximity of manifolds. As we mentioned earlier, this is motivated in our setting by the fact that computing the gradient in \eqref{eq:hmc-boundary-potential} has a negligible computational cost. In this direction, candidate samples are generated by Hamiltonian Monte Carlo (HMC) \cite{betancourt2017conceptual, girolami2011riemann}. We will elaborate on this choice, compared to other sampling mechanisms for the problem at hand, in Subsection \ref{subsec:alternatives-exploration}. From the set of generated samples, a space-filling subset is extracted for the next batch to be queried.

The framework presents a variety of practical advantages. First, by using gradient-based dynamics, it generates a large pool of candidates distributed in a tube around $\widetilde{\Gamma}$; furthermore, this pool is sought in a continuous search space, unlike margin-based schemes that are evaluating uncertainty estimates on fixed grids (or quasi-random variants thereof). Using gradient information, despite the increased computational cost, helps the adaptive sampling process to avoid function evaluations far away from $\widetilde{\Gamma}$ in order to accurately represent it; this aspect becomes increasingly important in higher-dimensional spaces, where memory constraints become more significant. Moreover, moving away from a discrete search space for the next batch to be queried is expected to provide benefits when the target boundary is complex and has high curvature. Secondly, the subsequent space-filling down-selection discourages repeatedly querying the same local portion of the boundary, which therefore mitigates the mode-collapse behavior often observed in naive uncertainty sampling. It also provides a very natural framework to exploit batch parallelism, as different Markov chains can be evolved in parallel to contain the computational overhead. Lastly, the parameter $\beta$ provides a single knob to trade off global exploration versus local refinement: smaller $\beta$ yields broader exploration in the normal direction (that can accelerate convergence when the initial boundary estimate is crude), whereas larger $\beta$ concentrates samples closer to $\widetilde{\Gamma}$ and effectively focuses the search on refining the interface location; $\beta$ could also be increased during active learning according to a simulated-annealing schedule \cite{van1987simulated, bertsimas1993simulated}.

In our implementation, because tuning an annealing schedule for $\beta$ is generally challenging, we select a constant value from a pilot run carried out on a toy two dimensional example. We discuss more on this in Section \ref{sec:examples}. Furthermore, because we are interested in the restriction of the interface to the bounded parametric domain $\mathcal{D}$, we enforce box constraints through reflective boundaries: whenever a leapfrog step proposes a point outside $\mathcal{D}$, the critical coordinates are mirrored back into the domain and the corresponding momentum component is flipped, yielding a simple, stable constrained dynamics. This is a computationally cheap alternative to reparameterization, which would require modifying all function and gradient evaluations.

The proposed active learning technique is summarized in pseudocode in Algorithm \ref{alg:hmc-proposed-algo}. After starting from a coarse initial dataset $\mathcal{P}_0$, the active learning loop starts: given the current labeled dataset, \texttt{EstimateSVMBoundary} trains a new support vector machine. Afterwards, given a pre-defined inverse-temperature $\beta = \beta_t$, \texttt{CollectHMCSamples} collects informative boundary samples using Hamiltonian Monte Carlo. Lastly, in \texttt{CollectDiverseSubset}, a greedy algorithm selects a diverse space-filling subset of size $N_{\text{add}}$.  Here, a standard Euclidean distance is used, but the framework naturally allows us to use more application-specific distances, such as \eqref{eq:kernel-normalized}. This batch is fed to the underlying computational model for the next queries, and the procedure is repeated until the computational budget is exhausted.

\begin{algorithm}[ht]
\LinesNotNumbered
\caption{Active learning for excursion set estimation with HMC-guided sampling}
\label{alg:hmc-proposed-algo}

\KwData{Total budget $N_{\mathrm{end}}$, number of initial points $N_{\mathrm{init}}$, batch size $N_{\mathrm{add}}$, labeling function $y: \mathcal{D} \mapsto \{-1, 1 \}$ and inverse-temperature schedule $\{ \beta_t \}_t$.}

\vspace{4pt}

Draw an initial sample $\mathcal{M}_0 \subset \mathcal{D}$ of size $N_{\text{init}}$ with a low-discrepancy sequence \\
Collect initial dataset $\mathcal{P}_0 = \{ \mathcal{M}_0, y(\mathcal{M}_0) \}$ \\[10pt]

$t = 0$ \\
\While{$\textnormal{size}(\mathcal{P}_{t}) \leq N_{\mathrm{end}}$}{
    $f_t \gets \texttt{EstimateSVMBoundary}(\mathcal{P}_{t})$ \\
    $\widetilde{\mathcal{M}}_t \gets \texttt{CollectHMCSamples}(\beta_t, f_t)$ \\
    $\mathcal{M}_t \gets \texttt{CollectDiverseSubset}(\widetilde{\mathcal{M}}_t, \mathcal{P}_t, N_{\text{add}})$ \\
    $\mathcal{P}_{t+1} \gets \mathcal{P}_{t} \cup \{ \mathcal{M}_t, y(\mathcal{M}_t) \}$ \\
    $t \gets t+1$ \\
}

\vspace{4pt}

\textbf{Output}: Final dataset $\mathcal{P}_t$
\end{algorithm}

\subsection{Hamiltonian Monte Carlo} \label{subsec:hmc}

Hamiltonian Monte Carlo (HMC) is a Markov Chain Monte Carlo (MCMC) method designed to sample from a generic target probability distribution of the form 
\begin{equation}
\label{eq:hmc-target}
    \pi(\mu) = \frac{\pi_u(\mu)}{\int \pi_u(\xi) d\xi}, \quad \text{where} \quad \pi_u(\mu) = e^{- U(\mu)},  
\end{equation}
with a potential energy function $U(\mu) = - \log \pi_u(\mu)$ that is finite and differentiable. To sample with HMC, one needs access to $U(\mu)$ and its gradient $\nabla U(\mu)$. In our setting, $U$ is induced by the current SVM decision function $f_t(\mu)$ at active learning iteration $t$ and takes the form $U(\mu) = \frac{\beta}{2} f_t(\mu)^2$, with $\nabla U_t(\mu) = \beta f_t(\mu)\, \nabla f_t(\mu)$ that can be computed analytically.

The success of HMC stems from augmenting the state space with an auxiliary momentum variable, which enables long-range, coherent proposals that follow approximate iso-probability contours of the target. This mitigates the diffusive random-walk behavior typical of many Metropolis-type samplers, especially in moderate to high dimensions and for strongly correlated targets. In its vanilla version, HMC works as follows: given a (constant) symmetric and positive-definite mass matrix $M$, it defines the Hamiltonian
\begin{equation}
\label{eq:hmc-hamiltonian}
    H(\mu, p) = U(\mu) + K(p),
\end{equation}
where $K(p) = \frac{1}{2} p^T M^{-1} p$ is the kinetic energy. The corresponding Hamiltonian dynamics reads as
\begin{equation}
\label{eq:hmc-dynamics}
    \begin{cases}
        \dot{\mu} = M^{-1} p \\
        \dot{p} = - \nabla U(\mu)
    \end{cases}.
\end{equation}
In practice, from a given configuration $(\mu, p)$, the dynamics \eqref{eq:hmc-dynamics} is integrated for $L$ leapfrog steps with step size $\varepsilon$, producing a proposal $(\mu', p')$. The pair $(\varepsilon, L)$ controls the trade-off between exploration and acceptance rate: too large a step size induces large integration error and low acceptance, while too small a step size leads to high acceptance but short effective moves. A Metropolis acceptance-rejection mechanism corrects the bias introduced by numerical integration errors and ensures that the resulting chain targets \eqref{eq:hmc-target}; that is, proposals are accepted with probability
\begin{equation}
\label{eq:hmc-metropolis}
    \alpha = \min \{ 1, \exp \{ H(\mu, p) - H(\mu', p') \} \}.
\end{equation}
The mass matrix $M$ can be interpreted as a linear preconditioner: choosing $M$ to match the scale and anisotropy of the target improves mixing and reduces the need for excessively small $\varepsilon$. When the geometry of the target distribution is particularly complex, mixing can be improved by taking a position-dependent mass matrix $M(\mu)$, for example, the Monge metric $\mathbf{I} + \alpha \nabla \pi(\mu) \nabla \pi(\mu)^T$ or the local negative Hessian $\nabla^2 U(\mu)$. When $M = M(\mu)$, the resulting Hamiltonian is no longer separable and time integration of \eqref{eq:hmc-dynamics} requires implicit solves and a higher computational cost; while remedies have been proposed, e.g. by using hierarchical schemes \cite{kailas2026adaptive}, in our experiments we guarantee quality of the generated samples, while containing the computational overhead, by:

\begin{itemize}
    \item \textit{Rescaled inputs.} Input parameters are rescaled to the unit hypercube $(-1, 1)^n$, helping to simplify the geometry of the target distribution. 
    \item \textit{Warm-start/boundary initialization.} We partially alleviate the need for long burn-in periods by initializing each chain close to the current
    estimate of the separating surface. This is done to avoid problematic initializations in regions where $\|\nabla f_t(\mu)\| \approx 0$ 
    (which can happen far away from the support vectors when using a Gaussian kernel). This is obtained by solving \eqref{eq:marzouk-proposal}
    with a standard Gauss-Newton scheme from a randomly initialized point in $\mathcal{D}$, or by applying clustering to a lightweight grid of points within the margin.
    \item \textit{Parallel HMC.} We run a larger number of shorter chains from different initializations to improve coverage and promote diversity; this also exposes embarrassingly parallel structure, as each chain can be evolved independently.
    \item \textit{Constant mass matrix.} Due to the above, we can guarantee the quality of the boundary coverage even when using a constant mass matrix $M$. This keeps leapfrog updates explicit and inexpensive, avoiding the implicit solves required by Riemann-manifold variants.
\end{itemize}

\subsection{Boundary exploration mechanism} \label{subsec:alternatives-exploration}

In this subsection, we further motivate the use of HMC compared to other alternatives for the boundary exploration mechanism. Our objective is to generate candidate samples that remain close to the current estimated boundary while efficiently spreading along it. To this end, we consider target distributions of the form
\begin{equation}
\label{eq:boundary-target-density}
\pi_t(\mu) \propto \exp\left(-\frac{\beta}{2} f_t^2(\mu) \right),
\end{equation}
which concentrates mass where $| f_t(\mu) | \lesssim 1/\sqrt{\beta}$. Here we use a slight abuse of notation, since $\mu$ is effectively constrained to be within the parameter space $\mathcal{D}$. Among possible schemes to generate candidate samples distributed according to \eqref{eq:boundary-target-density}, we mention:

\begin{enumerate}
    \item \textit{Unadjusted Langevin Algorithm (ULA)}. Langevin dynamics-based approaches \cite{bussi2007accurate, xifara2014langevin} propose to integrate the stochastic differential equation
    \begin{equation}
    \label{eq:ula}
        dX_s = - \beta f_t(\mu) \nabla f_t(X_s) ds + \sqrt{2} \, dB_s,
    \end{equation}
    e.g., with an Euler--Maruyama scheme. In \eqref{eq:ula}, $dB_s$ is a Brownian motion differential. Under a very general set of conditions on $f_t$, the numerical bias introduced by discretizing \eqref{eq:ula} vanishes and for long enough time horizons the samples are distributed according to \eqref{eq:boundary-target-density}. Such conditions can be summarized with $f_t$ being strongly-convex, which is not the case for a linear combination of Gaussian kernels; even in the case of a linear SVM, whose Hessian is rank deficient, this condition is not met unless regularization is added.

    \item  \textit{Metropolis-Adjusted Langevin Algorithm (MALA)}. To mitigate for the numerical bias introduced by the discretization of \eqref{eq:ula}, MALA uses a Langevin step as a proposal and adds a Metropolis acceptance-rejection mechanism. The (overdamped) Langevin proposal with step size $h > 0$ reads
    \begin{equation}
    \label{eq:mala-proposal}
        \mu' = \mu - h \beta_t f_t(\mu) \nabla f_t(\mu) + \sqrt{2 h} \, \xi, \qquad \xi \sim \mathcal{N}(0, I),
    \end{equation}
    which corresponds to a Gaussian proposal density $q(\mu'\mid \mu) = \mathcal{N}\big(\mu - h \beta_t f_t(\mu) \nabla f_t(\mu), \, 2 h I\big)$. The proposal is accepted with probability
    \begin{equation}
    \label{eq:mala-accept}
        \alpha(\mu, \mu') = \min\left\{1,\, \frac{\pi(\mu')\, q(\mu\mid \mu')}{\pi(\mu)\, q(\mu'\mid \mu)}\right\}
        = \min\left\{1,\, \exp\left(\frac{\beta_t}{2} (f_t^2(\mu) - f_t^2(\mu') \right)\, \frac{q(\mu\mid \mu')}{q(\mu'\mid \mu)}\right\}.
    \end{equation}
    For both MALA and ULA, as for HMC, the dynamics is constrained to remain in $\mathcal{D}$ through reflective boundaries. In our setting, we notice that the drift term vanishes as $f_t(\mu) \to 0$. Therefore, even with Metropolis correction, the proposal becomes dominated by Gaussian noise and the motion becomes effectively diffusive along the low-energy region, causing slow mixing. This issue is exacerbated in higher parameter space dimensions and when the estimated boundary has flat directions with vanishing gradients.

    \item \textit{Random-walk level set exploration}. In low dimensions, a computationally inexpensive alternative is to perform an explicit random walk along the level set $f_t(\mu) = 0$ and then generate adversarial samples in the normal direction to obtain candidate samples. A standard space-filling criterion can subsequently select a batch for query. Denoting $h(\mu)=\nabla f_t(\mu)$, the orthogonal projector onto the tangent space is
    \begin{equation}
    \label{eq:tangent-projector}
        P(\mu) = I - \frac{h(\mu) h(\mu)^T}{\|h(\mu)\|^2}.
    \end{equation}
    A simple tangential random-walk step with step-size $s$ can then be written as
    \begin{equation}
    \label{eq:levelset-rw}
        \mu_{k+1} = \mu_k + s \, \frac{P(\mu_k)\, \zeta_k}{\|P(\mu_k)\, \zeta_k\|}, \qquad \zeta_k \sim \mathcal{N}(0, I),
    \end{equation}
    possibly followed by a projection step to enforce $f_t(\mu_{k+1})\approx 0$. In two dimensions, the tangent space is one-dimensional, so (up to orientation) the unit tangent direction is uniquely determined by $g(\mu)$; a convenient choice is
    \begin{equation}
    \label{eq:tangent-at-x}
        t(\mu) = \frac{1}{\|h(\mu)\|} \left( -h_2(\mu),\, h_1(\mu) \right).
    \end{equation}
    Hence, the motion is fully deterministic except for choosing the sign $\pm t(\mu)$. Once a boundary point $\mu$ is obtained, adversarial samples in the normal direction can be proposed as
    \begin{equation}
    \label{eq:normal-adversarial}
        \mu^{\pm} = \mu \pm \Delta_h \, h(\mu),
    \end{equation}
    where $0 < \Delta_h \ll 1$ controls the offset from the currently estimated interface. When scaling this to higher dimensions, tangential exploration gets increasingly difficult, and constrained MCMC mechanisms need to be introduced to allow fast tangential mixing; this becomes basically equivalent (in terms of tangential sample spread and computational cost) to running HMC with a large value of $\beta$.
\end{enumerate}

\begin{figure}[ht]
    \centering
    \begin{subfigure}[t]{0.32\linewidth}
        \centering
        \includegraphics[width=\linewidth]{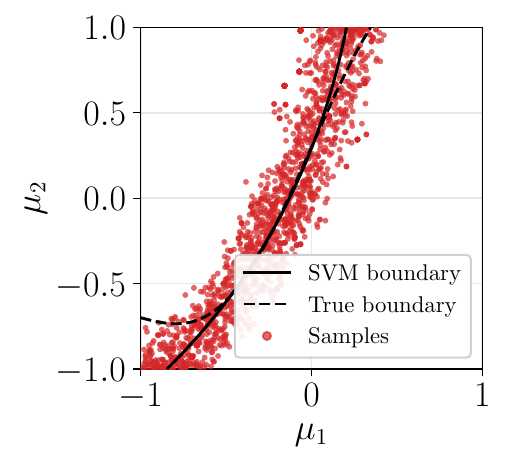}
        \caption{$\beta = 10$}
        \label{fig:hmc-ka-10}
    \end{subfigure}\hfill
    \begin{subfigure}[t]{0.32\linewidth}
        \centering
        \includegraphics[width=\linewidth]{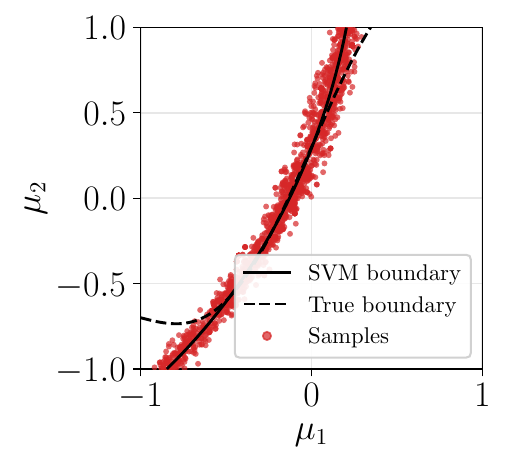}
        \caption{$\beta = 100$}
        \label{fig:hmc-ka-100}
    \end{subfigure}\hfill
    \begin{subfigure}[t]{0.32\linewidth}
        \centering
        \includegraphics[width=\linewidth]{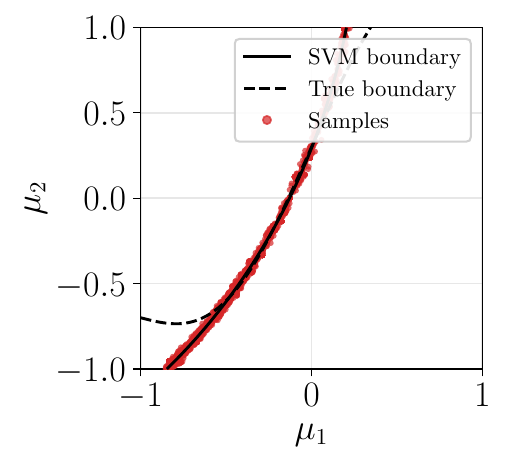}
        \caption{$\beta = 1000$}
        \label{fig:hmc-ka-1000}
    \end{subfigure}

    \caption{Illustration of the samples generated by Hamiltonian Monte Carlo for increasing values of the inverse-temperature $\beta$ on a simple two-dimensional example. The dashed line is the true boundary and the solid line is the SVM estimate that is used to define the potential energy. As $\beta$ increases, samples concentrate more closely to the boundary. Tangential momentum, leading to good boundary coverage, is not directly affected by the choice of $\beta$.}
    \label{fig:hmc-ka}
\end{figure}

Compared to the diffusive exploration of Langevin schemes and constrained random walks in higher dimensions, HMC introduces momentum, which enables long, 
coherent moves along low-energy regions of \eqref{eq:boundary-target-density}. A simple geometric picture can help to motivate the term \emph{tangential exploration}.
We assume that $f_t(\mu)$ is continuously differentiable and $\nabla f_t(\mu) \neq 0$ in a neighborhood of the estimated interface $\widetilde{\Gamma}$, so that $\widetilde{\Gamma}$
is locally a smooth $(n-1)$-dimensional manifold. This geometric condition is in practice satisfied most of the time: for a Gaussian kernel SVM, vanishing gradients happen only
far from the support vectors and hence from the boundary. This further motivates our initialization strategy described earlier.
At a point $\mu \in \widetilde{\Gamma}$, the normal direction is spanned by $\nabla f_t(\mu)$ and the tangent space is
\begin{equation}
\label{eq:tangent-space}
\mathrm{T}_{\mu}\widetilde{\Gamma} = \{v \in \mathbb{R}^n : v^T \nabla f_t(\mu) = 0\}.
\end{equation}
Since $\nabla U_t(\mu)=\beta f_t(\mu) \nabla f_t(\mu)$ is proportional to $\nabla f_t(\mu)$, the force term $-\nabla U_t(\mu)$ acts (to first order) in the normal direction, while tangential components of the momentum are not directly penalized. Consequently, once the chain has reached the low-energy region around $f_t(\mu)\approx 0$, leapfrog trajectories can propagate efficiently along the level sets of $f_t$ instead of relying on a random-walk exploration. Finally, the tangential component of the HMC dynamics is not directly affected by the choice of $\beta$: this parameter only controls the spread of the samples in the normal direction to the boundary, while tangential momentum can persist along the low-energy level sets. Figure \ref{fig:hmc-ka} shows an illustration of the generated samples for different values of $\beta$ on a simple two-dimensional example.

\subsection{Implementation details and computational cost} \label{subsec:hmc-cost}

In this subsection, we discuss the computational cost associated with the proposed algorithm. Compared with uncertainty-based criteria, our method introduces an additional candidate-generation step based on HMC sampling. The key point is that for regime boundary detection in real-world engineering systems this overhead is very small compared to a single high-fidelity model evaluation. Furthermore, we discussed in Subsection \ref{subsec:hmc} about some of the implementation details to contain this additional cost.

We now quantify the resulting wall-clock times on a dimension-dependent analytical example and show that, for moderate dimensions (the ones considered in this manuscript, and the ones frequently encountered in parametric computational models), the additional overhead remains small relative to typical simulation times. We consider the analytical regime boundary
\begin{equation}
\label{eq:dimension-dependent-discontinuity}
    g(\mu) = \mu_n - \sum_{i=1}^{n-1} \mu_i^3,
\end{equation}
which we will use again to study the performance of the proposed approach in Section \ref{sec:examples}. Here, we target the $\theta = 0$ level set of \eqref{eq:dimension-dependent-discontinuity}; our focus is purely on timing as a function of dimension and the number of support vectors. For each dimension $n = 2, 4, \ldots, 10$, we start from $N_{\mathrm{init}} = 30$ points and add batches of size $N_{\mathrm{add}} = 5$ until the budget $N_{\mathrm{end}} = 200$ is exhausted.

\begin{figure}[ht]

    \begin{subfigure}[t]{0.325\linewidth}
        \centering
        \includegraphics[width=\linewidth]{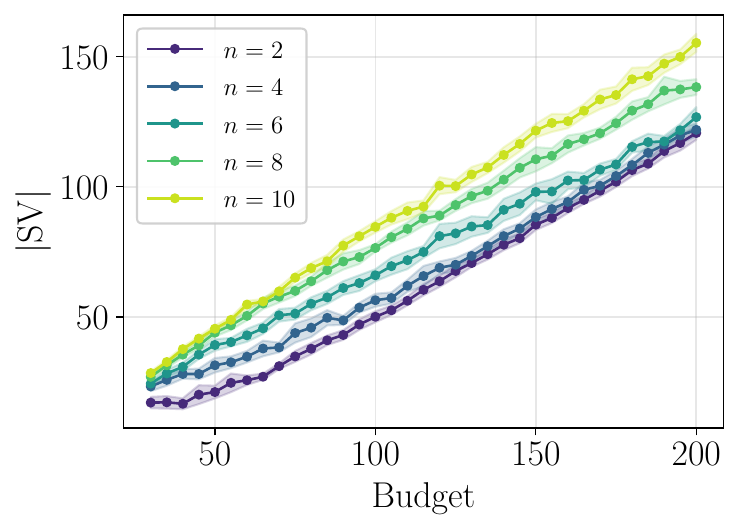}
        \label{fig:dimension_scaling_sv}
        \caption{}
    \end{subfigure}\hfill
    \begin{subfigure}[t]{0.325\linewidth}
        \centering
        \includegraphics[width=\linewidth]{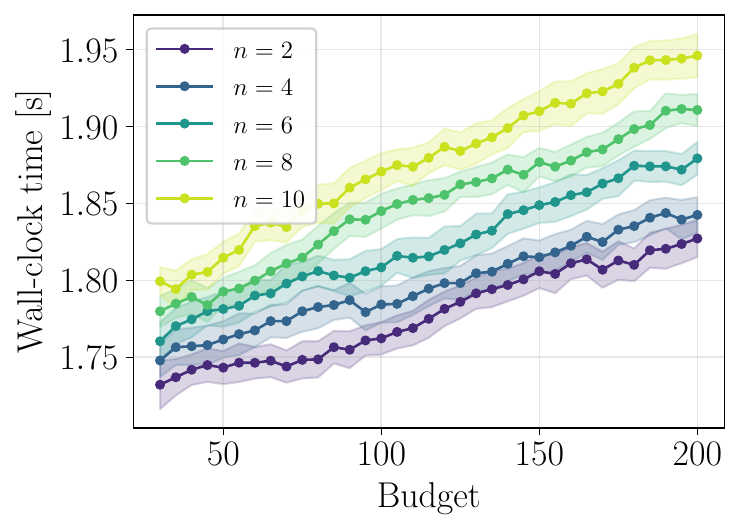}
        \label{fig:dimension_scaling_timing}
        \caption{}
    \end{subfigure}\hfill
    \begin{subfigure}[t]{0.325\linewidth}
        \centering
        \includegraphics[width=\linewidth]{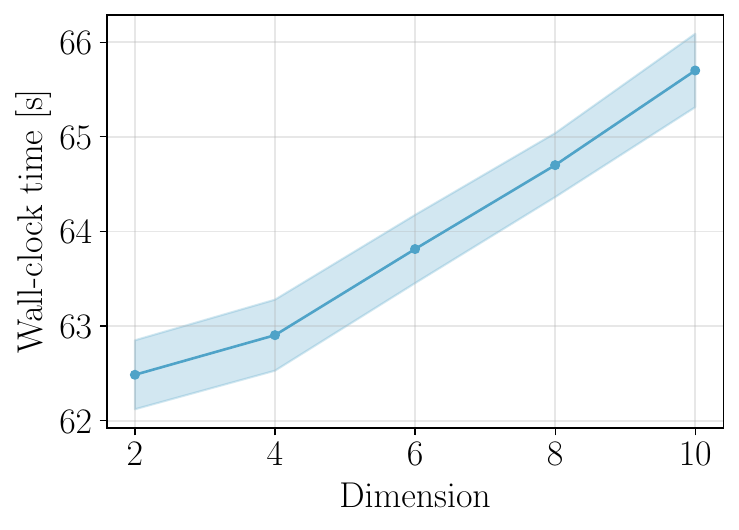}
        \label{fig:dimension_scaling_total_time}
        \caption{}
    \end{subfigure}

    \caption{Dimension scaling timings: (a) growth of $|SV|$; (b) serial wall-clock time growth during active learning; (c) total serial wall-clock time across dimensions. Shaded regions represent a $95 \%$ CI on the mean.}
    \label{fig:dimension_scaling}

\end{figure}

For candidate generation, we use $N_c = 10$ HMC chains with $N_b = 200$ burn-in steps, $L = 10$ leapfrog steps, initial step-size $\varepsilon = 10^{-3}$ and $N_s = 200$ retained samples per chain, yielding $N_c N_s = 2000$ candidate points per iteration; we then extract $N_{\mathrm{add}} = 5$ points using a greedy space-filling selection. We report wall-clock times averaged over $20$ random seeds for each dimension. Figure \ref{fig:dimension_scaling} shows various statistics obtained: despite the increased number of support vectors, the candidate generation time per iteration remains approximately constant throughout the active learning loop, and the total sampling wall-clock time over the entire active learning run remains on the order of one minute. The reported timings are sequential; distributing chains across multiple processors can further reduce the overhead. For example, if we were to parallelize over $5$ processors (consistently with the batch size $N_{\mathrm{add}} = 5$), then the total overhead would decrease to about $10$ seconds. Therefore, whenever each model evaluation is in the order of tens of seconds or minutes, even in the low-data regime, this additional CPU time is negligible. In Section \ref{sec:examples}, we show improved sample efficiency on analytical and real-world simulation models, which further justifies these additional costs.

%% file: sections/examples.tex
We now demonstrate the performance of the proposed approach on a variety of problems: analytical level sets in low dimension, boundaries of increasing dimensions, and finally real-world examples ranging from computational biology to hyperbolic PDEs.

This section is split as follows: we start by considering analytical examples. Such cases are smooth, so they could be solved more efficiently with regression-based estimation; however, our goal is just to compare active learning strategies for classification. Specifically, we compare our proposed approach with margin and cluster sampling, that are investigated as baselines, as well as random sampling (in which case, $N_{\mathrm{add}}$ space-filling points are selected from a low-discrepancy sequence agnostically to the zero level set location). Moreover, to isolate the effects of the different components of our proposed approach, that is the integration to sample from \eqref{eq:hmc-target} and the space-filling downselection, we consider a margin sampling scheme with diversity-promoting downselection, as well as a MALA integration scheme with the same downstream extraction. Later in the section, we then show real-world engineering systems with discontinuous simulator outputs, where classification outperforms regression in reconstructing the regime boundary. For regression, we consider the Straddle acquisition function \eqref{eq:ambiguity-straddle} as a competitive scheme for regression-based estimation and the maximum variance acquisition function \eqref{eq:variance-gp-acquisition-function} as a baseline; both are executed in batch version.

For all considered classification problems, if not specified otherwise, we use a Gaussian kernel. On the other hand, Gaussian process regression uses a Matérn kernel, which is more robust in non-smooth scenarios. In both cases, after every active learning iteration, we cross-validate the main parameters of the kernel representation using the same search space for all algorithms tested. In particular, for SVMs we finetune $\gamma$ appearing in \eqref{eq:svm-grad-cheap} and the parameter $C$ controlling the penalty for misclassifications; for GPs, we finetune the length scales and the parameter $\nu$ of the Matérn kernel. For operations with support vector machines, Gaussian processes, and $k$-means clustering, we use the implementations provided in \texttt{scikit-learn} \cite{kramer2016scikit}; on the other hand, HMC uses an in-house implementation. Lastly, all experiments were carried on the ETH cluster Euler. Among the implementation details:

\begin{itemize}
    \item For pool-based strategies, we generate an unlabeled candidate pool $\mathcal{U}$ of size $2000$ with Latin Hypercube sampling (LHS) in $\mathcal{D}$ and regenerate $\mathcal{U}$ at each active learning iteration.
    \item For the proposed algorithm, we tune $\beta$ on a toy two dimensional example and keep such value constant, that is we implement a constant schedule for the inverse-temperature $\beta$ throughout active learning iterations; more precisely, in all the examples, we use $\beta = 50.0$ without annealing. Importantly, for $\beta$ in the ranges $10^1-10^2$, we observed that the algorithm's performance was essentially unchanged. The HMC setup is the same described in Subsection \ref{subsec:hmc-cost}. The same setup, inverse-temperature, and conclusions hold for MALA.
    \item For cluster sampling and margin with diversity in active classification, we apply $k$-means and the space-filling greedy selection, respectively, on the set of candidate points with $|f_t(\mu)| \leq 1/\sqrt{\beta}$ for the same value of $\beta = 50.0$ used in HMC and MALA.
\end{itemize}

To compare the performance of the different algorithms, we average over $200$ seeds corresponding to different initializations and report the following quantities: (i) we compute the mean $F_1$ score on an independent test set of size $10^4$, unless specified otherwise; (ii) in the analytical examples, we compute additionally the Hausdorff distance $d_H(\Gamma, \widetilde{\Gamma})$ between true and approximated boundary by evaluating its discrete counterpart on a grid with size $\varepsilon_H = 5 \times 10^{-3}$, that is $\propto \varepsilon_H^{-(n-1)}$ points, and normalizing it as $\widehat{d}_H(\Gamma, \widetilde{\Gamma}) = \frac{d_H(\Gamma, \widetilde{\Gamma})}{\mathrm{diam}(\mathcal{D})}$ with $\mathrm{diam}(\mathcal{D}) = 2\sqrt{n}$; and (iii) we compute the percentage of seeds reaching a target $F_1$ score for a given computational budget, to draw conclusions on the robustness of the different approaches with respect to the initial design.

\subsection{Analytical two-dimensional functions} \label{subsec:analytical-discontinuities-results}

We start by fixing $\mathcal{D} = (-1, 1)^2$ and consider the sequence of increasingly complex level set functions
\begin{align}
    &g_1(\mu) = \mu_2 - 0.3 - 0.4 \sin(\pi \mu_1), \label{eq:f1-2d-discontinuities} \\
    &g_2(\mu) = \mu_2 - 0.3 - 0.4 \sin(2 \pi \mu_1) - \mu_1 \label{eq:f2-2d-discontinuities}, \\
    &g_3(\mu) = \mu_2 - 0.05 - 0.3 \sin(4 \pi \mu_1) - \mu_1 \label{eq:f3-2d-discontinuities},
\end{align}

where the scaling and translation factors are selected so that the target $\theta = 0$ level set of the above functions has only one connected component in $\mathcal{D}$. The three examples, of which the first two are taken from \cite{gorodetsky2014efficient}, have progressively increasing curvature. Therefore, for \eqref{eq:f1-2d-discontinuities}, we start with $N_{\mathrm{init}} = 10$ initial points and incrementally add batches of size $N_{\mathrm{add}} = 5$, until $N_{\mathrm{end}} = 100$ points; on the other hand, for \eqref{eq:f2-2d-discontinuities} and \eqref{eq:f3-2d-discontinuities}, from the same initialization, we add batches of size $N_{\mathrm{add}} = 10$ until the budget $N_{\mathrm{end}} = 150$ is exhausted.

\begin{figure}[H]

    \begin{subfigure}[b]{0.45\textwidth}
    \includegraphics[width=\linewidth]{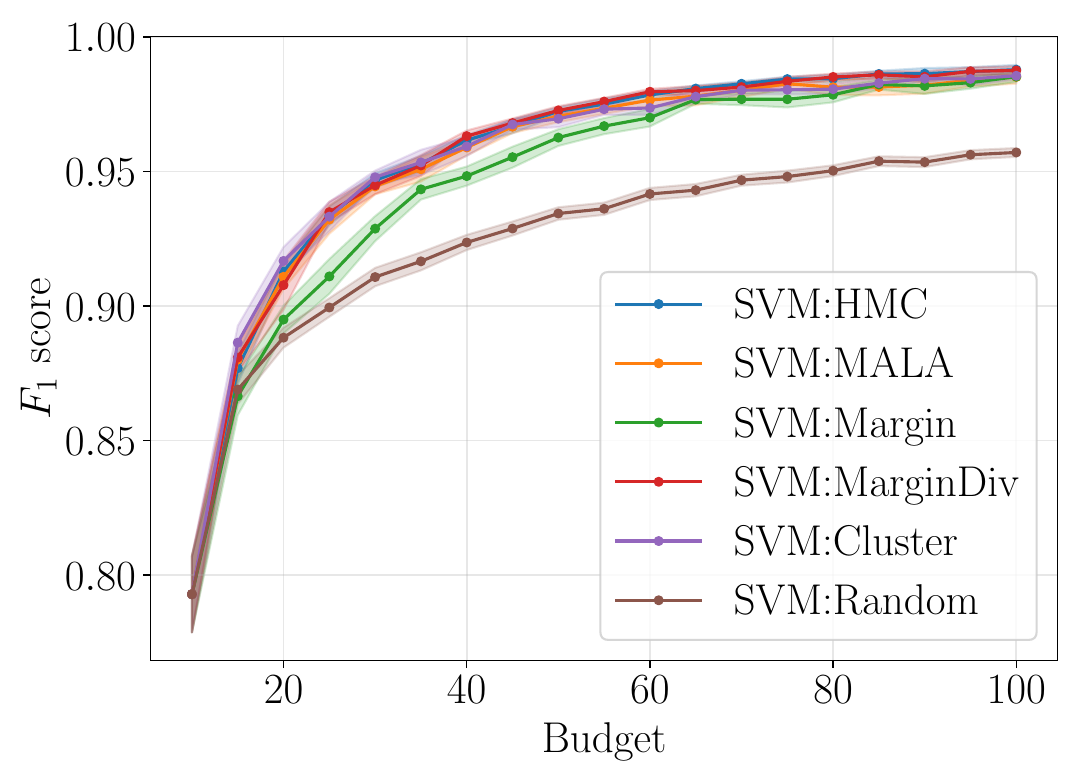}
    \caption{}
    \end{subfigure}
    \hfill
    \begin{subfigure}[b]{0.45\textwidth}
    \includegraphics[width=\linewidth]{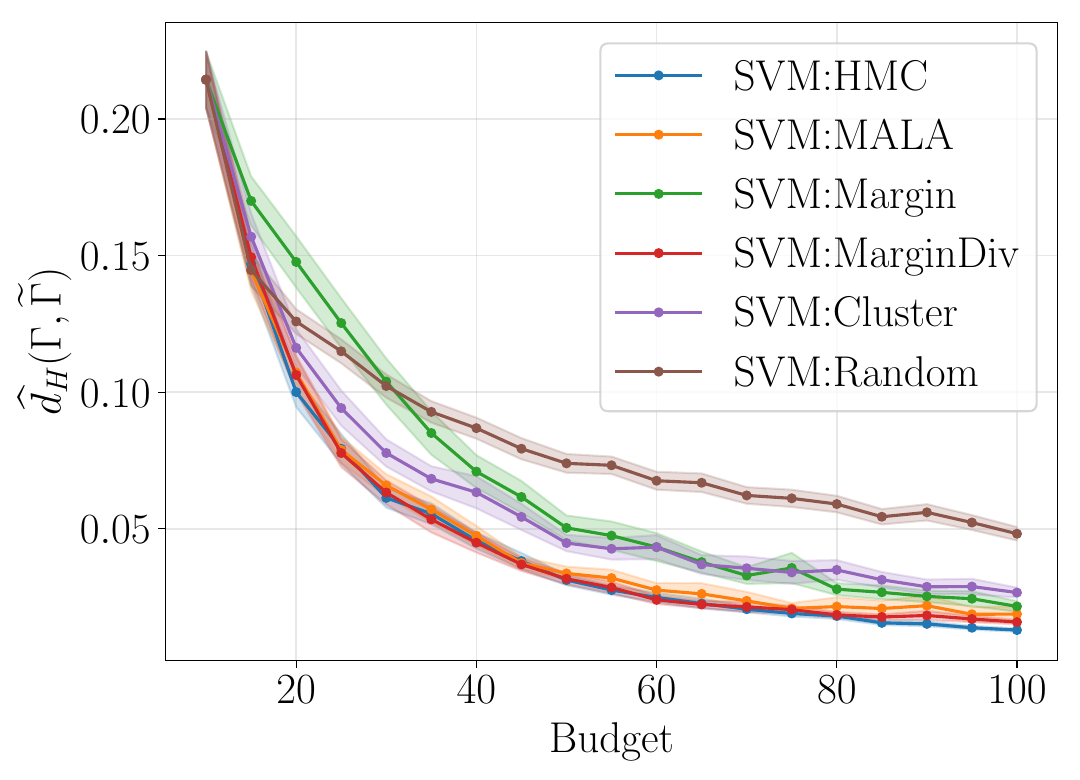}
    \caption{}
    \end{subfigure}
    
    \caption{Adaptive sampling results on \eqref{eq:f1-2d-discontinuities}. (a) Mean $F_1$ score; (b) mean Hausdorff distance. Shaded regions represent a $95 \%$ CI on the mean.}
    \label{fig:f1-2d-discontinuities}

\end{figure}

\begin{figure}[H]
    \centering
    \includegraphics[width=\linewidth]{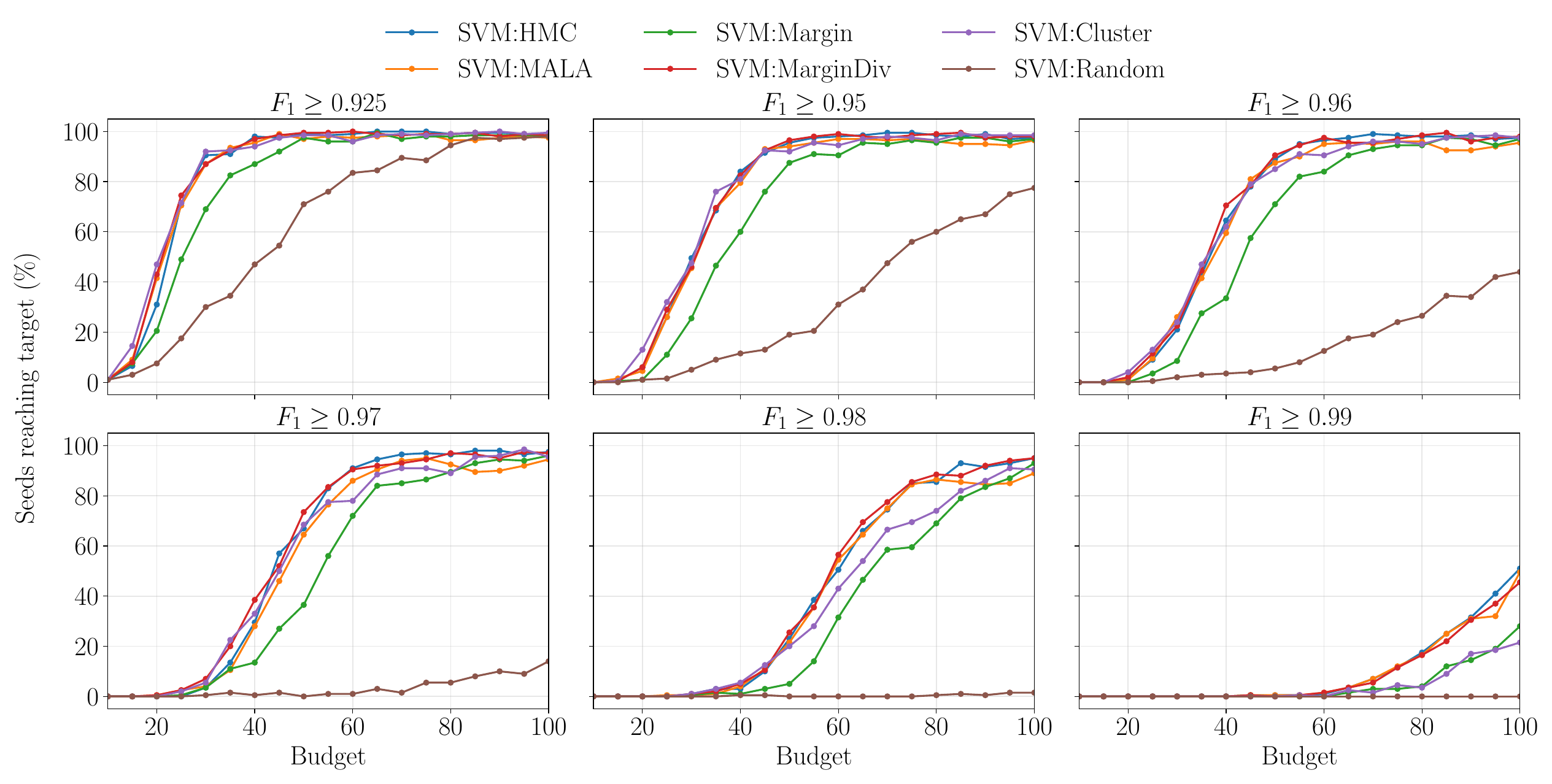}
    \caption{Percentage of seeds reaching different $F_1$ score targets on \eqref{eq:f1-2d-discontinuities}.}
    \label{fig:f1-seeds-at-target}
\end{figure}

We start by considering \eqref{eq:f1-2d-discontinuities}. In Figure \ref{fig:f1-2d-discontinuities} the adaptive sampling trend is depicted. In this case, the target boundary is simple enough so that all active classification algorithms converge to high $F_1$ scores before the computational budget is exhausted. The only noticeable difference stands in the gap between naive margin sampling and other algorithms in the low-data regime. The Hausdorff distance decay reveals a more interesting trend, as cluster sampling is suboptimal and we observe a clearer advantage when using algorithms performing a space-filling downselection. Among them, our proposed method has a slight advantage throughout the active learning iterations. In Figure \ref{fig:f1-seeds-at-target}, we report the percentage of seeds that reaches a certain target $F_1$ score for a given computational budget; this helps to compare the robustness of the algorithms to initialization from a different perspective with respect to the shaded regions in Figure \ref{fig:f1-2d-discontinuities}. The space-filling algorithms (HMC, MALA, margin with diversity) are essentially equivalent. We notice furthermore that for high target scores cluster sampling plateaus.

Lastly, Figure \ref{fig:f1_grid} shows the progression of the boundary approximation, for all the algorithms tested, on a representative seed: all space-filling algorithms produce an even distribution of samples along the estimated boundary, leading to improved accuracy (as already evidenced by the lower Hausdorff distance), while margin and cluster sampling tend to oversample the same regions, creating local clusters and promoting local overfitting.

\vspace{10pt}

% Paragraph break between rows of the 6x4 grids. The journal file tightens this.
\providecommand{\gridrowsep}{\par}
\begin{figure}[H]
    \centering
    % 4x2 grid (no lateral row labels). N labels are placed below the second row.
    \newcommand{\colcap}[1]{\makebox[0.238\linewidth]{\centering\small #1}}

    % Row 1
    \begin{subfigure}[t]{0.238\linewidth}
        \centering
        \includegraphics[width=\linewidth]{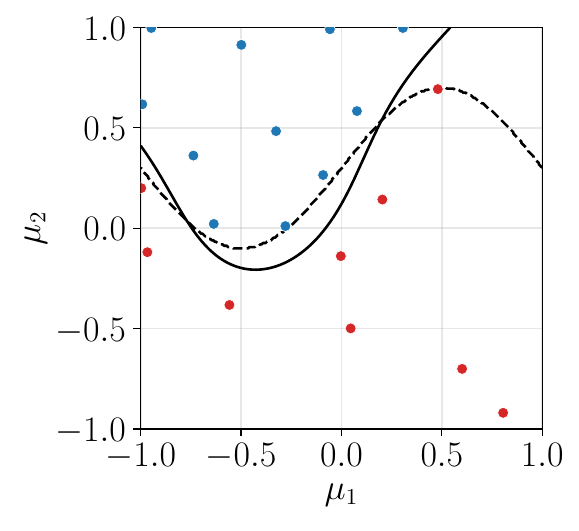}
        \label{fig:f1_hmc_n20}
    \end{subfigure}\hfill
    \begin{subfigure}[t]{0.238\linewidth}
        \centering
        \includegraphics[width=\linewidth]{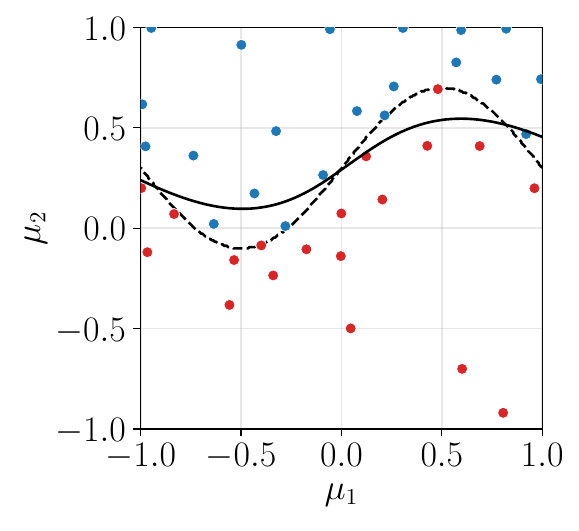}
        \label{fig:f1_hmc_n40}
    \end{subfigure}\hfill
    \begin{subfigure}[t]{0.238\linewidth}
        \centering
        \includegraphics[width=\linewidth]{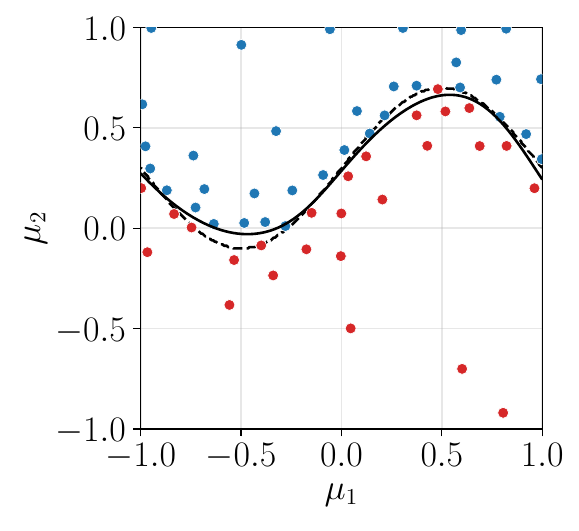}
        \label{fig:f1_hmc_n60}
    \end{subfigure}\hfill
    \begin{subfigure}[t]{0.238\linewidth}
        \centering
        \includegraphics[width=\linewidth]{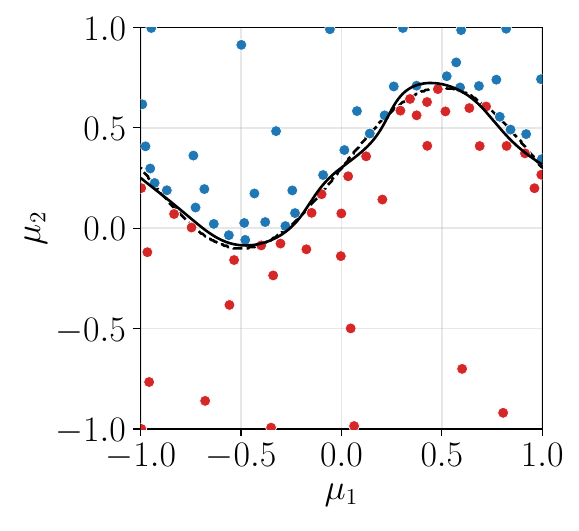}
        \label{fig:f1_hmc_n80}
    \end{subfigure}
    \gridrowsep
    % Row 2
    \begin{subfigure}[t]{0.238\linewidth}
        \centering
        \includegraphics[width=\linewidth]{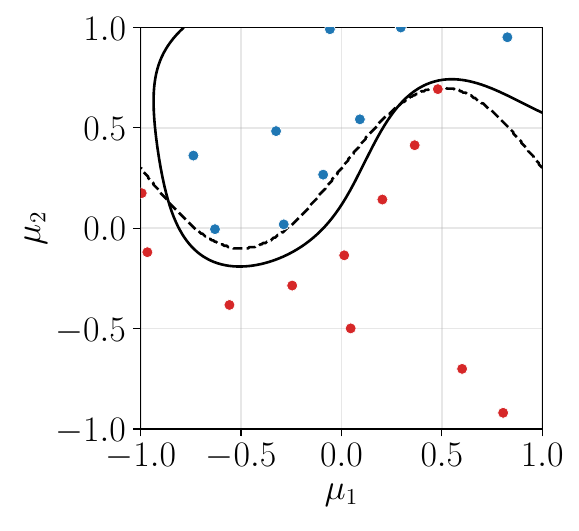}
        \label{fig:f1_mala_n20}
    \end{subfigure}\hfill
    \begin{subfigure}[t]{0.238\linewidth}
        \centering
        \includegraphics[width=\linewidth]{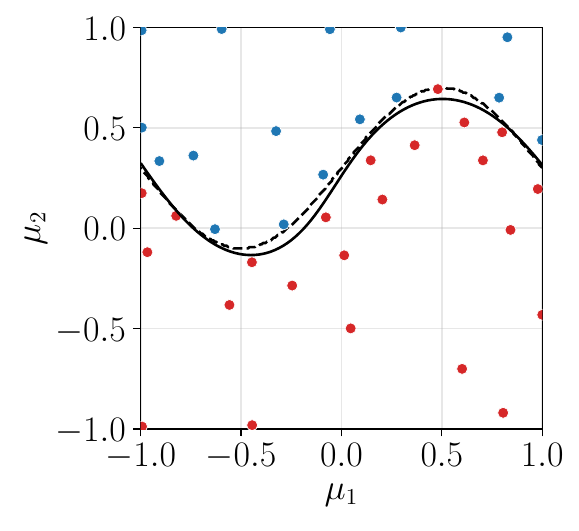}
        \label{fig:f1_mala_n40}
    \end{subfigure}\hfill
    \begin{subfigure}[t]{0.238\linewidth}
        \centering
        \includegraphics[width=\linewidth]{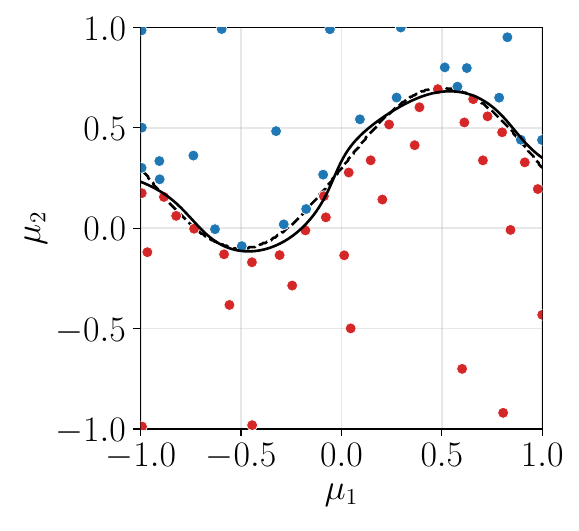}
        \label{fig:f1_mala_n60}
    \end{subfigure}\hfill
    \begin{subfigure}[t]{0.238\linewidth}
        \centering
        \includegraphics[width=\linewidth]{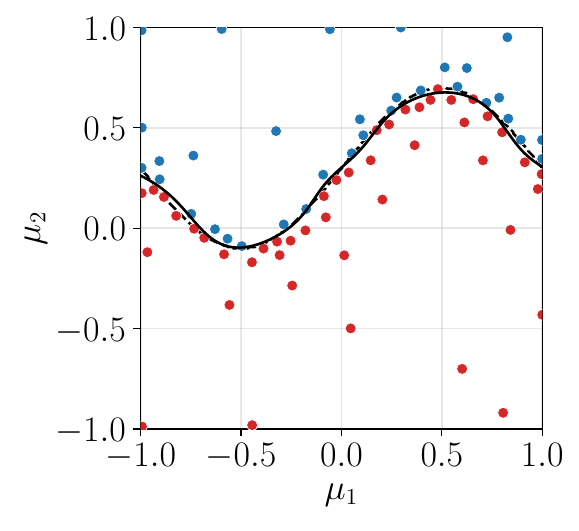}
        \label{fig:f1_mala_n80}
    \end{subfigure}
    \gridrowsep
    % Row 3
    \begin{subfigure}[t]{0.238\linewidth}
        \centering
        \includegraphics[width=\linewidth]{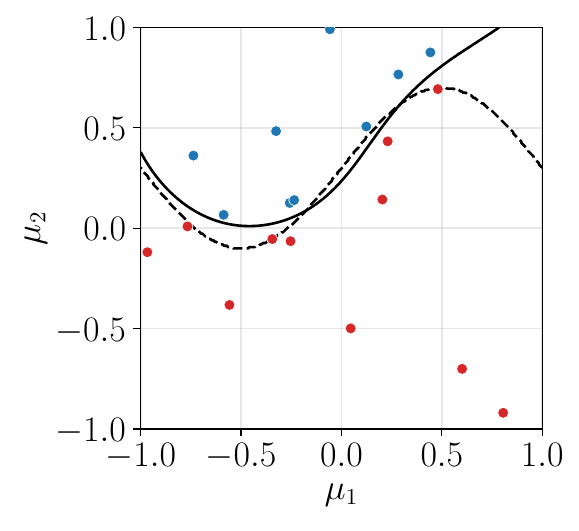}
        \label{fig:f1_margin_n20}
    \end{subfigure}\hfill
    \begin{subfigure}[t]{0.238\linewidth}
        \centering
        \includegraphics[width=\linewidth]{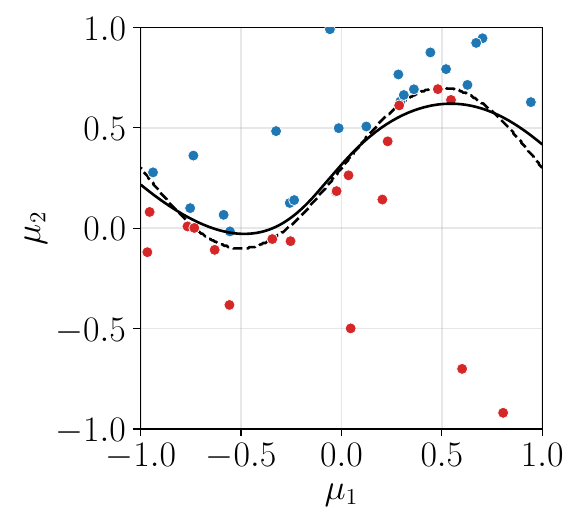}
        \label{fig:f1_margin_n40}
    \end{subfigure}\hfill
    \begin{subfigure}[t]{0.238\linewidth}
        \centering
        \includegraphics[width=\linewidth]{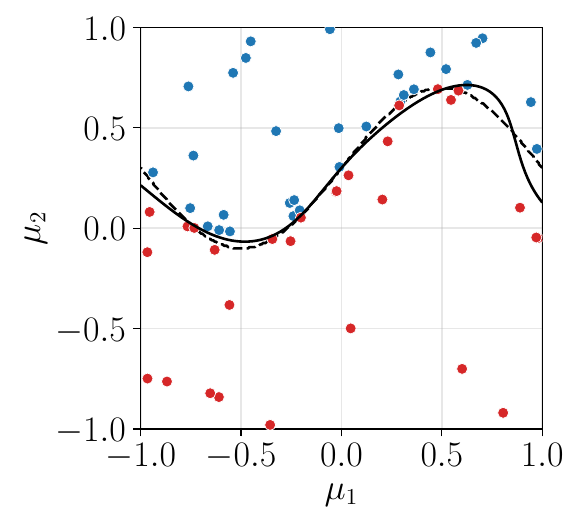}
        \label{fig:f1_margin_n60}
    \end{subfigure}\hfill
    \begin{subfigure}[t]{0.238\linewidth}
        \centering
        \includegraphics[width=\linewidth]{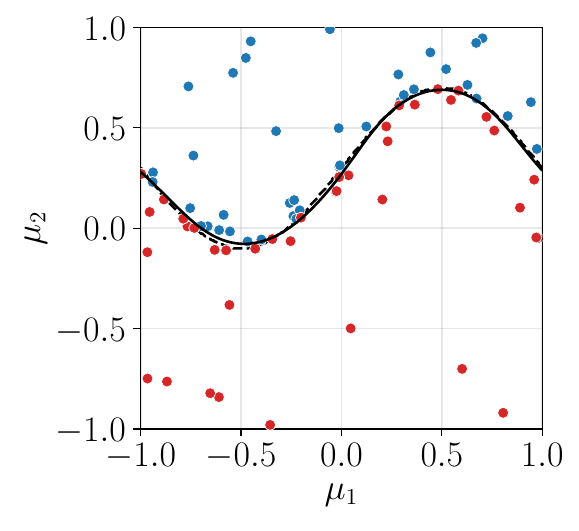}
        \label{fig:f1_margin_n80}
    \end{subfigure}
    \gridrowsep
    % Row 4
    \begin{subfigure}[t]{0.238\linewidth}
        \centering
        \includegraphics[width=\linewidth]{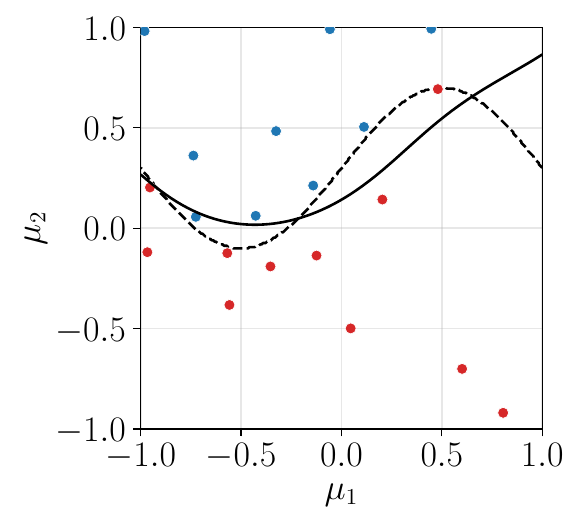}
        \label{fig:f1_margin_diversity_n20}
    \end{subfigure}\hfill
    \begin{subfigure}[t]{0.238\linewidth}
        \centering
        \includegraphics[width=\linewidth]{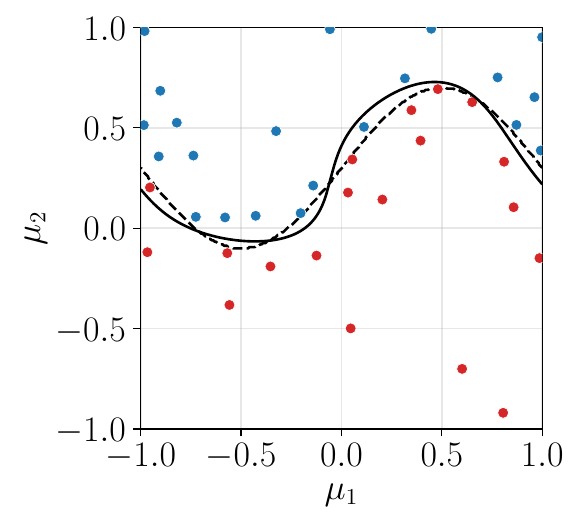}
        \label{fig:f1_margin_diversity_n40}
    \end{subfigure}\hfill
    \begin{subfigure}[t]{0.238\linewidth}
        \centering
        \includegraphics[width=\linewidth]{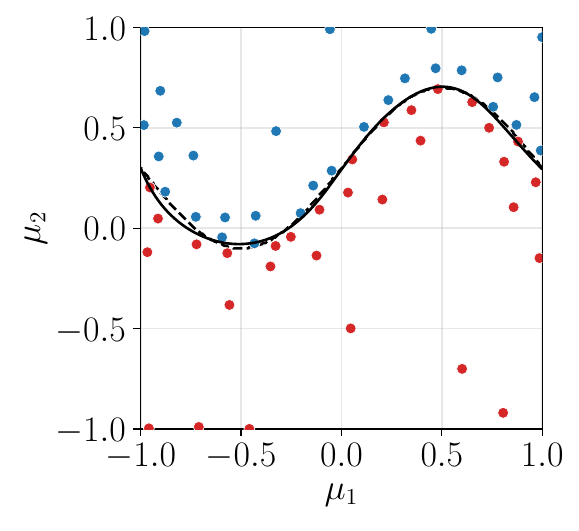}
        \label{fig:f1_margin_diversity_n60}
    \end{subfigure}\hfill
    \begin{subfigure}[t]{0.238\linewidth}
        \centering
        \includegraphics[width=\linewidth]{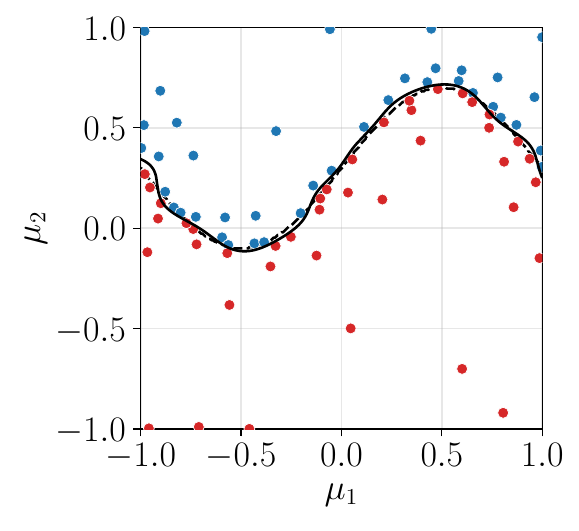}
        \label{fig:f1_margin_diversity_n80}
    \end{subfigure}
    \gridrowsep
    % Row 5
    \begin{subfigure}[t]{0.238\linewidth}
        \centering
        \includegraphics[width=\linewidth]{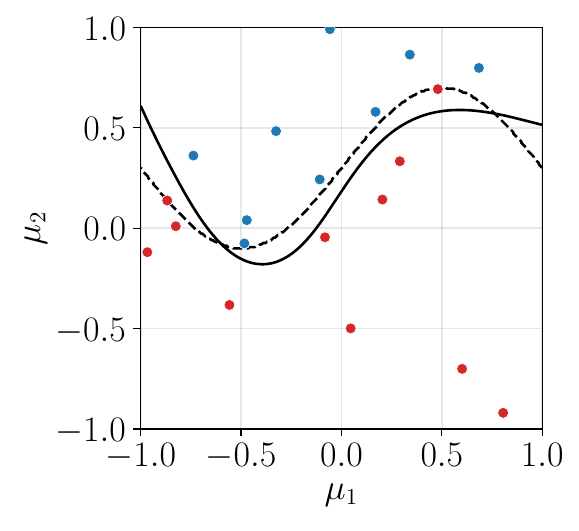}
        \label{fig:f1_cluster_n20}
    \end{subfigure}\hfill
    \begin{subfigure}[t]{0.238\linewidth}
        \centering
        \includegraphics[width=\linewidth]{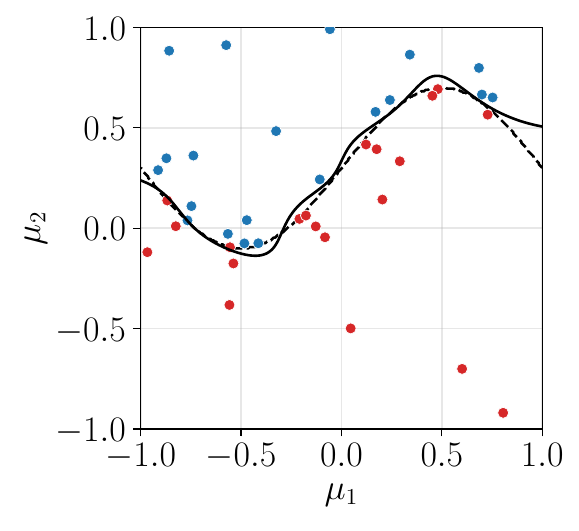}
        \label{fig:f1_cluster_n40}
    \end{subfigure}\hfill
    \begin{subfigure}[t]{0.238\linewidth}
        \centering
        \includegraphics[width=\linewidth]{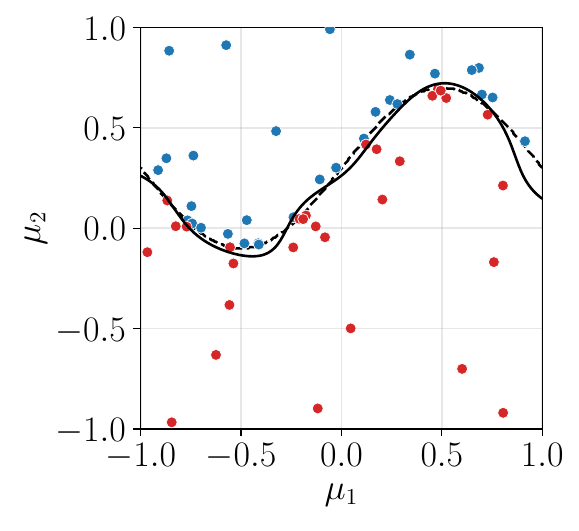}
        \label{fig:f1_cluster_n60}
    \end{subfigure}\hfill
    \begin{subfigure}[t]{0.238\linewidth}
        \centering
        \includegraphics[width=\linewidth]{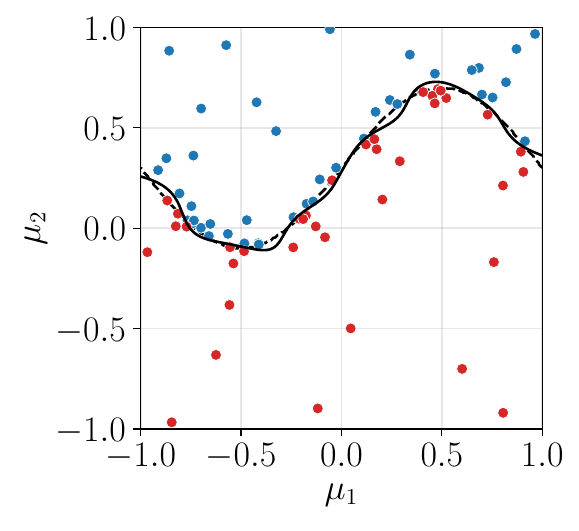}
        \label{fig:f1_cluster_n80}
    \end{subfigure}
    \gridrowsep
    % Row 6
    \begin{subfigure}[t]{0.238\linewidth}
        \centering
        \includegraphics[width=\linewidth]{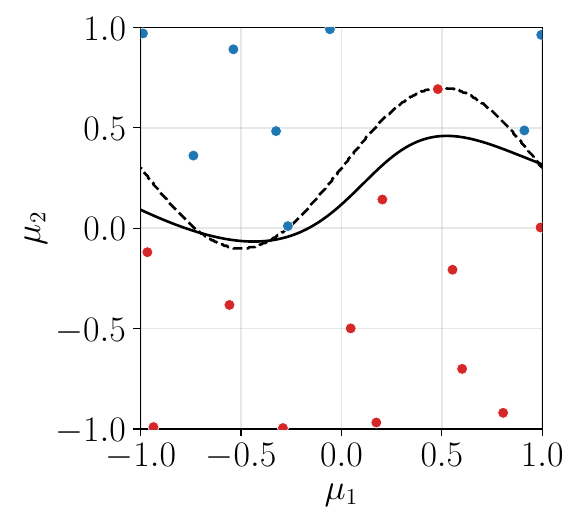}
        \caption{$N = 20.$}
        \label{fig:f1_random_n20}
    \end{subfigure}\hfill
    \begin{subfigure}[t]{0.238\linewidth}
        \centering
        \includegraphics[width=\linewidth]{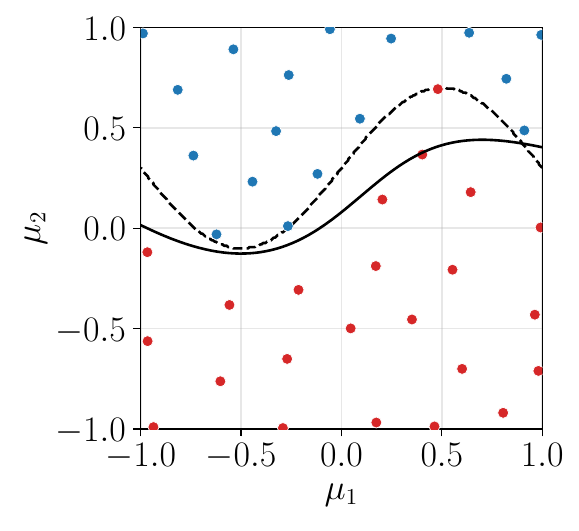}
        \caption{$N = 40.$}
        \label{fig:f1_random_n40}
    \end{subfigure}\hfill
    \begin{subfigure}[t]{0.238\linewidth}
        \centering
        \includegraphics[width=\linewidth]{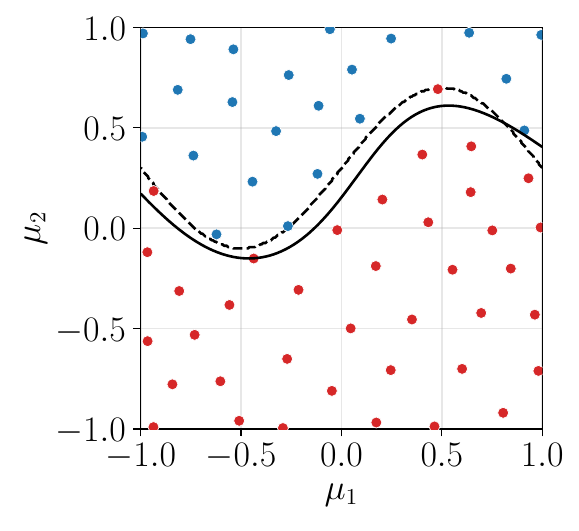}
        \caption{$N = 60.$}
        \label{fig:f1_random_n60}
    \end{subfigure}\hfill
    \begin{subfigure}[t]{0.238\linewidth}
        \centering
        \includegraphics[width=\linewidth]{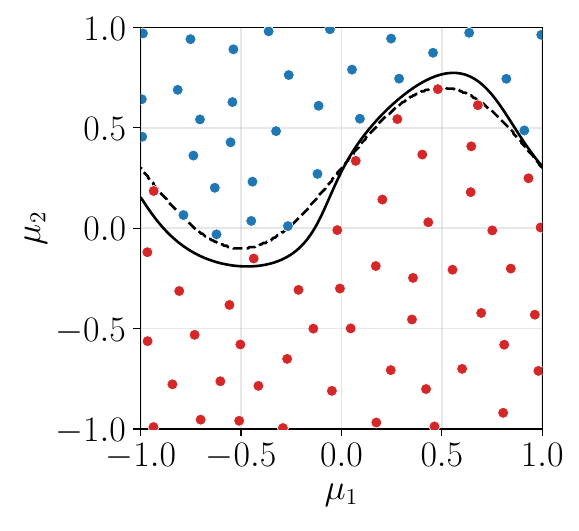}
        \caption{$N = 80.$}
        \label{fig:f1_random_n80}
    \end{subfigure}

    \caption{Adaptive sampling on \eqref{eq:f1-2d-discontinuities}. From top to bottom HMC, MALA, margin, margin + diversity, cluster and random sampling. Dashed/solid lines are the true/SVM boundary, respectively.}
    \label{fig:f1_grid}
\end{figure}

\begin{figure}[H]

    \begin{subfigure}[b]{0.45\textwidth}
    \includegraphics[width=\linewidth]{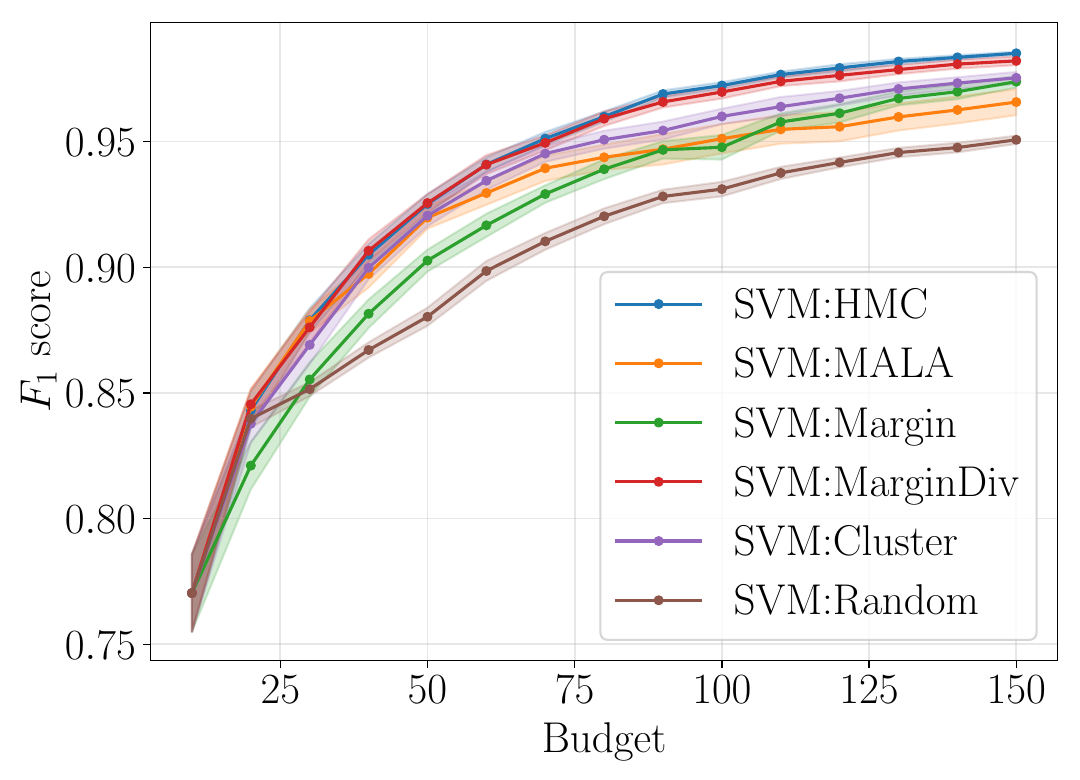}
    \caption{}
    \end{subfigure}
    \hfill
    \begin{subfigure}[b]{0.45\textwidth}
    \includegraphics[width=\linewidth]{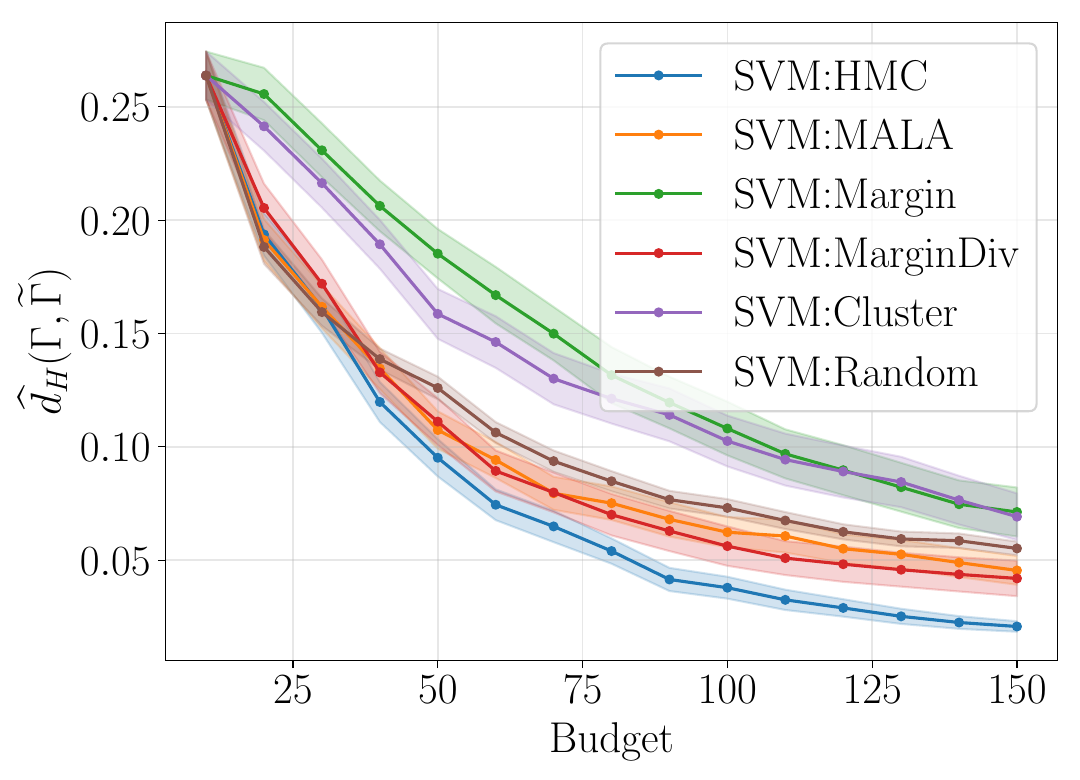}
    \caption{}
    \end{subfigure}
    
    \caption{Adaptive sampling results on \eqref{eq:f2-2d-discontinuities}. (a) Mean $F_1$ score; (b) mean Hausdorff distance. Shaded regions represent a $95 \%$ CI on the mean.}
    \label{fig:f2-2d-discontinuities}

\end{figure}

\begin{figure}[H]
    \centering
    \includegraphics[width=\linewidth]{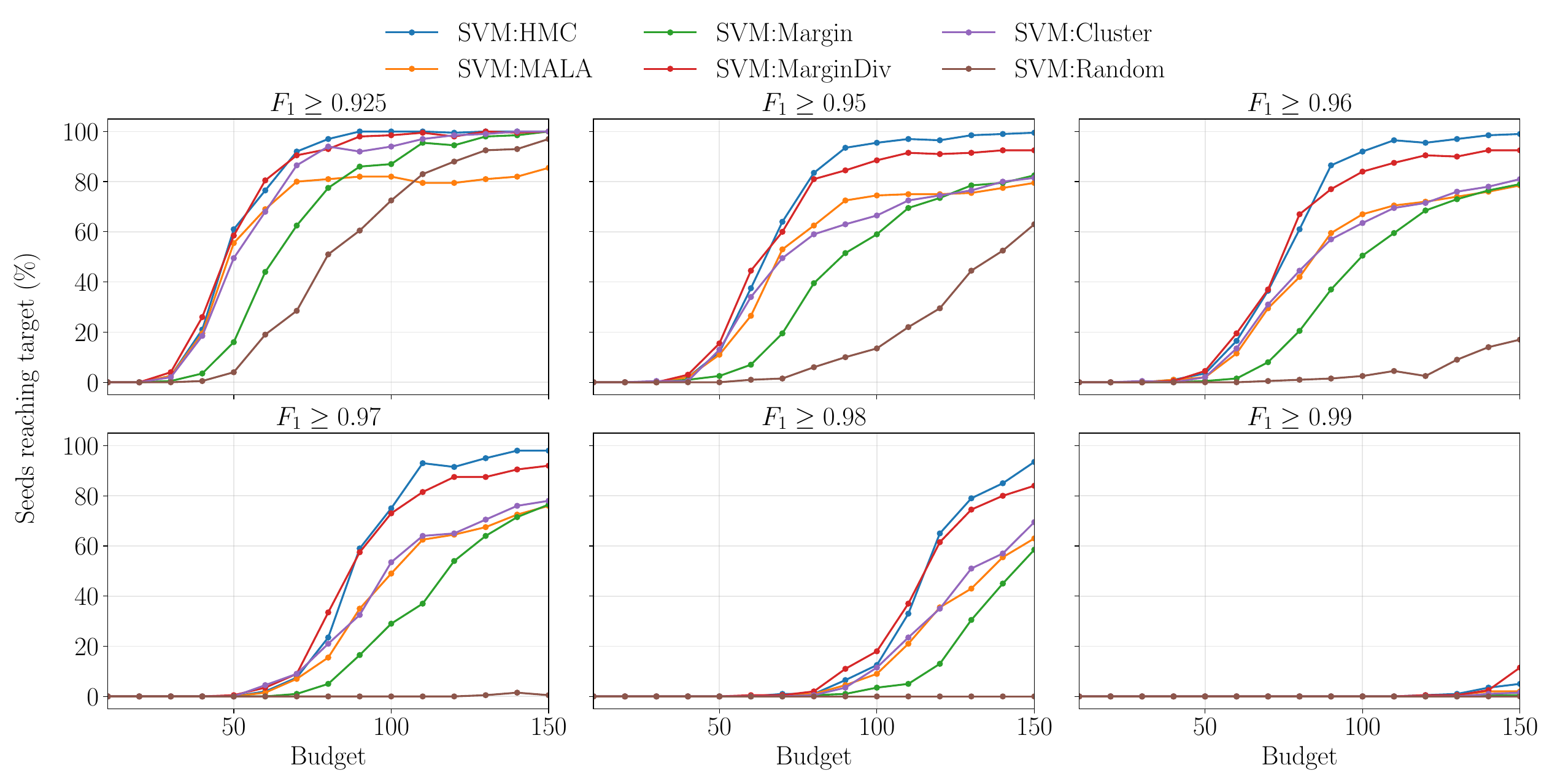}
    \caption{Percentage of seeds reaching different $F_1$ score targets on \eqref{eq:f2-2d-discontinuities}.}
    \label{fig:f2-seeds-at-target}
\end{figure}

\begin{figure}[H]
    \centering
    % 4x2 grid (no lateral row labels). N labels are placed below the second row.
    \newcommand{\colcap}[1]{\makebox[0.238\linewidth]{\centering\small #1}}

    % Row 1
    \begin{subfigure}[t]{0.238\linewidth}
        \centering
        \includegraphics[width=\linewidth]{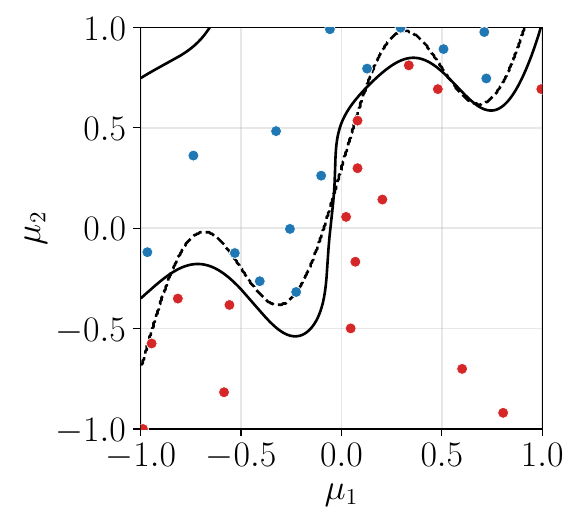}
        \label{fig:f2_hmc_n30}
    \end{subfigure}\hfill
    \begin{subfigure}[t]{0.238\linewidth}
        \centering
        \includegraphics[width=\linewidth]{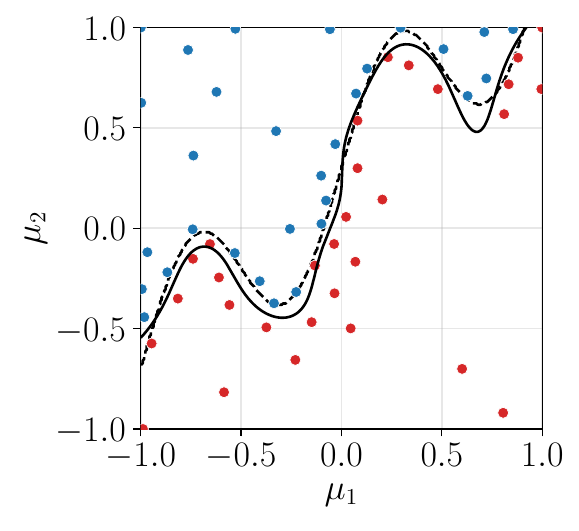}
        \label{fig:f2_hmc_n60}
    \end{subfigure}\hfill
    \begin{subfigure}[t]{0.238\linewidth}
        \centering
        \includegraphics[width=\linewidth]{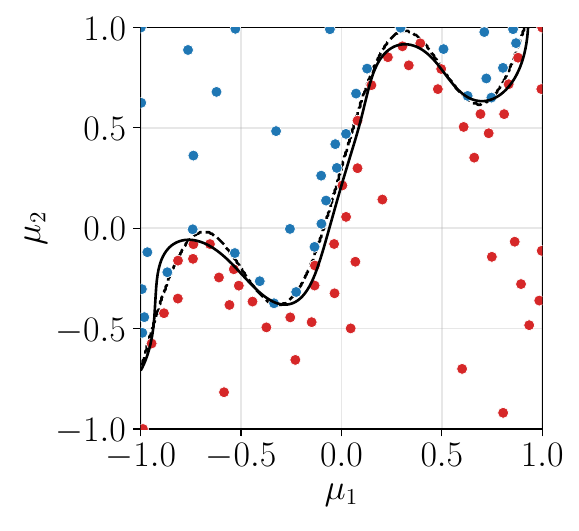}
        \label{fig:f2_hmc_n90}
    \end{subfigure}\hfill
    \begin{subfigure}[t]{0.238\linewidth}
        \centering
        \includegraphics[width=\linewidth]{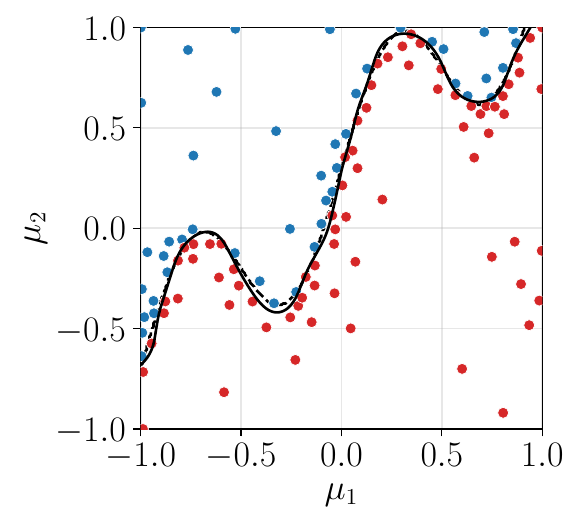}
        \label{fig:f2_hmc_n120}
    \end{subfigure}
    \gridrowsep
    % Row 2
    \begin{subfigure}[t]{0.238\linewidth}
        \centering
        \includegraphics[width=\linewidth]{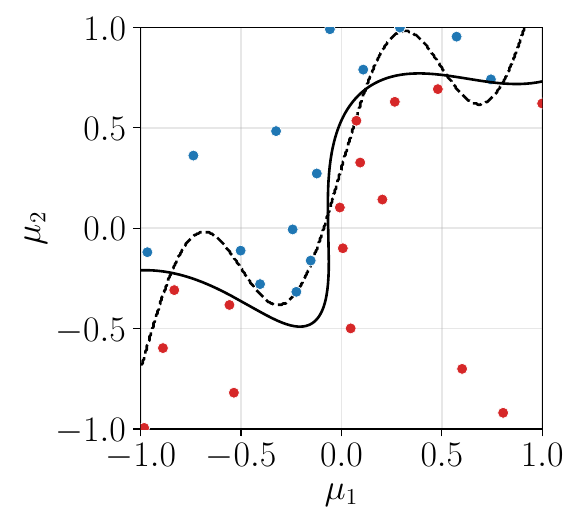}
        \label{fig:f2_mala_n30}
    \end{subfigure}\hfill
    \begin{subfigure}[t]{0.238\linewidth}
        \centering
        \includegraphics[width=\linewidth]{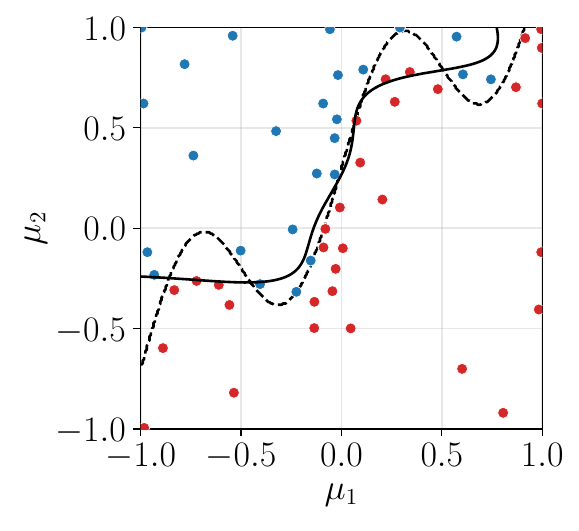}
        \label{fig:f2_mala_n60}
    \end{subfigure}\hfill
    \begin{subfigure}[t]{0.238\linewidth}
        \centering
        \includegraphics[width=\linewidth]{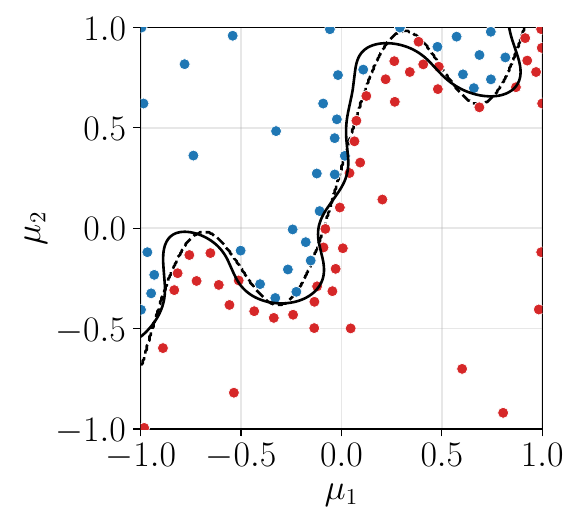}
        \label{fig:f2_mala_n90}
    \end{subfigure}\hfill
    \begin{subfigure}[t]{0.238\linewidth}
        \centering
        \includegraphics[width=\linewidth]{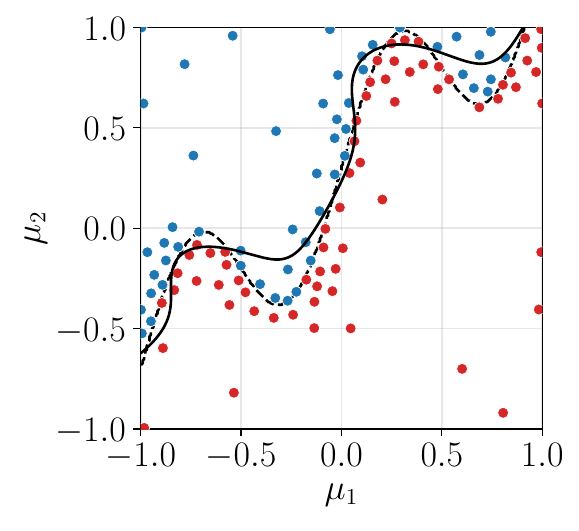}
        \label{fig:f2_mala_n120}
    \end{subfigure}
    \gridrowsep
    % Row 3
    \begin{subfigure}[t]{0.238\linewidth}
        \centering
        \includegraphics[width=\linewidth]{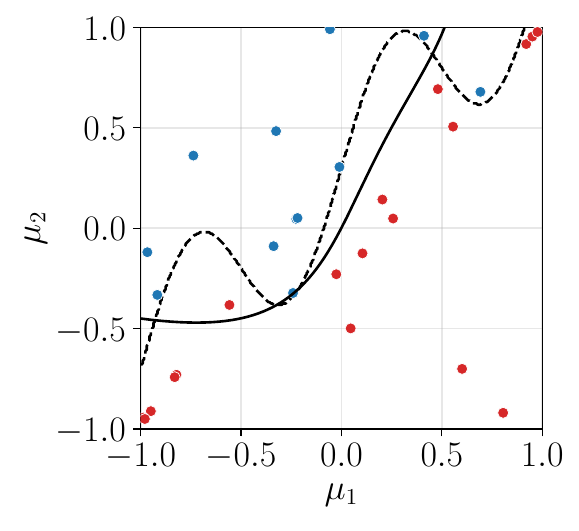}
        \label{fig:f2_margin_n30}
    \end{subfigure}\hfill
    \begin{subfigure}[t]{0.238\linewidth}
        \centering
        \includegraphics[width=\linewidth]{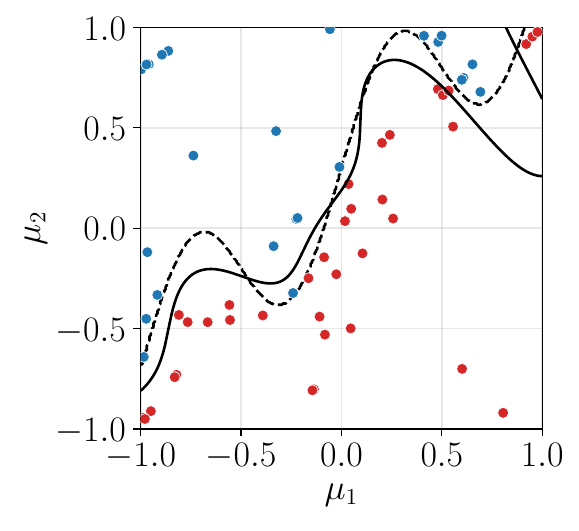}
        \label{fig:f2_margin_n60}
    \end{subfigure}\hfill
    \begin{subfigure}[t]{0.238\linewidth}
        \centering
        \includegraphics[width=\linewidth]{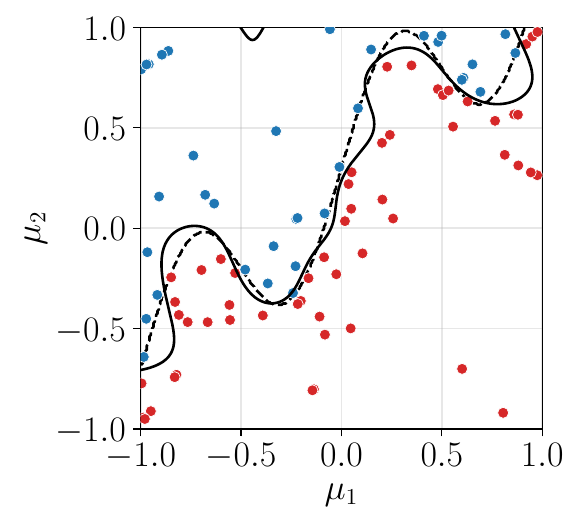}
        \label{fig:f2_margin_n90}
    \end{subfigure}\hfill
    \begin{subfigure}[t]{0.238\linewidth}
        \centering
        \includegraphics[width=\linewidth]{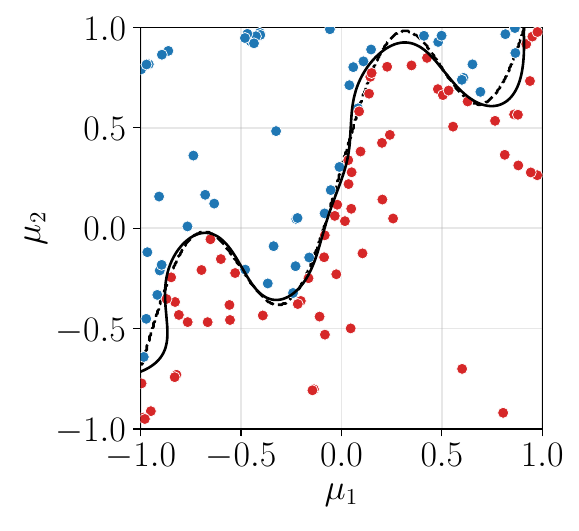}
        \label{fig:f2_margin_n120}
    \end{subfigure}
    \gridrowsep
    % Row 4
    \begin{subfigure}[t]{0.238\linewidth}
        \centering
        \includegraphics[width=\linewidth]{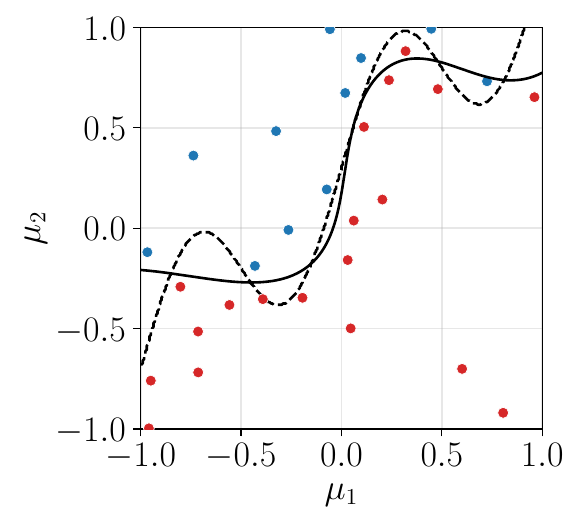}
        \label{fig:f2_margin_diversity_n30}
    \end{subfigure}\hfill
    \begin{subfigure}[t]{0.238\linewidth}
        \centering
        \includegraphics[width=\linewidth]{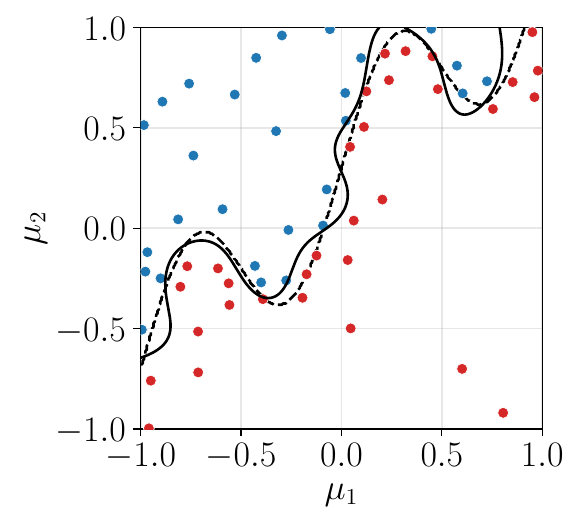}
        \label{fig:f2_margin_diversity_n60}
    \end{subfigure}\hfill
    \begin{subfigure}[t]{0.238\linewidth}
        \centering
        \includegraphics[width=\linewidth]{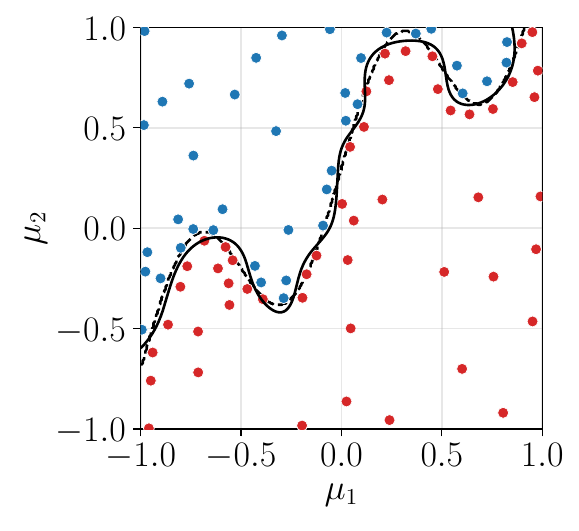}
        \label{fig:f2_margin_diversity_n90}
    \end{subfigure}\hfill
    \begin{subfigure}[t]{0.238\linewidth}
        \centering
        \includegraphics[width=\linewidth]{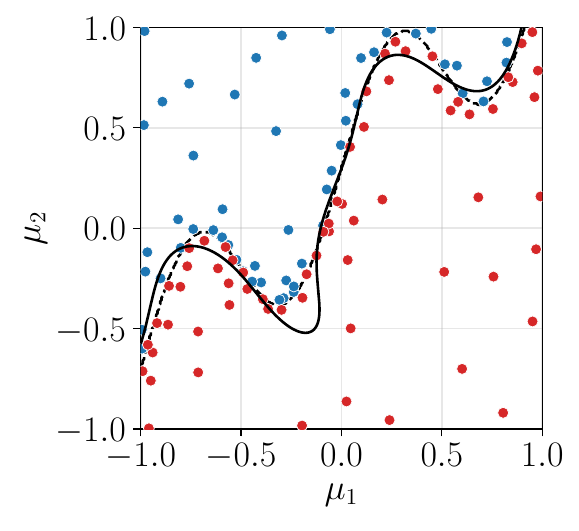}
        \label{fig:f2_margin_diversity_n120}
    \end{subfigure}
    \gridrowsep
    % Row 5
    \begin{subfigure}[t]{0.238\linewidth}
        \centering
        \includegraphics[width=\linewidth]{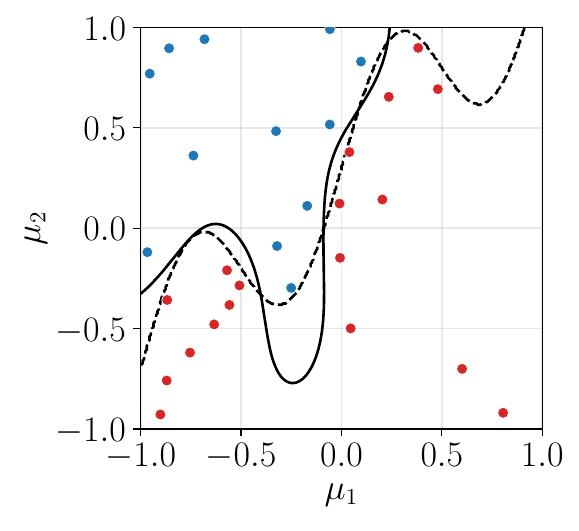}
        \label{fig:f2_cluster_n30}
    \end{subfigure}\hfill
    \begin{subfigure}[t]{0.238\linewidth}
        \centering
        \includegraphics[width=\linewidth]{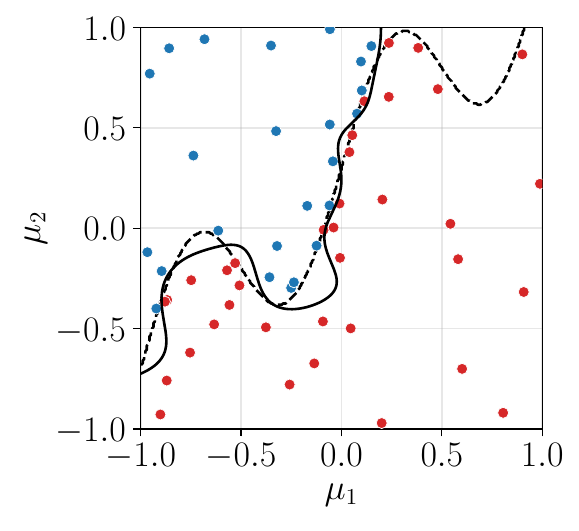}
        \label{fig:f2_cluster_n60}
    \end{subfigure}\hfill
    \begin{subfigure}[t]{0.238\linewidth}
        \centering
        \includegraphics[width=\linewidth]{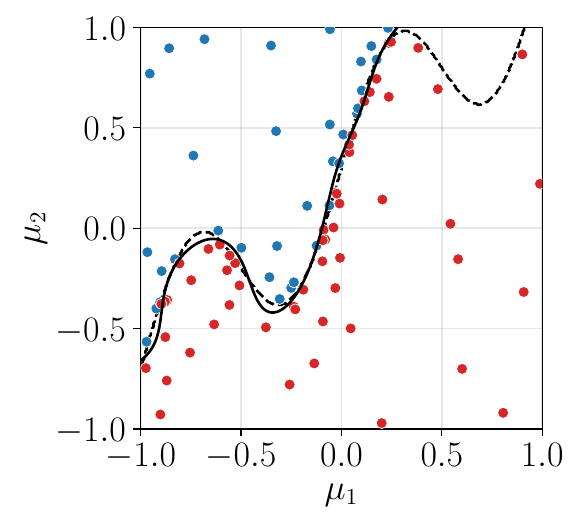}
        \label{fig:f2_cluster_n90}
    \end{subfigure}\hfill
    \begin{subfigure}[t]{0.238\linewidth}
        \centering
        \includegraphics[width=\linewidth]{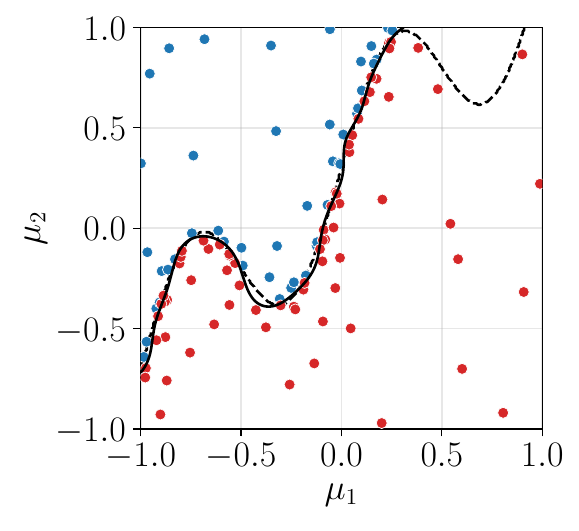}
        \label{fig:f2_cluster_n120}
    \end{subfigure}
    \gridrowsep
    % Row 6
    \begin{subfigure}[t]{0.238\linewidth}
        \centering
        \includegraphics[width=\linewidth]{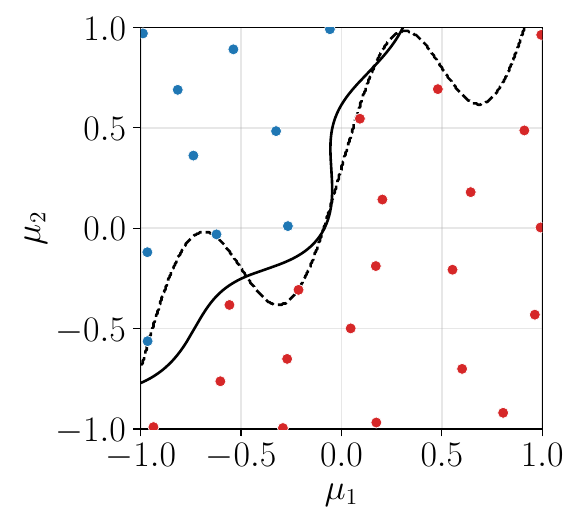}
        \caption{$N = 30.$}
        \label{fig:f2_random_n30}
    \end{subfigure}\hfill
    \begin{subfigure}[t]{0.238\linewidth}
        \centering
        \includegraphics[width=\linewidth]{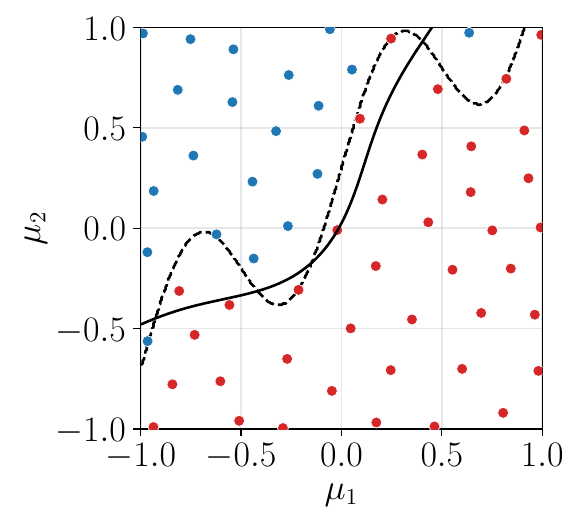}
        \caption{$N = 60.$}
        \label{fig:f2_random_n60}
    \end{subfigure}\hfill
    \begin{subfigure}[t]{0.238\linewidth}
        \centering
        \includegraphics[width=\linewidth]{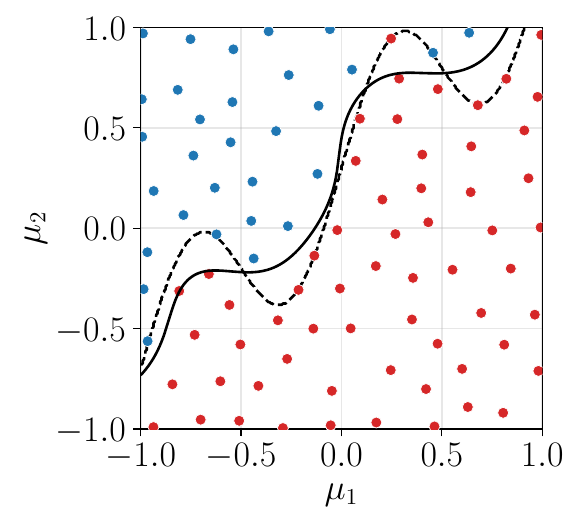}
        \caption{$N = 90.$}
        \label{fig:f2_random_n90}
    \end{subfigure}\hfill
    \begin{subfigure}[t]{0.238\linewidth}
        \centering
        \includegraphics[width=\linewidth]{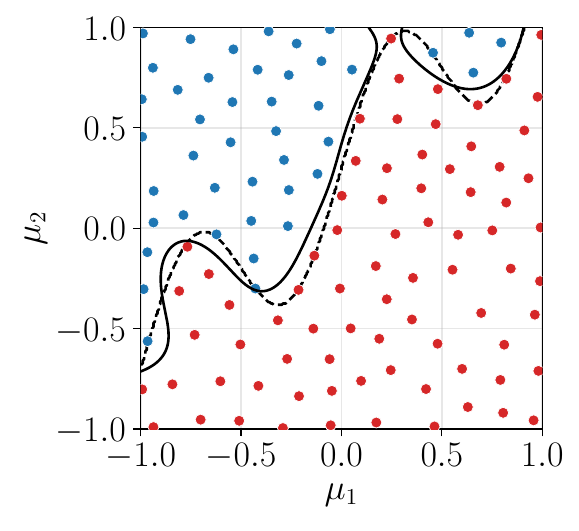}
        \caption{$N = 120.$}
        \label{fig:f2_random_n120}
    \end{subfigure}

    \caption{Adaptive sampling on \eqref{eq:f2-2d-discontinuities}. From top to bottom HMC, MALA, margin, margin + diversity, cluster and random sampling. Dashed/solid lines are the true/SVM boundary, respectively.}
    \label{fig:f2_grid}
\end{figure}

A clearer distinction between the space-filling methods can be appreciated by increasing the curvature of the target boundary. In Figure \ref{fig:f2-2d-discontinuities} we consider \eqref{eq:f2-2d-discontinuities} and observe a different trend. Margin and cluster sampling are both struggling to efficiently approximate the boundary (and are actually worse than random sampling on the Hausdorff distance, because they tend to miss entire regions of the boundary; see Figure \ref{fig:f2_grid}), while among the space-filling algorithms, MALA plateaus and HMC outperforms margin with diversity in both $F_1$ score and Hausdorff distance, with a larger gap in the latter metric. We also observe that the width of the confidence intervals is smaller for the proposed algorithm, indicating its superior robustness. In Figure \ref{fig:f2-seeds-at-target}, we equally notice that the proposed algorithm is consistently more likely to achieve very high $F_1$ scores given a fixed computational budget. This example provides useful insights: first, with an increasing complexity of the boundary, diversity-enforcing schemes have better sample efficiency. Secondly, the $F_1$ score and the Hausdorff distance are complementary metrics for boundary approximation quality, i.e. two methods can have similar $F_1$ scores but still infer very different boundaries; as such, the Hausdorff distance, when computable, gives a more thorough assessment of the performance of the different algorithms. The last observation, related to the aforementioned point, is that for complex boundaries continuous search spaces can be more effective than discrete ones, even in low dimensions.

\vspace{10pt}

\begin{figure}[H]

    \begin{subfigure}[b]{0.45\textwidth}
    \includegraphics[width=\linewidth]{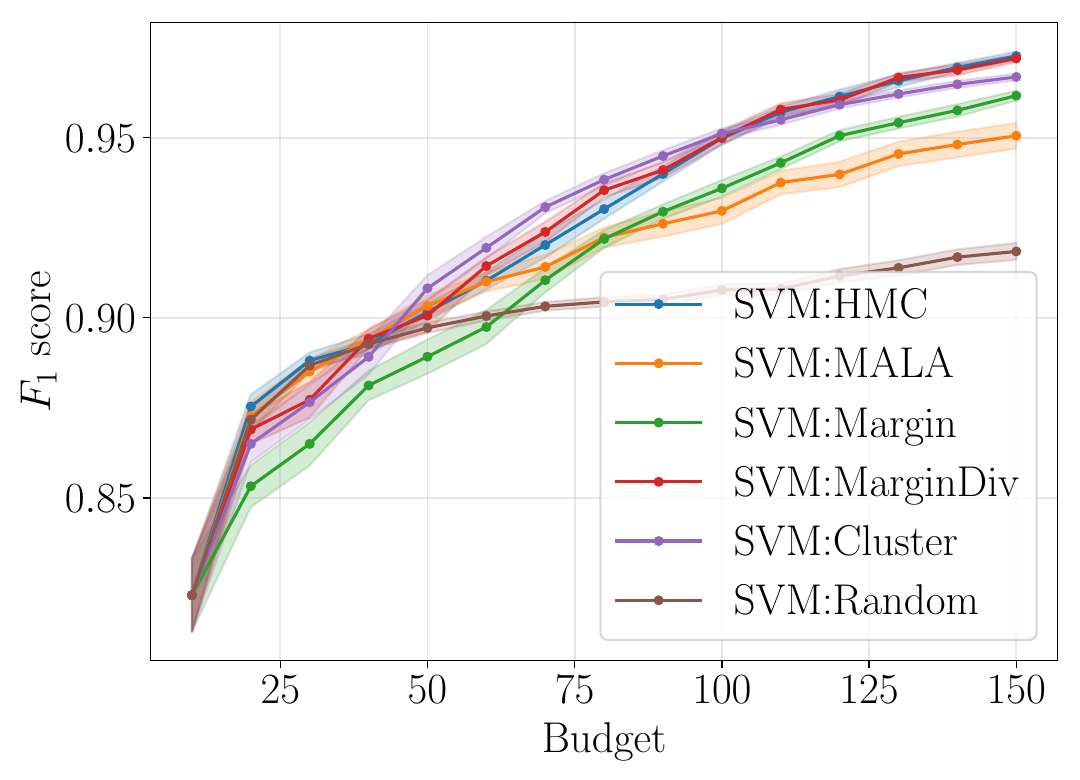}
    \caption{}
    \end{subfigure}
    \hfill
    \begin{subfigure}[b]{0.45\textwidth}
    \includegraphics[width=\linewidth]{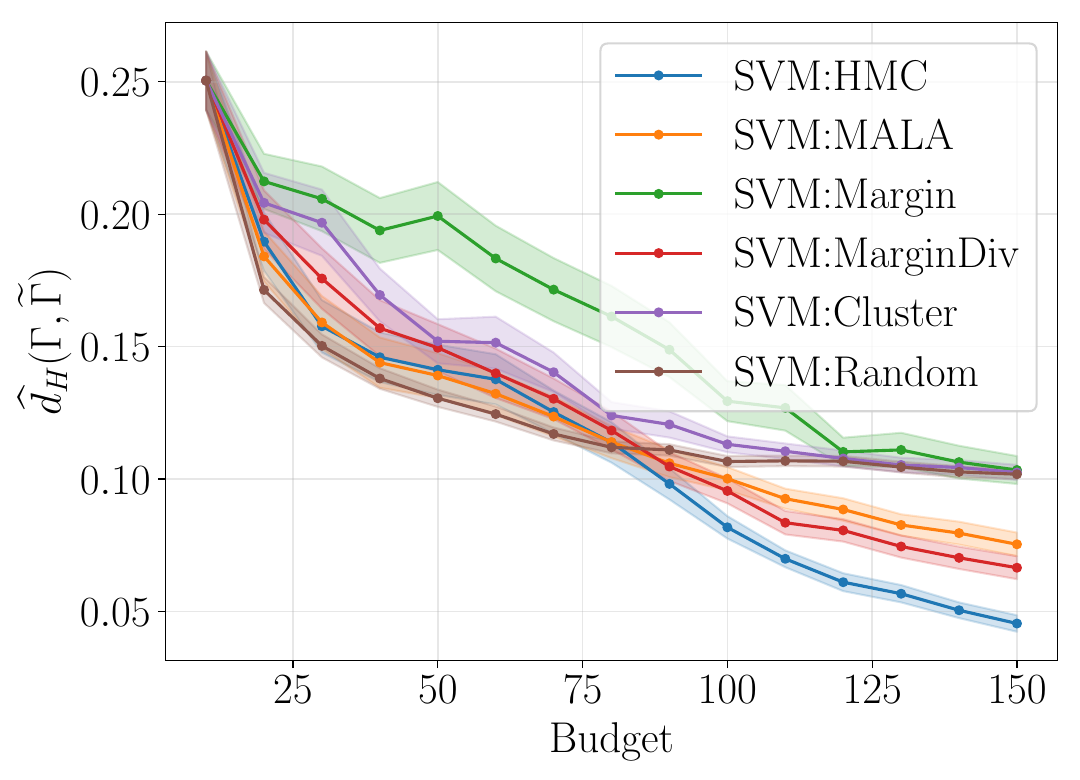}
    \caption{}
    \end{subfigure}
    
    \caption{Adaptive sampling results on \eqref{eq:f3-2d-discontinuities}. (a) Mean $F_1$ score; (b) mean Hausdorff distance. Shaded regions represent a $95 \%$ CI on the mean.}
    \label{fig:f3-2d-discontinuities}

\end{figure}

\begin{figure}[H]
    \centering
    % 4x2 grid (no lateral row labels). N labels are placed below the second row.
    \newcommand{\colcap}[1]{\makebox[0.238\linewidth]{\centering\small #1}}

    % Row 1
    \begin{subfigure}[t]{0.238\linewidth}
        \centering
        \includegraphics[width=\linewidth]{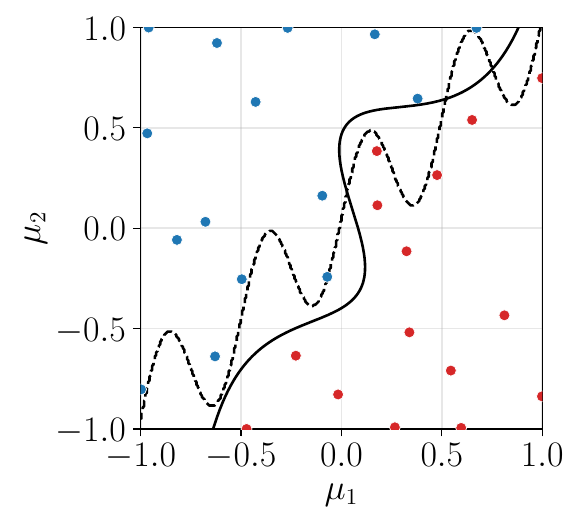}
        \label{fig:f3_hmc_n30}
    \end{subfigure}\hfill
    \begin{subfigure}[t]{0.238\linewidth}
        \centering
        \includegraphics[width=\linewidth]{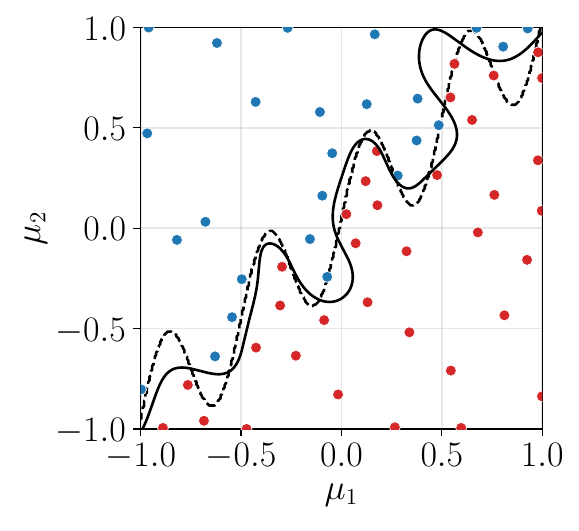}
        \label{fig:f3_hmc_n60}
    \end{subfigure}\hfill
    \begin{subfigure}[t]{0.238\linewidth}
        \centering
        \includegraphics[width=\linewidth]{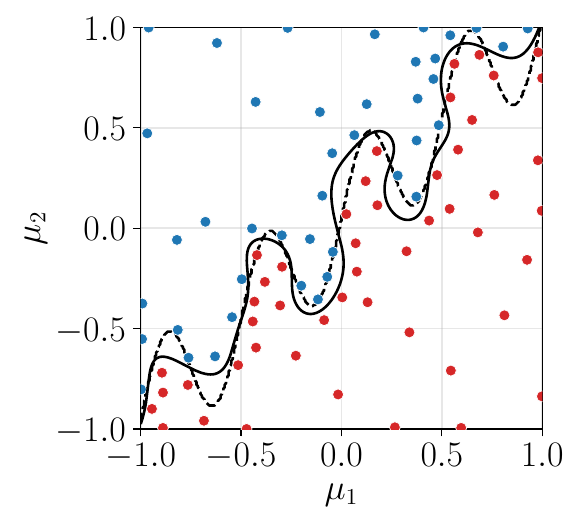}
        \label{fig:f3_hmc_n90}
    \end{subfigure}\hfill
    \begin{subfigure}[t]{0.238\linewidth}
        \centering
        \includegraphics[width=\linewidth]{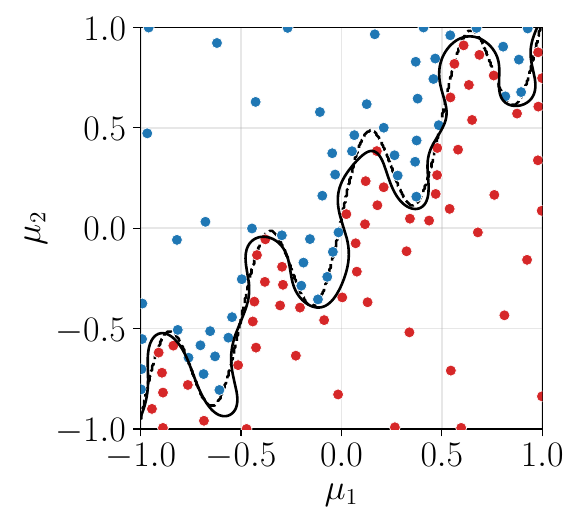}
        \label{fig:f3_hmc_n120}
    \end{subfigure}
    \gridrowsep
    % Row 2
    \begin{subfigure}[t]{0.238\linewidth}
        \centering
        \includegraphics[width=\linewidth]{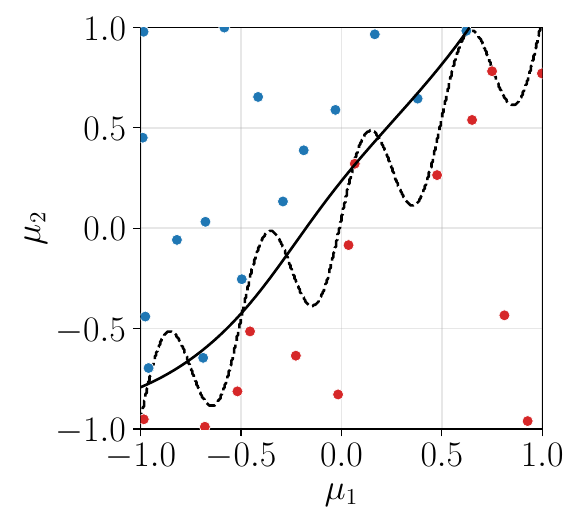}
        \label{fig:f3_mala_n30}
    \end{subfigure}\hfill
    \begin{subfigure}[t]{0.238\linewidth}
        \centering
        \includegraphics[width=\linewidth]{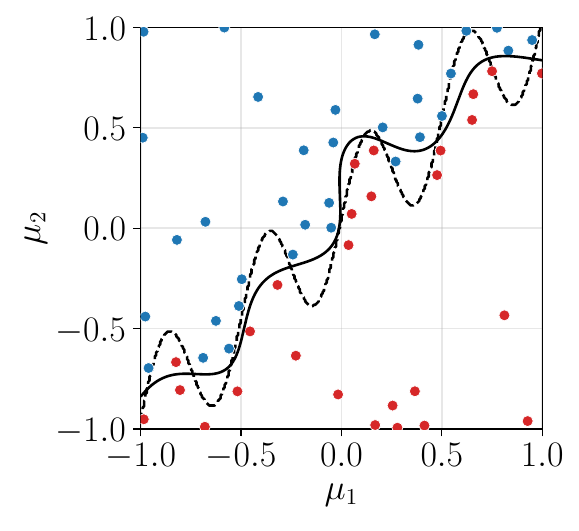}
        \label{fig:f3_mala_n60}
    \end{subfigure}\hfill
    \begin{subfigure}[t]{0.238\linewidth}
        \centering
        \includegraphics[width=\linewidth]{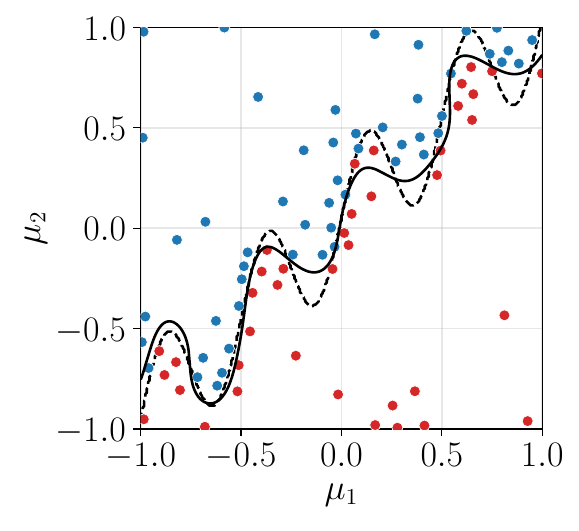}
        \label{fig:f3_mala_n90}
    \end{subfigure}\hfill
    \begin{subfigure}[t]{0.238\linewidth}
        \centering
        \includegraphics[width=\linewidth]{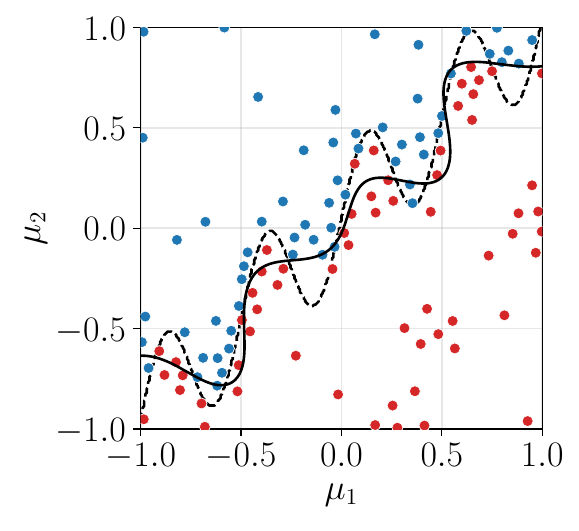}
        \label{fig:f3_mala_n120}
    \end{subfigure}
    \gridrowsep
    % Row 3
    \begin{subfigure}[t]{0.238\linewidth}
        \centering
        \includegraphics[width=\linewidth]{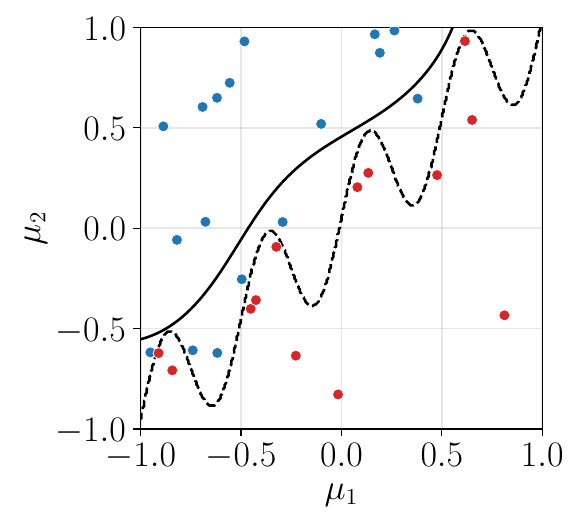}
        \label{fig:f3_margin_n30}
    \end{subfigure}\hfill
    \begin{subfigure}[t]{0.238\linewidth}
        \centering
        \includegraphics[width=\linewidth]{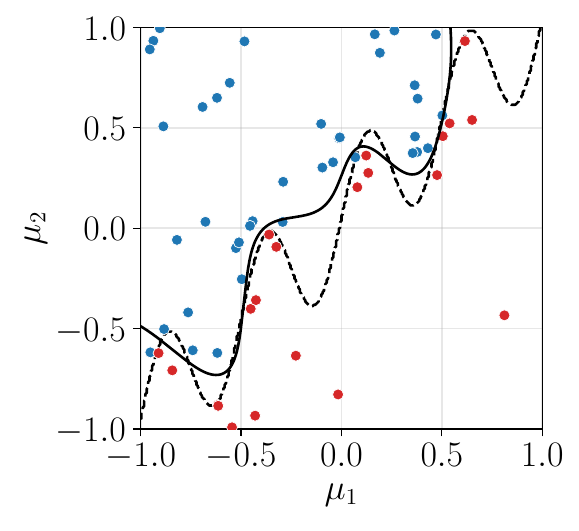}
        \label{fig:f3_margin_n60}
    \end{subfigure}\hfill
    \begin{subfigure}[t]{0.238\linewidth}
        \centering
        \includegraphics[width=\linewidth]{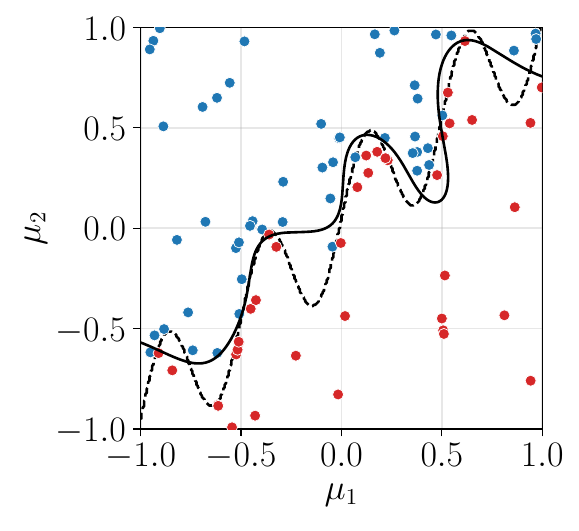}
        \label{fig:f3_margin_n90}
    \end{subfigure}\hfill
    \begin{subfigure}[t]{0.238\linewidth}
        \centering
        \includegraphics[width=\linewidth]{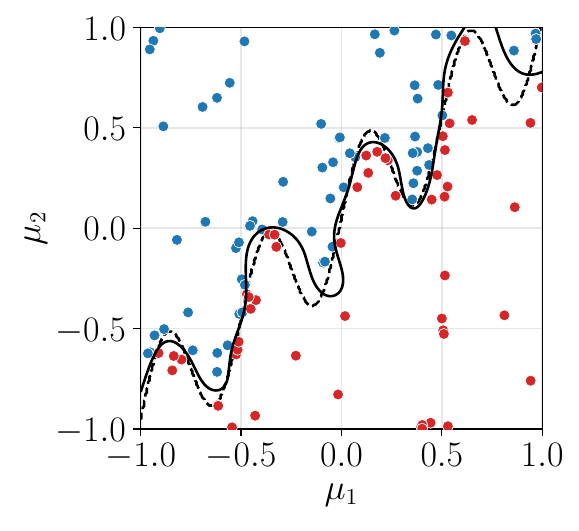}
        \label{fig:f3_margin_n120}
    \end{subfigure}
    \gridrowsep
    % Row 4
    \begin{subfigure}[t]{0.238\linewidth}
        \centering
        \includegraphics[width=\linewidth]{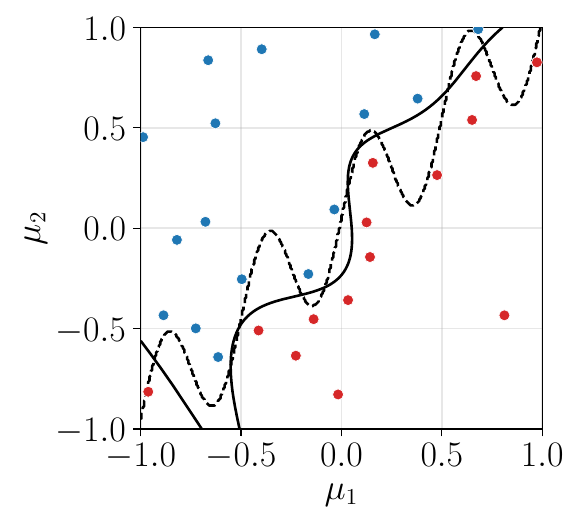}
        \label{fig:f3_margin_diversity_n30}
    \end{subfigure}\hfill
    \begin{subfigure}[t]{0.238\linewidth}
        \centering
        \includegraphics[width=\linewidth]{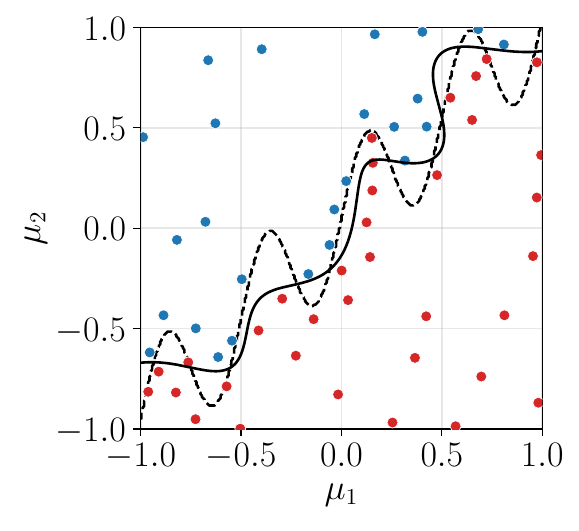}
        \label{fig:f3_margin_diversity_n60}
    \end{subfigure}\hfill
    \begin{subfigure}[t]{0.238\linewidth}
        \centering
        \includegraphics[width=\linewidth]{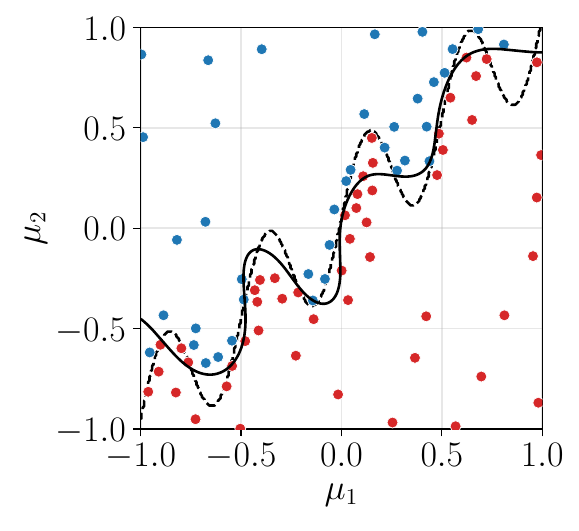}
        \label{fig:f3_margin_diversity_n90}
    \end{subfigure}\hfill
    \begin{subfigure}[t]{0.238\linewidth}
        \centering
        \includegraphics[width=\linewidth]{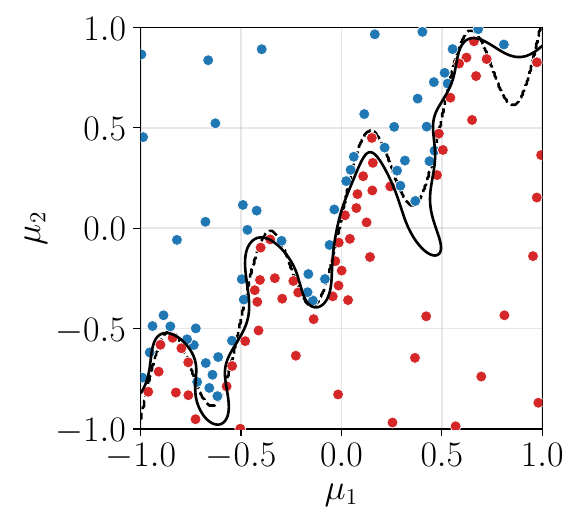}
        \label{fig:f3_margin_diversity_n120}
    \end{subfigure}
    \gridrowsep
    % Row 5
    \begin{subfigure}[t]{0.238\linewidth}
        \centering
        \includegraphics[width=\linewidth]{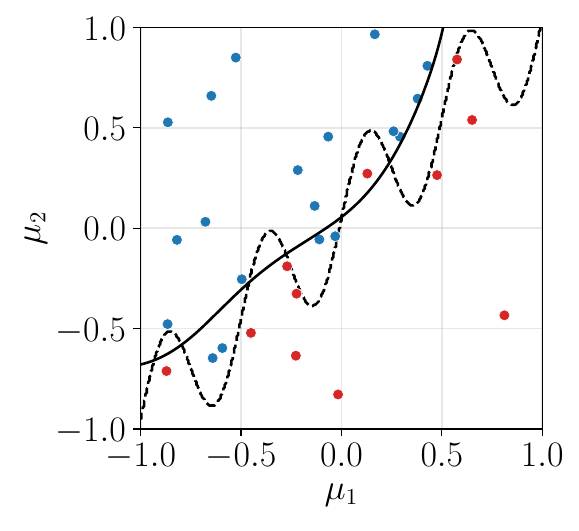}
        \label{fig:f3_cluster_n30}
    \end{subfigure}\hfill
    \begin{subfigure}[t]{0.238\linewidth}
        \centering
        \includegraphics[width=\linewidth]{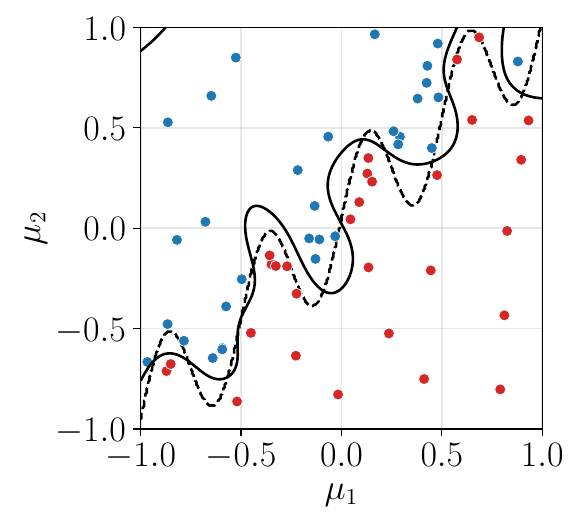}
        \label{fig:f3_cluster_n60}
    \end{subfigure}\hfill
    \begin{subfigure}[t]{0.238\linewidth}
        \centering
        \includegraphics[width=\linewidth]{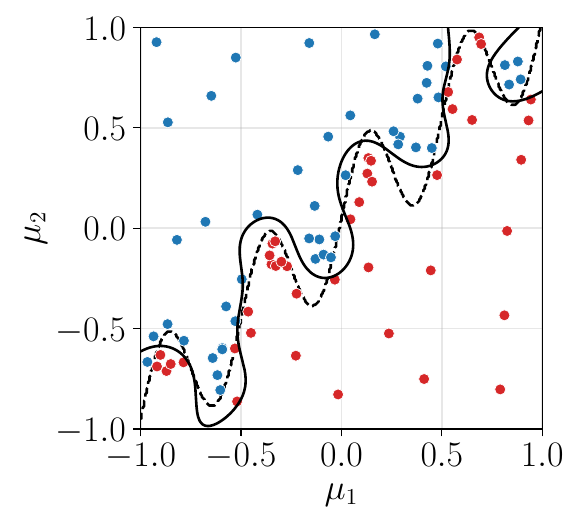}
        \label{fig:f3_cluster_n90}
    \end{subfigure}\hfill
    \begin{subfigure}[t]{0.238\linewidth}
        \centering
        \includegraphics[width=\linewidth]{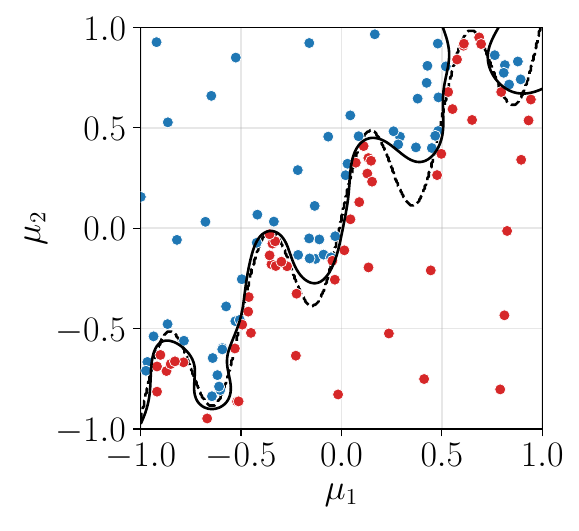}
        \label{fig:f3_cluster_n120}
    \end{subfigure}
    \gridrowsep
    % Row 6
    \begin{subfigure}[t]{0.238\linewidth}
        \centering
        \includegraphics[width=\linewidth]{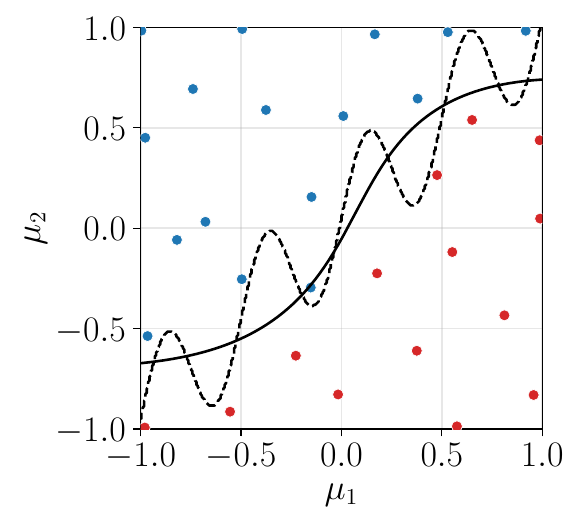}
        \caption{$N = 30.$}
        \label{fig:f3_random_n30}
    \end{subfigure}\hfill
    \begin{subfigure}[t]{0.238\linewidth}
        \centering
        \includegraphics[width=\linewidth]{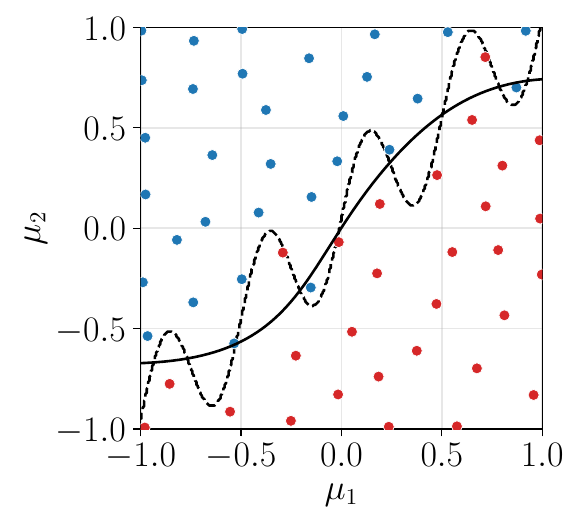}
        \caption{$N = 60.$}
        \label{fig:f3_random_n60}
    \end{subfigure}\hfill
    \begin{subfigure}[t]{0.238\linewidth}
        \centering
        \includegraphics[width=\linewidth]{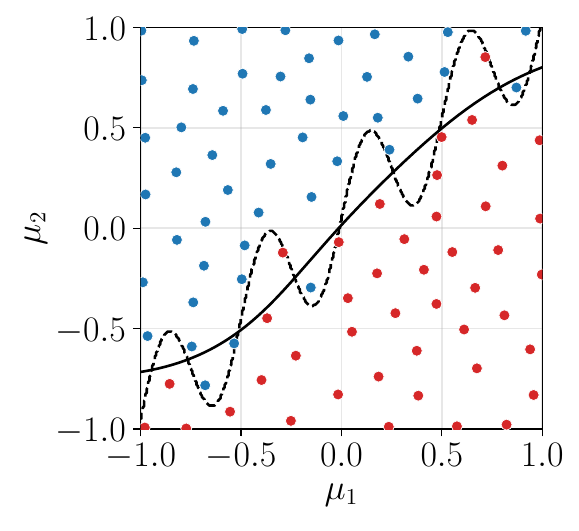}
        \caption{$N = 90.$}
        \label{fig:f3_random_n90}
    \end{subfigure}\hfill
    \begin{subfigure}[t]{0.238\linewidth}
        \centering
        \includegraphics[width=\linewidth]{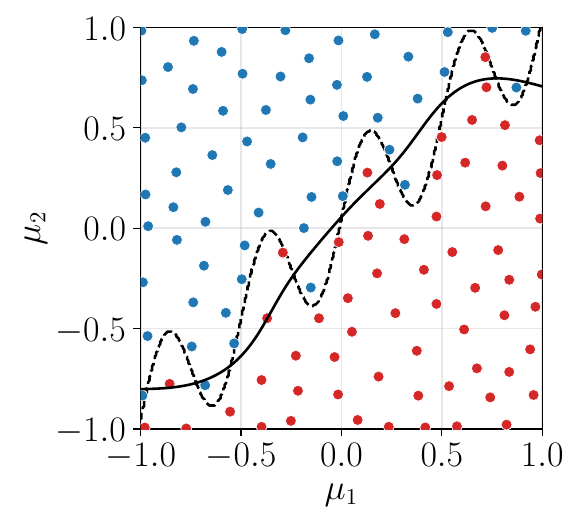}
        \caption{$N = 120.$}
        \label{fig:f3_random_n120}
    \end{subfigure}
    
    \caption{Adaptive sampling on \eqref{eq:f3-2d-discontinuities}. 
    From top to bottom HMC, MALA, margin, margin + diversity, cluster and random sampling.
    Dashed/solid lines are the true/SVM boundary, respectively.
    }
    \label{fig:f3_grid}
\end{figure}

To confirm the trend, we further increase the complexity of the boundary to be inferred by considering \eqref{eq:f3-2d-discontinuities}. It exhibits very high curvature, albeit while remaining a single connected component. Figure \ref{fig:f3-2d-discontinuities} shows a trend similar to that observed for \eqref{eq:f2-2d-discontinuities}: schemes with a space-filling downstream selection are the only ones that outperform random sampling in Hausdorff distance, and the $F_1$ score is again somewhat deceiving, as cluster sampling starts off with the highest score in the early data regime, but its Hausdorff distance is always suboptimal. HMC and margin with diversity are again the two methods that offer the best performance, and our proposed approach shows once again a clear improvement in Hausdorff distance throughout the entire adaptive sampling iterations and, therefore, superior boundary approximation quality. This can be equally appreciated in Figure \ref{fig:f3_grid} where we illustrate the progression of all algorithms on a representative seed. A final, but important observation is the following: Figure \ref{fig:f3-2d-discontinuities} shows that random sampling is the best scheme for both metrics in the early stages. This signals that if such complexity is known a priori, then the initial design could be enlarged and adaptive sampling could be started later on. The framework proposed in this work would allow for a similar approach by imposing an annealing schedule on $\beta$; this could also improve stability and mitigate the early plateauing of both HMC and MALA before they eventually recover a positive trend. However, as mentioned, picking a sound schedule is generally complex.

\vspace{10pt}

\begin{figure}[H]
    \centering
    \includegraphics[width=\linewidth]{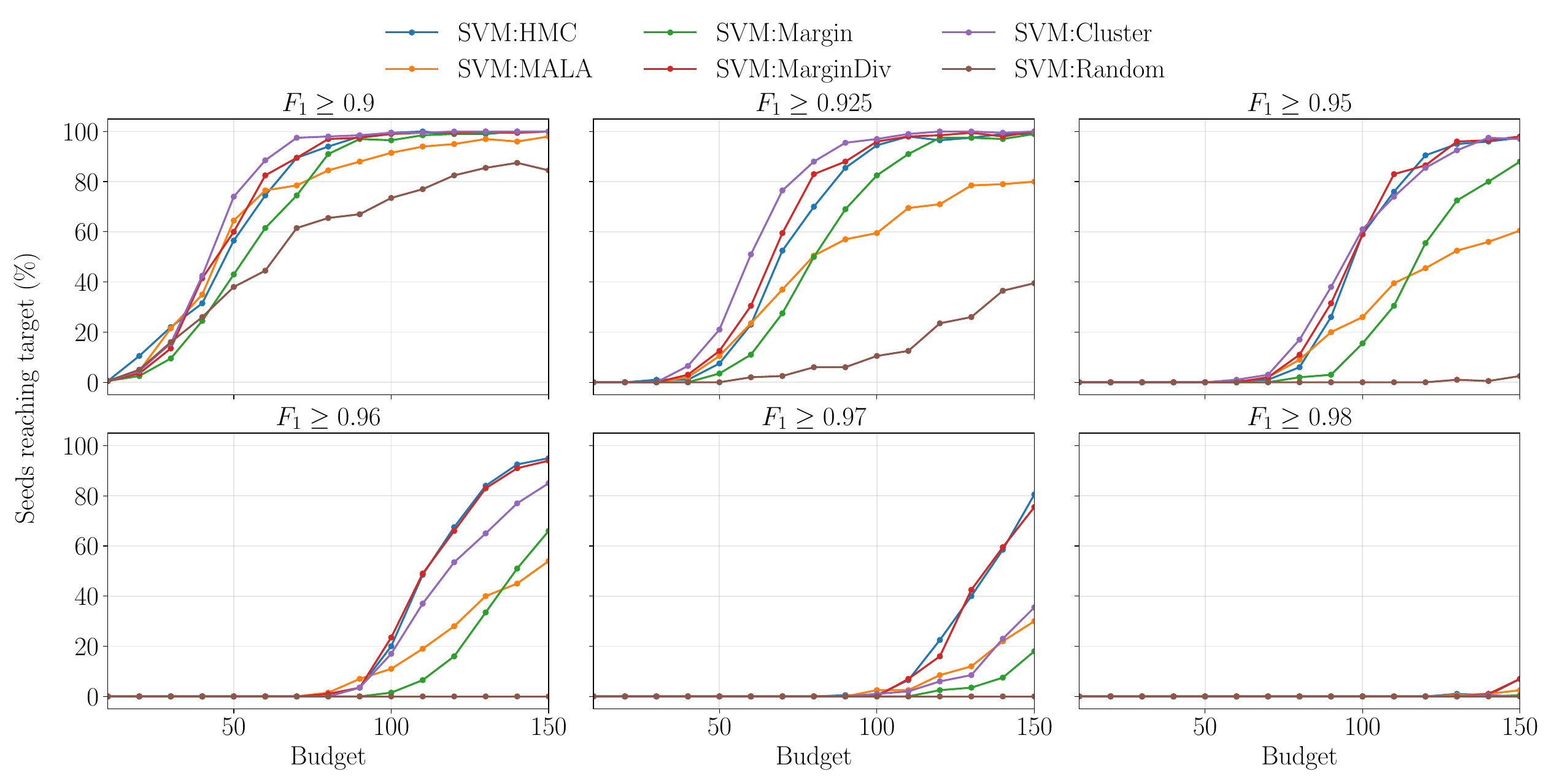}
    \caption{Percentage of seeds reaching different $F_1$ score targets on \eqref{eq:f3-2d-discontinuities}.}
    \label{fig:f3-seeds-at-target}
\end{figure}

\subsection{Dimension scaling}
\label{subsec:dimension-scaling}

The two-dimensional analytical examples discussed in Subsection \ref{subsec:analytical-discontinuities-results} have shown that space-filling schemes are generally outperforming margin and cluster sampling. Therefore, in this example and throughout the rest of the results section, we only compare HMC, MALA and margin with diversity, and retain random sampling as a baseline. Our goal is now to study the scaling of the different active learning strategies with the dimensionality of the parameter space; to this aim, we consider the analytical regime boundary \eqref{eq:dimension-dependent-discontinuity} that we have already discussed in Section \ref{subsec:hmc-cost}, and again targets its $\theta = 0$ level set.

\begin{figure}[ht]

    \begin{subfigure}[b]{0.325\textwidth}
    \includegraphics[width=\linewidth]{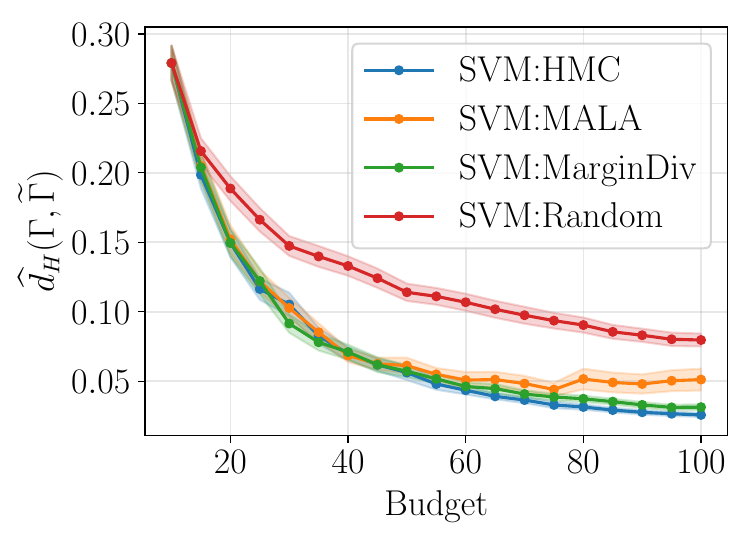}
    \caption{$n = 2$.}
    \end{subfigure}
    \hfill
    \begin{subfigure}[b]{0.325\textwidth}
    \includegraphics[width=\linewidth]{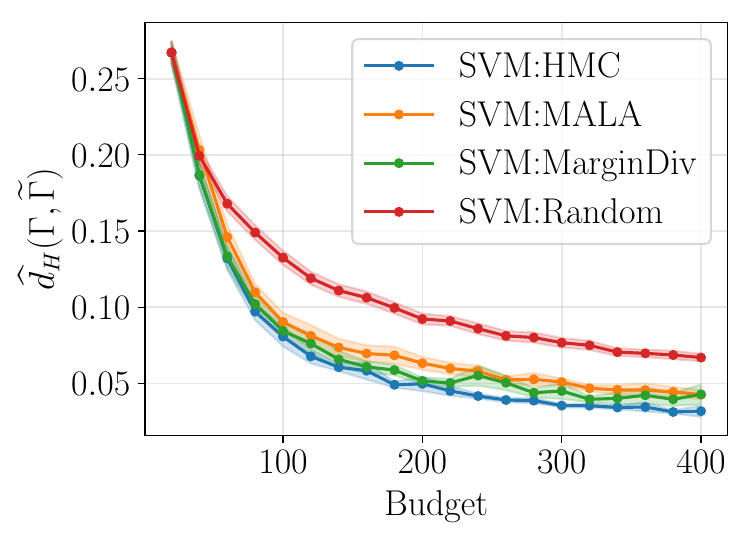}
    \caption{$n = 3$.}
    \end{subfigure}
    \hfill
    \begin{subfigure}[b]{0.325\textwidth}
    \includegraphics[width=\linewidth]{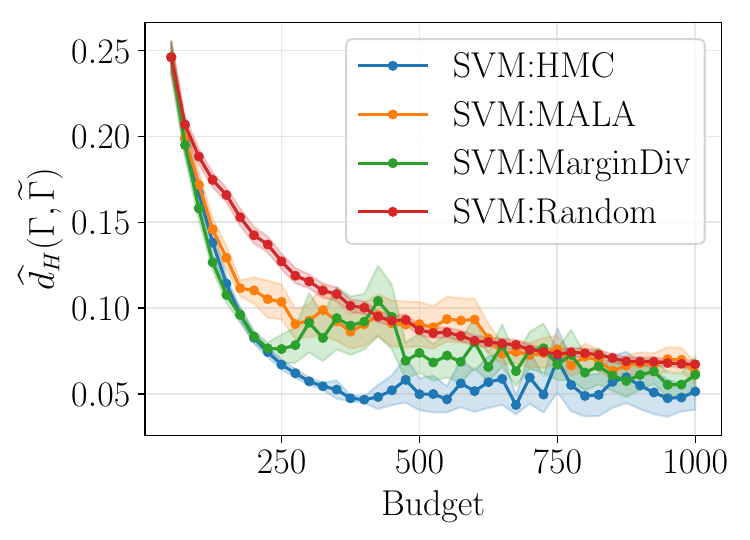}
    \caption{$n = 4$.}
    \end{subfigure}
    
    \caption{Hausdorff distance decay on \eqref{eq:dimension-dependent-discontinuity} for different diversity-enforcing schemes in different dimensions. Shaded regions represent a $95 \%$ CI on the mean.}
    \label{fig:dimension-scaling-hausdorff}

\end{figure}

\begin{figure}[H]
    \centering
    \providecommand{\seedsplotwidth}{\linewidth}
    \includegraphics[width=\seedsplotwidth]{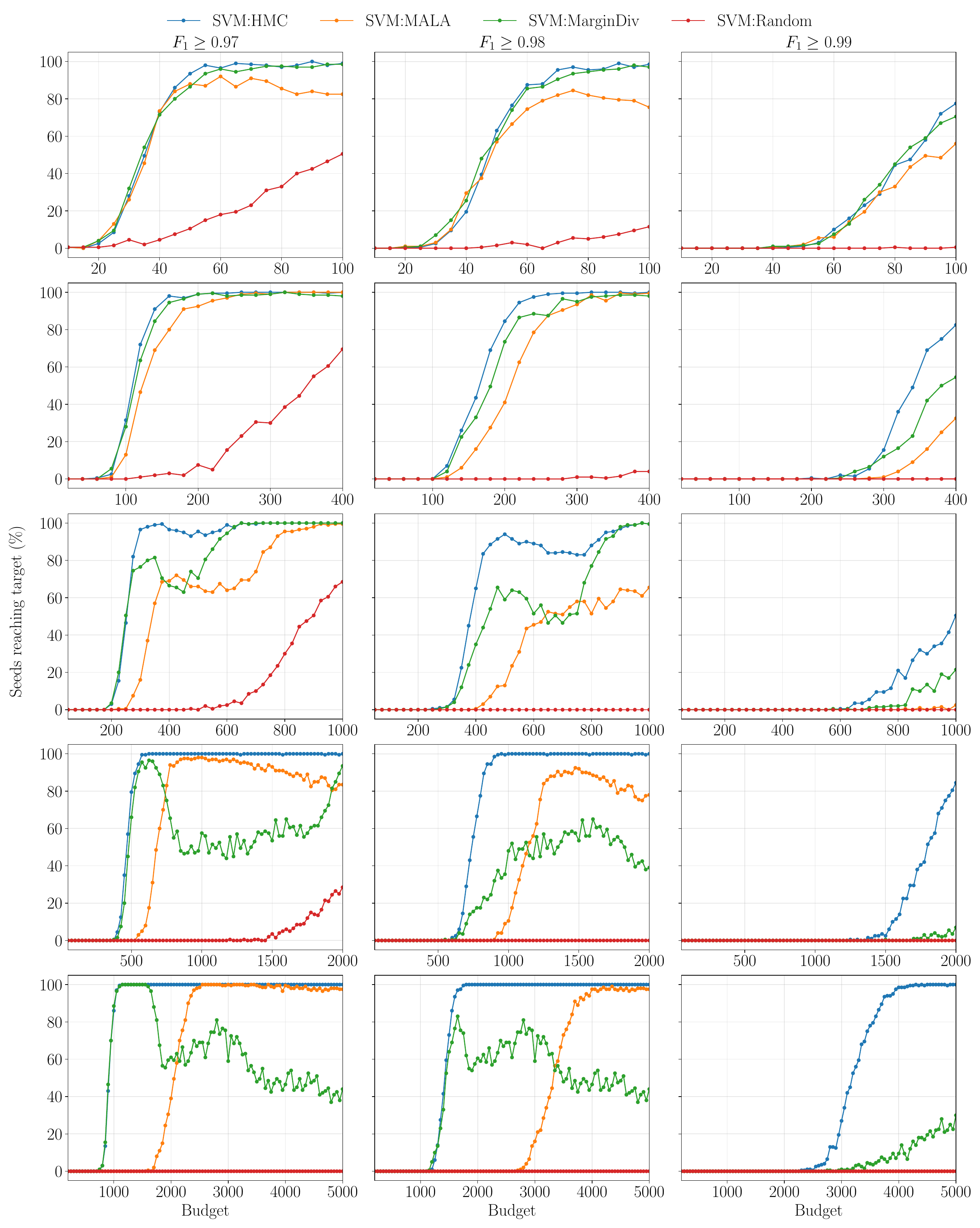}
    \caption{Percentage of seeds reaching different $F_1$ score targets on \eqref{eq:dimension-dependent-discontinuity} for different dimensions, from top to bottom $n = 2, 3, \ldots, 6$.}
    \label{fig:dimension-scaling-seeds-at-target}
\end{figure}

When it is feasible computationally to get converged measurements of the Hausdorff distance (in our case, up to dimension $n = 4$), we show in Figure \ref{fig:dimension-scaling-hausdorff} that with an increase in the dimension the gap between our proposed approach and competing methods gets larger and larger; this shows improved robustness and scaling of the concentration of samples in a tube around the boundary, but also the effect of their efficient spread tangentially. Indeed, while also MALA tries to concentrate samples in a tube around the boundary, its sample efficiency is comparable and in some cases even worse than that of the margin with diversity approach. On the other hand, in Figure \ref{fig:dimension-scaling-seeds-at-target} we show for an increasing dimension of the parameter space $n = 2, \ldots, 6$ the percentage of seeds in which each adaptive sampling strategy reaches a certain $F_1$ score target ($0.97, 0.98, 0.99$) within the prescribed computational budgets.
While in lower dimensions the three approaches have similar performance, from $n = 4$ onward margin with diversity
tends to be unstable and struggles to reach high scores, while MALA is generally more stable but is largely trailing HMC.
This effect becomes more pronounced as the dimension increases, highlighting the greater reliability and robustness of the proposed approach.

It is worth mentioning that these considerations hold for a fixed sample size (in this case, $2000$ points); while the latter could be enlarged and this would definitely lead to better results for margin-based schemes, it wouldn't fundamentally resolve the curse of dimensionality and is likely to lead to memory issues and increased computational costs. On the other hand, HMC is by construction more robust in higher dimensions for a given sample size, because it uses gradient-based dynamics to focus on a tube around the boundary while being excellent at traversing low-energy regions, therefore guaranteeing a more even coverage of the boundary before the space-filling downselection is performed. Furthermore, while to increase predictive accuracy with margin-based schemes we have little choice but to increase the sample size, with the proposed approach we observed that performance could be improved by simply increasing the number of chains $N_c$, while keeping fixed the total number of generated candidates.

\subsection{Rolling ball on a Gaussian terrain} \label{subsec:rolling-ball}

\begin{figure}[H]
    \begin{subfigure}[c]{0.45\textwidth}
    \includegraphics[width=\linewidth]{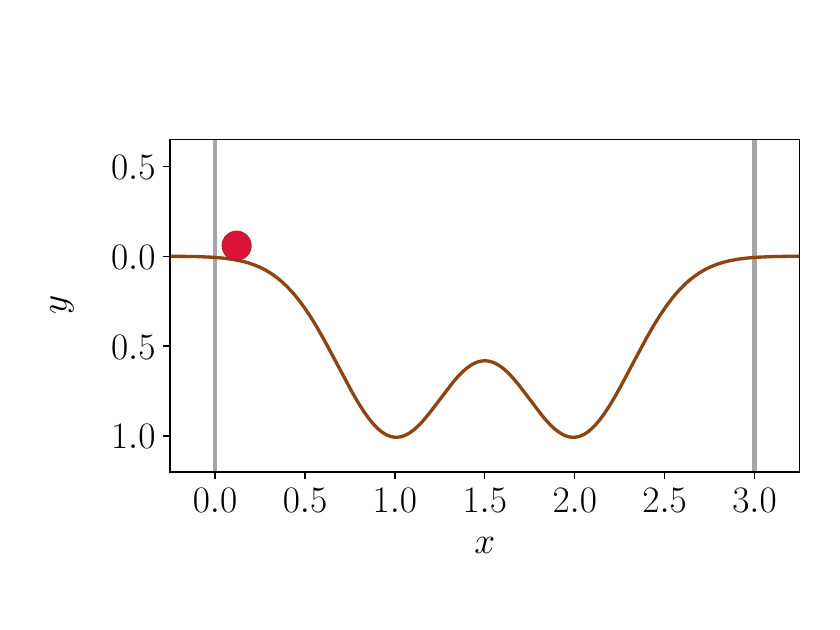}
    \caption{}
    \label{fig:bouncing-ball-illustration}
    \end{subfigure}
    \hfill
    \begin{subfigure}[c]{0.45\textwidth}
    \includegraphics[width=\linewidth]{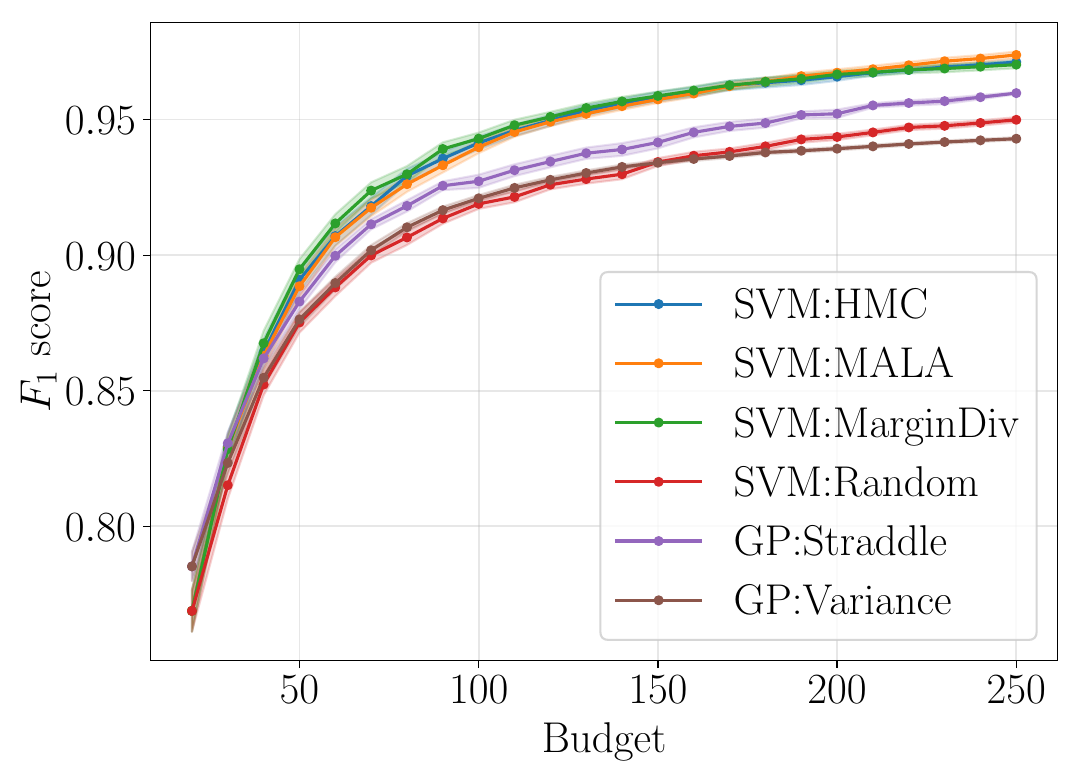}
    \caption{}
    \label{fig:bouncing-ball-acc}
    \end{subfigure}
    \caption{Rolling ball example: (a) problem setting; (b) mean $F_1$ score. Shaded regions represent a $95 \%$ CI on the mean.}
    \label{fig:bouncing-ball}
\end{figure}

In this example, we consider a rigid disk of radius $R = 0.08 \, \mathrm{m}$ rolling on a one-dimensional terrain formed by two inverted Gaussian wells
\begin{equation}
\label{eq:rolling-ball-terrain}
    h(x) = - H_1 \exp \left(
    -\left(\frac{x-C_1}{\sigma_1}\right)^{2}
  \right) - H_2 \exp \left(
    -\left(\frac{x-C_2}{\sigma_2}\right)^{2}
  \right),
\end{equation}
with fixed centers $C_1 = 1 \, \mathrm{m}, C_2 = 2 \, \mathrm{m}$, widths $\sigma_1 = \sigma_2 = 0.45 \, \mathrm{m}$ and heights $H_1 = H_2 = 1 \, \mathrm{m}$; see Figure \ref{fig:bouncing-ball-illustration} for an illustration. The ball is launched from the initial position $x_0 = 0$ with initial velocity $v_0$ along the terrain tangent and is constrained to remain on the surface; the tangential motion is governed by rolling inertia, gravity, and rolling friction with a coefficient $\mu_{\mathrm{roll}}$. Two reflective walls placed at $x = 0 \, \mathrm{m}$ and $x = 3 \, \mathrm{m}$ have a shared restitution coefficient $e$. Given the parametric triplet $Z := (e, v_0, \mu_{\mathrm{roll}})$ with ranges defined in Table \ref{tab:rolling-ball-params}, the considered discontinuous output is the final horizontal position of the ball $x(t=\infty, Z)$, with a threshold value corresponding to the peak of the terrain $\theta = \frac{C_1 + C_2}{2}$.

\begin{table}[H]
    \centering
    \caption{Parameters range for the rolling ball example.}
    \label{tab:rolling-ball-params}
    \begin{tabular}{lc}
        \hline
        Parameter & Range \\
        \hline
        Reflective walls restitution coefficient ($e$, dimensionless) & $(0.5, \, 0.8)$ \\
        Initial velocity ($v_0$, in $\mathrm{m/s}$) & $(2, \, 4)$ \\
        Rolling friction coefficient ($\mu_{\mathrm{roll}}$, dimensionless) & $(0.04, \, 0.08)$ \\
        \hline
    \end{tabular}
\end{table}

\begin{figure}[H]
    \centering
    \includegraphics[width=\linewidth]{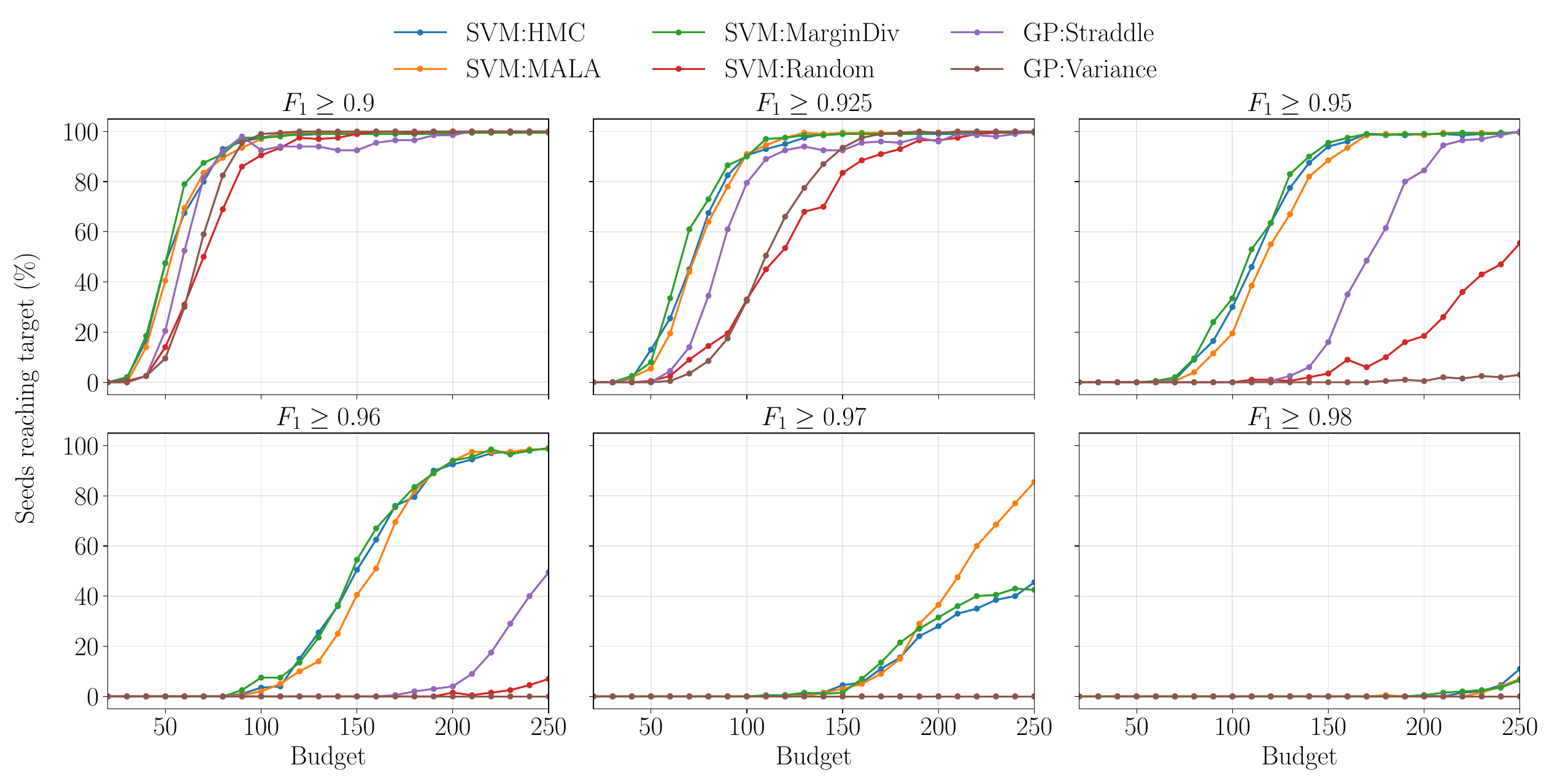}
    \caption{Percentage of seeds reaching different $F_1$ score targets on the rolling ball example depicted in Figure \ref{fig:bouncing-ball-illustration}.}
    \label{fig:bouncing-ball-seeds-at-target}
\end{figure}

In this example, we start from $N_{\text{init}} = 20$ points and add batches of size $N_{\text{add}} = 10$ until the computational budget $N_{\text{end}} = 250$ is exhausted. Because the final position can only take discrete values and is therefore discontinuous with respect to the parameters, we observe in Figure \ref{fig:bouncing-ball-acc} that regression-based level set estimation has a good performance in low-data regime but then trails classification-based approaches when higher scores are targeted. In particular, Figure \ref{fig:bouncing-ball-seeds-at-target} shows that Straddle never reaches $F_1 \geq 0.97$, which is likely a consequence of the fact that regression cannot accurately resolve the discontinuity front. On the other hand, classification methods are able to target high scores more efficiently. Due to the low dimensionality of the problem, for a given budget, the space-filling classification algorithms are basically equivalent, with HMC and margin leading for most target scores and MALA generally trailing (except for $F_1 \geq 0.97$).

\subsection{Genetic toggle switch} \label{subsec:genetic-toggle-switch}

We now consider a genetic toggle switch model proposed in \cite{gardner2000construction}, which has been already used for parameter inference \cite{marzouk2009stochastic} and discontinuity detection \cite{jakeman2011characterization}. This is a differential-algebraic model of a genetic circuit implemented in \textit{E. coli plasmids}, for which the equations of motion read as follows:
\begin{equation}
\label{eq:genetic-toggle-switch}
\begin{cases}
    \frac{du}{dt} = \frac{\alpha_1}{1 + v^\beta} - u \\
    \frac{dv}{dt} = \frac{\alpha_2}{1 + w^\gamma} - v \\
    w = \frac{u}{(1 + [\mathrm{IPTG}]/K)^\eta}
\end{cases}.
\end{equation}
In \eqref{eq:genetic-toggle-switch}, $u$ and $v$ represent the concentration of the first and 
second repressor, respectively; $\alpha_1$ and $\alpha_2$ are the effective rates of synthesis 
of the repressors; $\gamma$ and $\beta$ represent the cooperativity of repression of the two 
promoters, respectively; finally, $[\mathrm{IPTG}]$ is the concentration of the chemical compound
that induces the switch. The parameters $K$ and $\eta$ describe the binding of $[\mathrm{IPTG}]$ 
to the first repressor. In this experiment, we fix $[\mathrm{IPTG}] = 4 \times 10^{-5}$, $\gamma = 2$,
and $\beta = 2.5$, as well as the initial condition $u_0 = v_0 = 0.1$ for time integration 
of \eqref{eq:genetic-toggle-switch}. We consider the four parameters 
$Z := (\alpha_1, \alpha_2, \eta, K)$, which vary within a hypercube centered around the 
nominal configuration $Z_0 := (45, 15, 2.0015, 6.7 \times 10^{-5})$, with a C.o.V. of $\pm 15 \%$.
As the parameters change, the system exhibits a bistable regime characterized by two alternative
stable states, determined by the dominant repressor concentration at steady state. The quantity of 
interest is therefore the $\theta = 0$ level set of the function $g(Z) = v(t=\infty, Z) - u(t=\infty, Z)$,
which is discontinuous in $Z$.

\begin{figure}[H]
    \begin{subfigure}[b]{0.45\textwidth}
    \includegraphics[width=\linewidth]{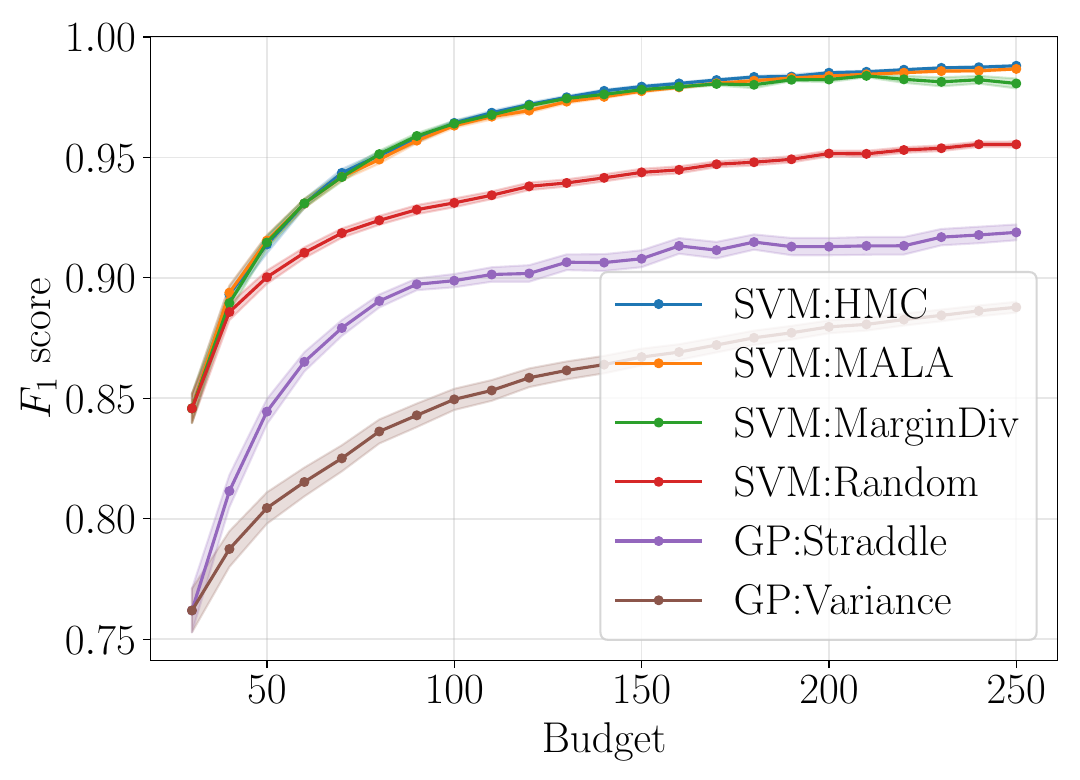}
    \caption{}
    \label{fig:genetic-toggle-switch}
    \end{subfigure}
    \hfill
    \begin{subfigure}[b]{0.45\textwidth}
    \includegraphics[width=\linewidth]{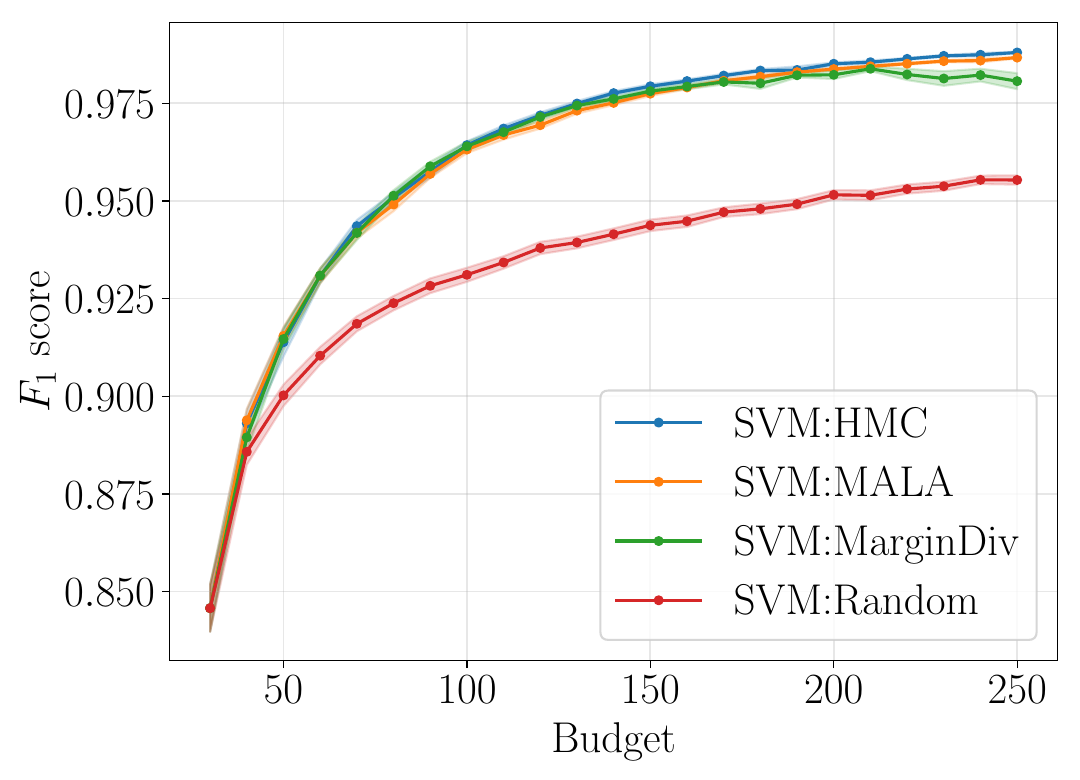}
    \caption{}
    \label{fig:genetic-toggle-switch-only-classification}
    \end{subfigure}
    
    \caption{Adaptive sampling results on the genetic toggle switch \eqref{eq:genetic-toggle-switch}. (a) Mean $F_1$ score; (b) mean $F_1$ score, only classification methods. Shaded regions represent a $95 \%$ CI on the mean.}
    \label{fig:genetic-toggle-switch-acc}
\end{figure}

\begin{figure}[H]
    \centering
    \includegraphics[width=\linewidth]{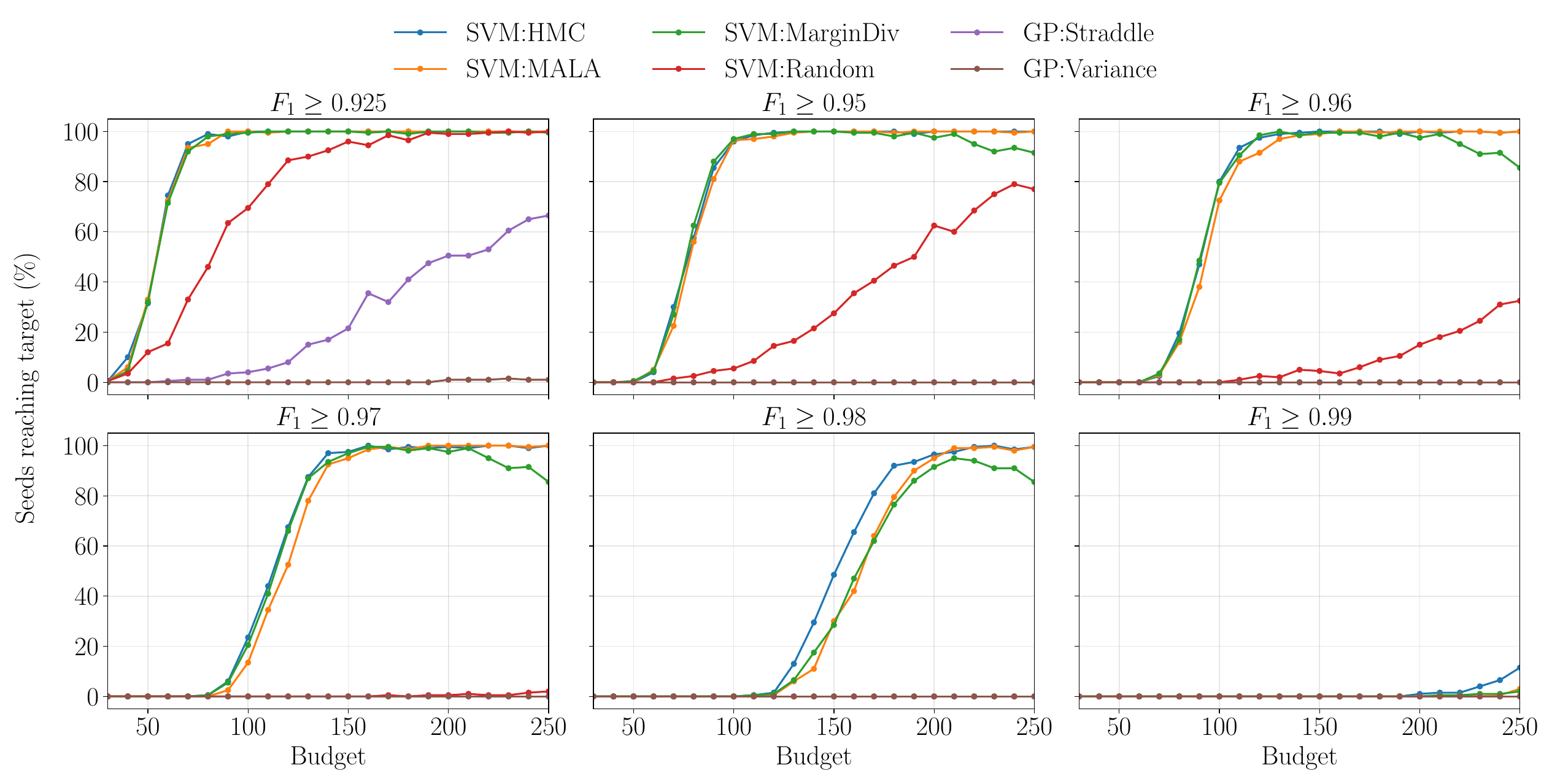}
    \caption{Percentage of seeds reaching different $F_1$ score targets on the genetic toggle switch \eqref{eq:genetic-toggle-switch}.}
    \label{fig:genetic-toggle-switch-seeds-at-target}
\end{figure}

We start from $N_{\text{init}} = 30$ points and enrich the training dataset with $N_{\text{add}} = 10$ points at each iteration until we reach
the training size $N_{\text{end}} = 250$. In Figure \ref{fig:genetic-toggle-switch}, we observe that
classification largely outperforms regression; indeed, closer inspection reveals that the latter
never reaches a $0.95$ $F_1$ score on any of the seeds. On the other hand,
Figures \ref{fig:genetic-toggle-switch-only-classification} and \ref{fig:genetic-toggle-switch-seeds-at-target}
show that among classification algorithms, the proposed approach is the most robust and efficient
for a given computational budget and with respect to initialization, respectively. 
In particular, within the given budget, it is the only algorithm that reaches $0.99 \, F_1$ score
on a non-negligible percentage of seeds. On the other hand, MALA performs well but is generally
less sample-efficient and margin with diversity shows deterioration at higher target scores. 
Contrarily to the previous example, therefore, here we observe a clearer separation between the
different algorithms; this could be justified by the larger parameter space dimension.

\vspace{-15pt}

\subsection{Dam-break shallow water equation} \label{subsec:dam-break-swe}

We take as final example a two-dimensional dam-break shallow water equation \cite{ndengna2024high, dutykh2009relevance}, which is a widely utilized reference for numerical schemes and model order reduction \cite{mizan2026parametric}. The hyperbolic equations describing the water height $h(x, y)$ and the velocity $\mathbf{u}(x, y) = (u_x(x, y), \: u_y(x, y))$ are given in conservative form by
\begin{equation}
\label{eq:swe-eqs}
    \begin{cases}
        \frac{\partial h}{\partial t} + \frac{\partial q_x}{\partial x} + \frac{\partial q_y}{\partial y} = 0, \\
        \frac{\partial q_x}{\partial t} + \frac{\partial}{\partial x} \left( \frac{q_x^2}{h} + \frac{1}{2} g h^2 \right) + \frac{\partial}{\partial y} \left( \frac{q_x q_y}{h} \right) = - g h \frac{\partial Z_b(x, y)}{\partial x} - \tau_x, \\
        \frac{\partial q_y}{\partial t} + \frac{\partial}{\partial x} \left( \frac{q_x q_y}{h} \right) + \frac{\partial}{\partial y} \left( \frac{q_y^2}{h} + \frac{1}{2} g h^2 \right) = - g h \frac{\partial Z_b(x, y)}{\partial y} - \tau_y, \\
    \end{cases}
\end{equation}
where $q_x = h u_x$ and $q_y = h u_y$. In \eqref{eq:swe-eqs}, we consider a Gaussian hill bottom topography
\begin{equation}
\label{eq:gaussian-hill}
    Z_b(x, y) = H \exp\left( - \frac{(x-x_h)^2}{W_x^2} - \frac{(y-y_h)^2}{W_y^2} \right),
\end{equation}
and Manning friction
\begin{equation}
\label{eq:manning-friction}
    \tau_x = g n^2 \frac{u_x \sqrt{u_x^2 + u_y^2}}{h^{1/3}}, \quad \tau_y = g n^2 \frac{u_y \sqrt{u_x^2 + u_y^2}}{h^{1/3}}.
\end{equation}
Finally, $g$ is the acceleration due to gravity. The equations above are defined in the spatial domain $\Omega = (0, L_x) \times (0, L_y)$ with $L_x = L_y = 200 \, \mathrm{m}$ and are endowed with the initial conditions
\begin{equation}
\label{eq:swe-ic}
    \eta(x, y, 0) = \begin{cases}
        \eta_L, \: \: x < L_x/2, \\
        \eta_R, \: \: x \geq L_x/2
    \end{cases}, \quad h(x, y, 0) = \eta(x, y, 0) - Z_b(x, y),
\end{equation}
and $q_x(x, y, 0) = q_y(x, y, 0) = 0$, where $\eta(x, y, t) = h(x, y, t) + Z_b(x, y)$ is the absolute height of the water free surface. Lastly, we use outflow boundary conditions on the boundaries $x = 0$ and $x = L_x$ and reflective boundary conditions at $y = 0$ and $y = L_y$.

\medskip

The equations \eqref{eq:swe-eqs} are solved numerically using a cell-averaged finite-volume method. We discretize $\Omega$ by a Cartesian grid with $N_x = 400$ and $N_y = 100$ finite volume cells, with a finer discretization along $x$ because we consider wet-bed simulations ($\eta_R > H$) and therefore a shock develops in that direction. The discretization has been chosen to get converged results in the downstream computation of quantities of interest. For interface fluxes, a local Lax-Friedrichs scheme \cite{riestiana2021numerical} is employed. The terms $\frac{\partial Z_b(x, y)}{\partial x}$ and $\frac{\partial Z_b(x, y)}{\partial y}$ are computed analytically by differentiating \eqref{eq:gaussian-hill}. Finally, we use a Heun time-stepping scheme subject to a CFL condition and time-march up to $T_f = 10 \, \mathrm{s}$.

\medskip

Regarding the parametric configuration, we fix $y_h = 100 \, \mathrm{m}$, $H = 2 \, \mathrm{m}, W_x = W_y = 20 \, \mathrm{m}$. The variation of the four parameters $\mu := (\eta_L, \eta_R, n, x_h)$ that influence the flow characteristics is summarized in Table \ref{tab:swe-params}.

\begin{table}[ht]
    \centering
    \caption{Parameters range for the dam-break shallow water.}
    \label{tab:swe-params}
    \begin{tabular}{lc}
        \hline
        Parameter & Range \\
        \hline
        Initial left $\eta$ ($\eta_L$, in $\mathrm{m}$) & $(8, \, 14)$ \\
        Initial right $\eta$ ($\eta_R$, in $\mathrm{m}$) & $(2, \, 6)$ \\
        Bottom roughness coefficient ($n$, dimensionless) & $(0, \, 0.4)$ \\
        Gaussian hill peak position ($x_h$, in $\mathrm{m}$) & $(100, 150)$ \\
        \hline
    \end{tabular}
\end{table}

Our objective is to determine the persistence of a supercritical flow region during the time interval $[0, T_f]$. To do this, we compute the critical water height
\begin{equation}
\label{eq:critical-height}
    h_{\mathrm{cr}}(t, \mu) = \left( \frac{h^2(t, \mu) (u_x^2(t, \mu) + u_y^2(t, \mu))}{g} \right)^{1/3}
\end{equation}
The field quantity $\phi_\mathrm{cr}(t, \mu) = h(t, \mu) - h_\mathrm{cr}(t, \mu)$ is negative if the flow is subcritical and signals that the flow is locally supercritical otherwise. To track the time persistence of supercritical flows, we can therefore compute a time-dependent quantity given by
\begin{equation}
\label{eq:swe-time-dependent-integral}
    \Phi(t, \mu) := \int_{\Omega_{\mathrm{cr}}} (\phi_{\mathrm{cr}}(t, \mu))_+,
\end{equation}
where $\Omega_{\mathrm{cr}} = (x_h, L_x) \times (0, L_y)$ is the region starting from the downstream slope of the Gaussian hill and the area where the flow is most likely to be supercritical, and $(y)_+ = \max(0, y)$. For safety and design considerations, we are interested in the amount of time that such an integrand is positive, that is,
\begin{equation}
\label{eq:swe-functional}
    g(\mu) := \max(\mathcal{T}(\mu)) - \min(\mathcal{T}(\mu)), \quad \text{where} \quad \mathcal{T}(\mu) = \{ t \in (0, T_f]: \Phi(t, \mu) > 0 \}.
\end{equation}
We then define \eqref{eq:swe-functional} as our simulator output, together with a critical operational time $\theta = 0.2$ seconds.

\medskip

\begin{figure}[H]

    \begin{subfigure}[b]{0.45\textwidth}
    \includegraphics[width=\linewidth]{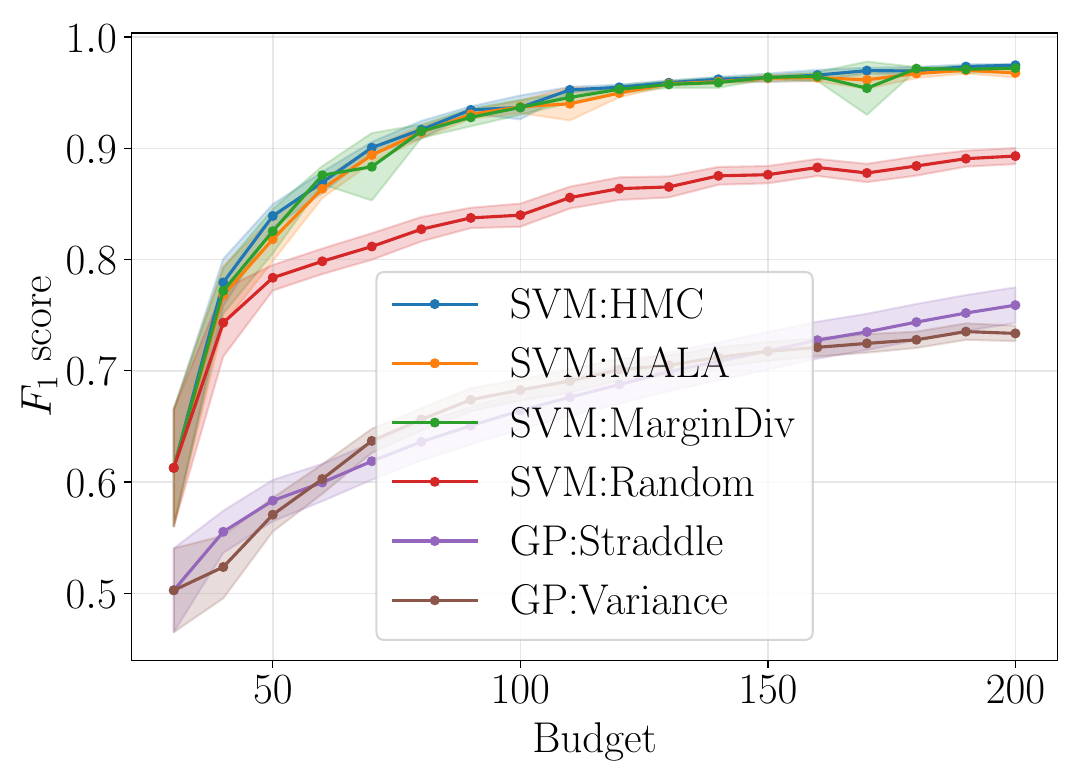}
    \caption{}
    \label{fig:swe}
    \end{subfigure}
    \hfill
    \begin{subfigure}[b]{0.45\textwidth}
    \includegraphics[width=\linewidth]{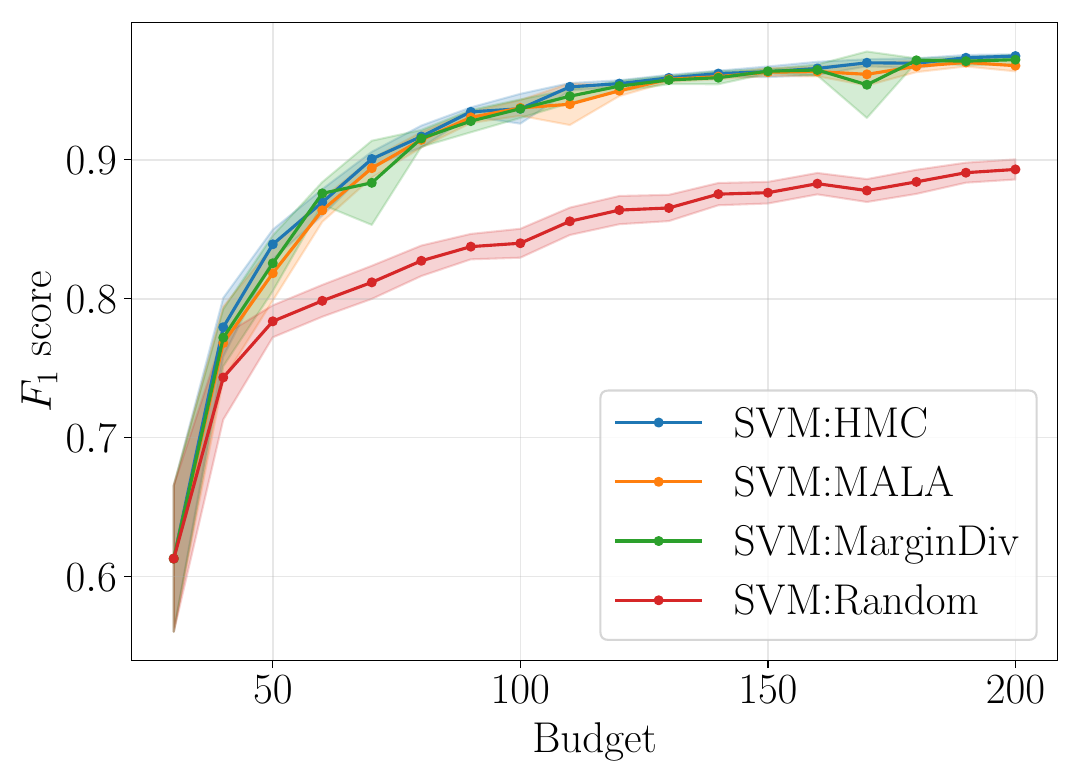}
    \caption{}
    \label{fig:swe-only-classification}
    \end{subfigure}
    
    \caption{Adaptive sampling results on the dam-break shallow water \eqref{eq:swe-eqs}. (a) Mean $F_1$ score; (b) mean $F_1$ score, only classification methods. Shaded regions represent a $95 \%$ CI on the mean.}
    \label{fig:swe-acc}
\end{figure}

\begin{figure}[H]
    \centering
    \includegraphics[width=\linewidth]{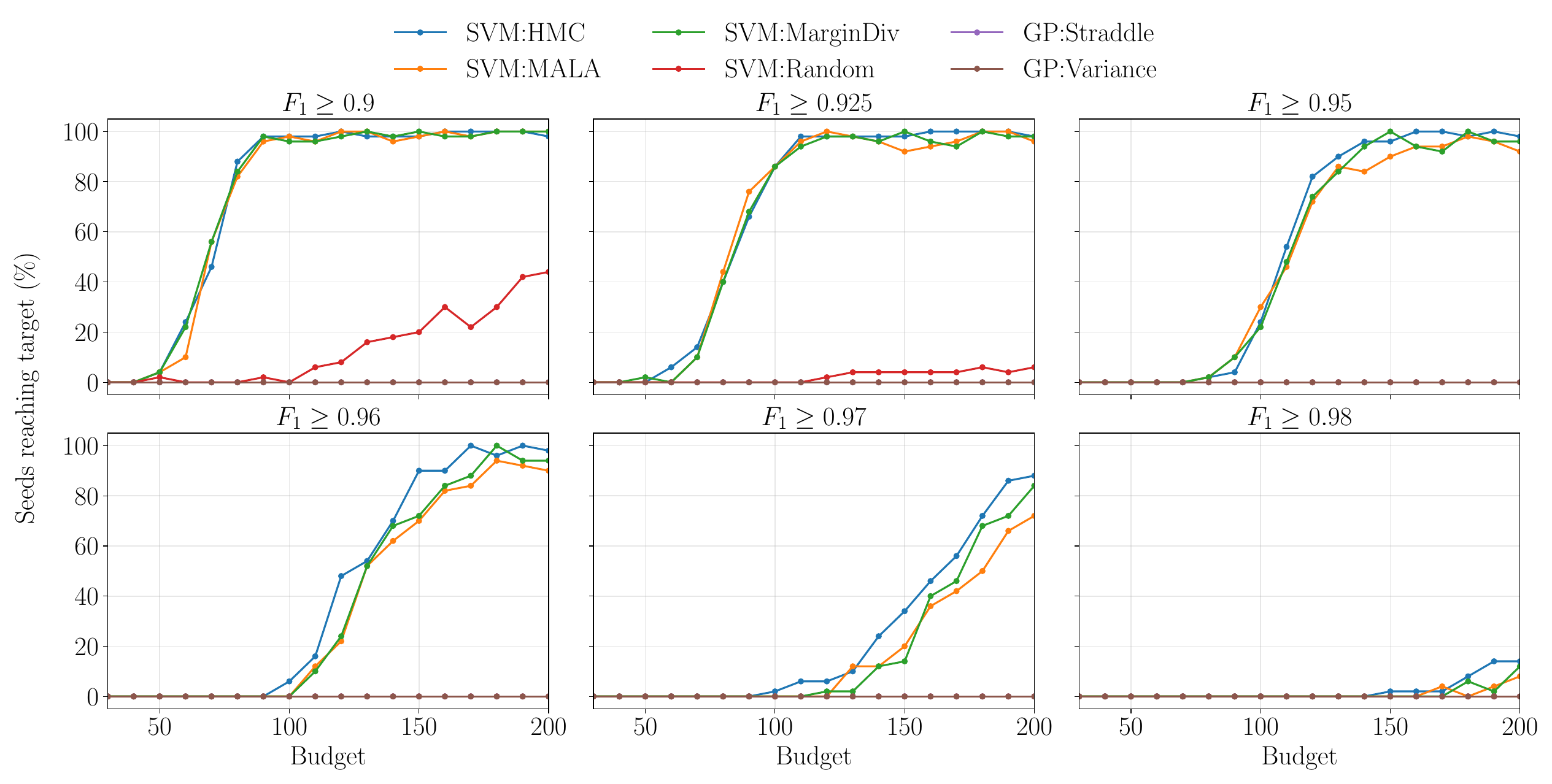}
    \caption{Percentage of seeds reaching different $F_1$ score targets on the dam-break shallow water \eqref{eq:swe-eqs}.}
    \label{fig:swe-seeds-at-target}
\end{figure}

The setup is similar to the previous example: we start from $N_{\text{init}} = 30$ quasi-random points and add batches of size $N_{\text{add}} = 10$ until the computational budget $N_{\text{end}} = 200$ is exhausted. In this example, we report the mean $F_1$ score computed on an independent set of $5 \times 10^3$ uniformly drawn samples and averaged over $50$ seeds. Because the transition between subcritical and supercritical flows features steep changes of the quantity of interest \eqref{eq:swe-functional}, regression-based strategies struggle to converge to acceptable accuracies, as we infer from Figure \ref{fig:swe}. Once again, on the other hand, Figure \ref{fig:swe-only-classification} shows that HMC remains the most sample-efficient method and generally exhibits a more robust trend, whereas margin with diversity yields oscillatory scores and MALA suffers from diffusive tangential motion. The improved sample efficiency with HMC is confirmed also by looking at the percentage of seeds reaching higher targets in Figure \ref{fig:swe-seeds-at-target}.

\vspace{-12pt}

%% file: sections/conclusions.tex
In this work, we proposed a novel methodology for regime boundary detection in discontinuous simulation models. For such instances, very often the regime switch happens across the boundary of an excursion set. We argued that since the simulator output defining such excursion set is discontinuous or non-smooth, regression-based schemes for level set estimation are inadequate and solving a related binary classification problem provides superior sample efficiency. To this aim, boundary estimation was carried out with support vector machines, and we further proposed a novel active learning strategy to mitigate for the mode-collapse behavior of uncertainty-based sampling. Our approach moves away from discrete search spaces on fixed grids and instead uses gradient-based dynamics to explore the estimated boundary tangentially. To be more precise, we derived a potential energy function from the estimated separating surface and utilized Hamiltonian Monte Carlo to foster coherent tangential motion along its boundary. Owing to the compact representation offered by support vector machines, function and gradient evaluations that need to be performed in this procedure remain computationally cheap; furthermore, we discussed how one can improve mixing, therefore allowing the additional overhead to be near-negligible compared to typical model evaluation costs for complex computational models.

In the results section, we started by showing on a series of analytical examples that the proposed approach has better sample efficiency with respect to both uncertainty-based sampling and an overly diffusive gradient-based exploration, even when both are improved by a diversity-promoting downselection. We presented examples in which this was observed in as low dimension as two (when the target boundary features high curvature), and with an increasing dimension of the parameter space, where the proposed approach can better contrast the curse of dimensionality. Subsequently, we considered a variety of increasingly complex simulation models with discontinuous outputs, and we showed that regression-based approaches are inaccurate; on the other hand, among the active classification strategies, the proposed algorithm was, in the majority of the examples, the most robust with respect to initialization and the most likely to reach a target $F_1$ score given a fixed computational budget.

To conclude, the limitations of the proposed approach should be mentioned. In very high dimensions, the computational overhead may become too large due to the increased number of support vectors and costly gradient evaluations. Slower mixing could also become an issue, although it could be mitigated with preconditioners coupled with hierarchical schemes that mitigate the increased computational cost. Finally, while we improve boundary coverage with gradient-based dynamics, the algorithm proposed in this work improves sample efficiency in the refinement of a connected component that was already identified with the initial static design, but in general it is not obvious why it would “discover” another connected component. While the framework allows for annealing/tempering on the inverse-temperature and this could provide benefits, we discussed that tuning such a parameter can be cumbersome unless a priori knowledge of some features of the true target boundary is available. Nevertheless, for many discontinuous engineering problems with expensive queries, the proposed algorithm remains computationally cheap and allows for accurate regime boundary detection with improved robustness and sample efficiency compared to existing approaches.